\documentclass[fleqn,usenatbib]{mnras}
\usepackage{newtxtext,newtxmath}
\usepackage[T1]{fontenc}
\DeclareRobustCommand{\VAN}[3]{#2}
\let\VANthebibliography\thebibliography
\def\thebibliography{\DeclareRobustCommand{\VAN}[3]{##3}\VANthebibliography}

\usepackage{flushend}
\usepackage{cuted}
\usepackage{graphicx}
\usepackage{amsmath}
\usepackage{xspace}
\usepackage{wrapfig}
\usepackage{dcolumn}
\usepackage{natbib}
\usepackage{float}
\usepackage{caption}
\usepackage{placeins}
\usepackage{rotating}
\usepackage{enumitem}
\usepackage{lastpage}
\usepackage{subcaption}
\usepackage[none]{hyphenat}
\usepackage{threeparttable}
\usepackage{multicol,amsmath}
\usepackage{newtxtext,newtxmath}
\usepackage[table,x11names]{xcolor}
\usepackage{multirow, longtable, supertabular, adjustbox}
\usepackage{tablefootnote}
\newcommand{\swift}{{\em Swift}\xspace}

\newcommand{\tninty}{{T$_{90}$}\xspace}
\newcommand{\Ep}{E$_{\rm peak}$\xspace}
\newcommand{\sw}[1]{\texttt{#1}}
\title[A Comparative Study of UVOT and XRT Data]{Investigating Temporal Features in \swift GRB Afterglows: A Comparative Study of UVOT and XRT Data}

\author[Amit K. Ror et al.]{
Amit K. Ror,$^{1,2}$\thanks{E-mail: mitturor77894@gmail.com}
S. B. Pandey$^{1}$,
S. R. Oates$^{3}$,
R. Gupta$^{1,4,5}$,
A. Aryan$^{1,6}$,
A. J. Castro-Tirado$^{7,8}$,
Sudhir Kumar$^{2}$\\
$^{1}$Aryabhatta Research Institute of Observational Sciences (ARIES), Manora Peak, Nainital-263002\\
$^{2}$Department of Applied Physics/Physics, Mahatma Jyotiba Phule Rohilkhand University, Bareilly-243006\\
$^{3}$Physics Department, Lancaster University, Bailrigg, Lancaster LA1 4YB, UK\\
$^{4}$Astrophysics Science Division, NASA Goddard Space Flight Center, Mail Code 661, Greenbelt, MD 20771, USA\\
$^{5}$NASA Postdoctoral Program Fellow \\
$^{6}$Graduate Institute of Astronomy, National Central University, 300 Jhongda Road, 32001 Jhongli, Taiwan\\
$^{7}$Instituto de Astrof\'isica de Andaluc\'ia (IAA-CSIC), Glorieta de la Astronom\'ia s/n, E-18008, Granada, Spain\\
$^{8}$Ingeniería de Sistemas y Autom\'atica, Universidad de M\'alaga, Unidad Asociada al CSIC por el IAA, Escuela de Ingenier\'ias Industriales,\\ Arquitecto Francisco Pe\~nalosa, 6, Campanillas, 29071 M\'alaga, Spain
}

\date{Accepted XXX. Received YYY; in original form ZZZ}
\pubyear{\the\year{}}

\begin{document}
\label{firstpage}
\pagerange{\pageref{firstpage}--\pageref{lastpage}}
\maketitle

\begin{abstract}
This study presents a statistical analysis of optical light curves (LCs) of 200 UVOT-detected GRBs from 2005 to 2018. We have categorised these LCs based on their distinct morphological features, including early flares, bumps, breaks, plateaus, etc. Additionally, to compare features across different wavelengths, we have also included XRT LCs in our sample. The early observation capability of UVOT has allowed us to identify very early flares in 21 GRBs preceding the normal decay or bump, consistent with predictions of external reverse or internal shock. The decay indices of optical LCs following a simple power-law (PL) are shallower than corresponding X-ray LCs, indicative of a spectral break between two wavelengths. Not all LCs with PL decay align with the forward shock model and require additional components such as energy injection or a structured jet. Further, plateaus in the optical LCs are primarily consistent with energy injection from the central engine to the external medium. However, in four cases, plateaus followed by steep decay may have an internal origin. The optical luminosity observed during the plateau is tightly correlated with the break time, indicative of a magnetar as their possible central engine. For LCs with early bumps, the peak position, correlations between the parameters, and observed achromaticity allowed us to constrain their origin as the onset of afterglow, off-axis jet, late re-brightening, etc. In conclusion, the ensemble of observed features is explained through diverse physical mechanisms or emissions observed from different outflow locations and, in turn, diversity among possible progenitors.
\end{abstract}


\section{Introduction} \label{sec:intro}
Gamma-ray bursts (GRBs) are extremely energetic (up to 10$^{55}$\,erg) and fascinating cosmic ($0 < z \lesssim 20$; \citealt{2000ApJ...536....1L}) events currently having the highest confirmed redshift of $z=9.4$, thanks to the efficient instruments onboard \swift \citep{2011ApJ...736....7C, Gehrels_2008}. GRBs are characterised by two distinct emission phases, i.e., prompt emission and afterglow \citep[see for review,][]{2015PhR...561....1K}. The initial burst, or prompt emission phase, involves a sudden release of energy, mostly in soft $\gamma$-rays or hard X-rays, originating within an ultra-relativistic jet driven by a not-yet-understood central engine (Black hole or Magnetar; \citealt{1993ApJ...405..273W, 1997ApJ...482L..29M, 2009MNRAS.396.2038B, 2025MNRAS.538L.100C}). The jet's compositions, whether baryonic or dominated by magnetic flux, are still uncertain, adding challenges in understanding the emission mechanisms responsible for the prompt emission phase \citep{2015AdAst2015E..22P, 2018NatAs...2...69Z, Gupta_2024, 2023MNRAS.519.3201C}. Theoretically, a baryonic-dominated jet is assumed to consist of multiple shells with varying Lorentz factors, and the collisions between these shells generate non-thermal prompt emission through internal shock processes \citep{1994ApJ...430L..93R, 1997ApJ...490...92K}. In a Poynting flux-dominated jet, the Internal-Collision-induced Magnetic Reconnection and Turbulence (ICMART) is one of the leading mechanisms proposed to explain the prompt emission phase \citep{1994MNRAS.267.1035U, 2002ApJ...566..712Z}. The duration of the bursting phase (\tninty) lasts from milliseconds to several minutes. The \tninty parameter provides a measure of the duration, defined as the time over which 90\% of the total fluence is collected. Within a few seconds, the transition from prompt emission to the broadband afterglow phase occurs. The afterglow emission occurs when the jet slows down due to its interaction with the surrounding medium, causing an external shock in the medium. The shock-accelerated electrons interact with the magnetic field in the surrounding medium to emit synchrotron photons \citep{1992MNRAS.258P..41R, 1997ApJ...482L..29M, 1997ApJ...489L..37S}. Two types of external shock are possible when a fireball intersects with the surrounding medium. Along with the forward shock, which moves through the surrounding medium and generates accelerated electrons in the medium itself, a reverse shock is also produced, which travels backward into the fireball ejecta, acting to decelerate the fireball. The forward shock model generates a light curve (LC) that initially rises and then decays, typically as a power-law or a series of power-laws \citep{Sari_1998, sbpandey, 2009MNRAS.395..490O}. The reverse shock, on the other hand, produces sharper features in early ($<$ 1000\,s) afterglow LCs across a broad spectral range (radio to optical-NIR) \citep{2003ApJ...582L..75K, 2007ApJ...655..391K, 2014ApJ...785...84J, 2023arXiv231216265G}. Some authors have also proposed the cannonball model to explain the prompt and afterglow emission, but it is not widely accepted within the GRB community \citep{2009AIPC.1111..333D}.

Before the launch of the \textit{Neil Gehrels Swift Observatory} (henceforth \swift; \citealt{Gehrels_2004}), afterglow observations were only possible at late times, typically a few hours post-burst, capturing only the late afterglow behaviour. At these times, the external forward shock model predicts that the afterglow LCs and spectra generally decay following a power-law (PL) $F$($t$,$\nu$) $\propto$ $t^{-\alpha}$ $\nu^{-\beta}$, where $\alpha$ $\sim$ 1 and $\beta$ $\sim$ 1, indicating a continuous decrease in brightness over time and frequency \citep{2002ApJ...571..779P, 2006ApJ...637..889Z}. \swift, launched in 2004, with its three instruments, Burst Alert Telescope (BAT, 15 - 150\,keV; \citealt{2005SSRv..120..143B}), X-Ray Telescope (XRT, 0.3 - 10\,keV; \citealt{2005SSRv..120..165B}) and Ultraviolet/Optical Telescope (UVOT, 1700 - 6500\,\AA;~\citealt{2005SSRv..120...95R}), can detect both prompt (in soft $\gamma$-rays or hard X-rays) and afterglow (from X-ray to ultraviolet/optical wavelengths) emission phases within 100\,s after the trigger time; the time at which BAT first detected the gamma-ray emission.
 
\subsection{X-ray light curves}
Unlike the smoothly decaying late afterglow emission, the early afterglow LCs observed in the \swift era have consistently posed challenges to existing afterglow models. The X-ray LCs observed by \swift-XRT typically exhibit canonical five-components \citep{2006ApJ...642..354Z, 2006ApJ...642..389N, 2006ApJ...647.1213O}. However, not all LCs display all five components. These features include:
(1) \sw{Steep decay phase}, characterised by a decay index \(\alpha_x \gtrsim 3\), is thought to be due to the tail of prompt emission from high latitudes relative to the observer's line of sight \citep{2005Natur.436..985T, 2007ApJ...666.1002Z, 2009ApJ...690L..10Z}. (2) \sw{Plateau or shallow decay phase} decays with \(0 \lesssim \alpha_x \lesssim 0.7\), i.e the flux observed remains almost constant \citep{1998ApJ...503..314P, 2007ApJ...662.1093W}. \cite{2007ApJ...670..565L} studied the plateau in 53 X-ray LC with a mean value of \(\alpha_{\rm x} \sim 0.35\). Similarly, in our previous work, we have found that 75/230 (33\%) \swift detected long GRBs with known redshift have plateaus in their X-ray LC \citep{2024ApJ...971..163R}. A potential explanation for this phenomenon is that the central engine may remain active during this phase, continuously injecting energy into the external shock to sustain the flux \citep{2007ApJ...670..565L, 2006A&A...460..415P, 2018ApJ...869..155S, 2024BSRSL..93..709R, 2024A&A...692A...3S}. However, in some GRBs, a plateau phase followed by a steep decay has been observed, believed to originate within the relativistic jet through internal shock, referred to as internal plateaus \citep{2007ApJ...665..599T}. One explanation for these internal plateaus is that they are caused by a magnetar central engine. It's a two-stage process; first, a magnetar is formed as the central engine of the burst, from which magnetic dipole radiation provides energy to the fireball to maintain the plateau phase. As a magnetar loses its energy, radiation pressure and centrifugal force can not support gravitational pull, it subsequently collapses into a black hole, which we observe as a steep decay in the LC \citep{2020ApJ...896...42Z}. Observations of both internal and external plateaus in the X-ray LCs of GRB 070802, GRB 090111, and GRB 120213A have supported the subsequent formation of magnetar and black hole central engine \citep{2020ApJ...896...42Z}. (3) \sw{Normal decay phase}, defined by \(0.7 \lesssim \alpha_x \lesssim 1.5\), corresponds to well-studied afterglow emission phase from the external forward shock in the surrounding medium \citep{1997ApJ...482L..29M, Sari_1998}. (4) \sw{Steep decay phase following normal decay phase}, characterised by \(\alpha_x \gtrsim 1.5\), generally corresponds to a jet break \citep{1999Sci...283.2069C, 1999Natur.398..389K}. However, in some cases, chromatic breaks in the observed multi-wavelength LCs challenge the jet break model \citep{2008ApJ...675..528L}. (5) \sw{Flares}, about 50\% of XRT LCs include flares, which are characterised by a sharp rise and decay. Flares are defined as the ($\delta t/t_{p} \ll 1$), where $\delta t$ is the width of the flare at full width at half maximum (FWHM) and $t_{p}$ is the peak time of the flare. Flares have spectral and temporal properties similar to the prompt emission, indicating their internal origin\citep{2005Sci...309.1833B, 2006ApJ...646..351L}. The decay indices conventions described above for the different segments of X-ray LCs will be applied throughout the paper to define the segments in the optical LCs as well.

\subsection{Optical light curves}
The optical LCs observed by \swift-UVOT are relatively less complex compared to the X-ray LCs. X-ray LCs typically exhibit an early steep decay phase, and 50\% of them are followed by multiple overlapping flares, which are often attributed to residual activity of the central engine. In contrast, these features are generally less prominent in the early optical LCs \citep{2013ApJ...774....2S}. However, early optical observations can still reveal noticeable features within the first few hours post-burst, such as flares due to internal or reverse shocks and bumps due to the onset of the afterglow, etc, providing valuable clues about the afterglow emission. In addition to the five components observed in X-ray LCs (plateaus, jet breaks, and flares are often observed in optical LCs, but early steep decays are rare), about 42\% of optical LCs include a rise at an early time \citep{2013A&A...557A..12Z}. Depending on their appearance and occurrence, there can be several possible reasons for the observed rise in early optical LCs, such as the onset of afterglow, passage of synchrotron characteristic frequency ($\nu_{m}$), structured jet, density fluctuation, flares, etc. \citep{2007AA...469L..13M, 2009MNRAS.395..490O, 2010ApJ...720.1513K, 2010ApJ...725.2209L, 2012ApJ...758...27L, 2013IJMPS..23..228L, 2013ApJ...774...13L, 2023ApJ...942...34R, 2023MNRAS.525.3262B, 2022ApJ...940..169D, 2024MNRAS.tmp.1527D}.

According to the fireball model \citep{1999ApJ...520..641S, 2007ApJ...655..973K}, the bump (the terms ``bump" and ``smooth bump" are used interchangeably in the paper, and both refer to the same phenomenon and defined as the smooth rise and decay in the afterglow LCs, without any noticeable spike) due to the onset of afterglow occurs when the hot fireball starts to interact with the surrounding medium. As the jet ploughs into the external medium, it sweeps up electrons, causing the emission to increase and produce a rise in the early afterglow LC. Initially, the bulk Lorentz factor ($\Gamma_{0}$) of the fireball remains constant, known as the coasting phase, but when the inertia of the collected matter becomes significant, i.e., its rest mass energy equals the kinetic energy of the fireball \citep{1999ApJ...520..641S}. At this stage, the Lorentz factor of the fireball reduces to half of its initial value, which also marks the peak time (\( t_p \)) in the early afterglow LC. Therefore, this phenomenon can be utilised to determine the Lorentz factor of the fireball \citep{2007AA...469L..13M}. The distance from the central engine at which this deceleration occurs is known as the deceleration radius (\( R_{dec} \)). The fireball then enters the self-similar solution phase \citep{1976PhFl...19.1130B}, and the optical/NIR LC starts to decay following a PL, following the predictions of the external forward shock. Early optical bumps are more common than X-ray bumps, possibly because early X-ray emission is dominated by the prompt emission tail during the steep decay phase, masking the bump in the X-ray LC \citep{2009MNRAS.395..490O, 2013ApJ...774...13L, 2024arXiv240904871G}.

In some GRBs, optical LCs exhibit late re-brightening bumps, which share a similar morphological shape with the onset bumps but occur much later, typically after 1 hour \citep{2013ApJ...774...13L}. These late re-brightenings are thought to have distinct origins, such as late energy injection from the central engine or density fluctuations in the surrounding medium \citep{2002A&A...396L...5L, 2007AdSpR..40.1186Z}. Simultaneous late-time rebrightening in the optical and X-ray LCs of 13 out of 26 GRBs, studied by \cite{2013ApJ...774...13L}, suggests a common external origin. In contrast, the remaining 13 GRBs exhibit chromatic rebrightening, which does not support a common origin. \cite{2014ApJ...785..113H} and \cite{2022MNRAS.513.2777K} have identified the late re-brightening in GRB 120326A and GRB 210204A as caused by the central engine activity, the longest known to date.

Earlier studies indicate that 26\% of optical LCs exhibit a plateau \citep{2012ApJ...758...27L}, a significantly lower percentage compared to X-ray LCs \citep{2013A&A...557A..12Z, eva09, 2024ApJ...971..163R}. This implies that not all optical LCs display plateaus corresponding to plateaus in the X-ray LCs in the same GRB afterglow. Even if both the optical and X-ray have plateau phases, they do not always show achromatic breaks with X-ray plateaus, challenge the energy injection scenario, and suggest a different origin for the observed plateaus in these GRB afterglows \citep{2012ApJ...758...27L}.

Additionally, optical LCs exhibit flares with characteristics similar to those seen in $\gamma$/X-ray LCs. \cite{2013ApJ...774....2S} found that 34\% of optical LCs consist of one or more flares, which is relatively less compared to the 50\% in X-ray LCs \citep{2006ApJ...647.1213O}. The early ($<$ 1000\,s) flares in optical LCs can be explained by considering either external reverse shocks or internal shocks. Flares correlated with prompt $\gamma$-ray and X-ray emissions are due to the erratic behaviour of the central engine and share a common internal origin. Otherwise, early flares in the optical LCs are primarily dominated by reverse shock emission \citep{2003ApJ...582L..75K, 2007ApJ...655..391K, 2014ApJ...785...84J, 2015ApJ...810..160G, 2015AdAst2015E..13G}. Combining the above-mentioned features observed in optical LCs, a synthetic optical LC with eight components was provided by \cite{2012ApJ...758...27L}.\\

Based on the above introduction, near-simultaneous observations in optical and X-rays are crucial for understanding the dynamics of the GRB fireball, its Lorentz factor, energy injection episodes, and the types of medium surrounding the burst. Therefore, in this paper, utilising the optical and X-ray observations from the \swift satellite, our aim is to understand the role of forward and reverse shocks in shaping the observed afterglow LCs. The early afterglow emissions in both optical and X-ray wavelengths are also not isolated from internal shock emission. Further, this paper also aims to study the contribution of internal shock emission in the afterglow LCs. Additionally, we examine the role of energy injection from the central engine in shaping afterglow LCs. Specifically, we explore whether late-time activity from the central engine is always required to explain observed afterglow features or whether fluctuations in the ambient medium density can produce similar effects.\\

Twenty years after the launch of \swift, it is important to review afterglow features to test existing afterglow models. A comparative analysis of several features observed by \swift-XRT has been conducted in a series of papers by \cite{2007ApJ...670..565L, 2009ApJ...707..328L, 2012ApJ...758...27L, 2013ApJ...774...13L, 2013EAS....61..203P, 2019ApJS..245....1T}. Optical afterglow has a much lower detection rate of (29\%) as compared to X-ray afterglow (96\%) \citep{2007SPIE.6686E..07B, 2009ApJ...690..163R}. Therefore, in the optical domain, earlier studies have been limited either by a small set of GRBs or by compiling afterglow LCs from various published papers that include observations from different space- and ground-based telescopes \citep{2010ApJ...720.1513K, 2013A&A...557A..12Z}. Due to the limited set of observations and sample sizes, these studies typically focused on one or two afterglow components. However, with its unique capabilities, \swift-UVOT provides unprecedented temporal resolution and coverage, enabling the exploration of features in optical LCs that were previously unattainable in the pre-\swift era \citep{2009MNRAS.395..490O, 2012MNRAS.426L..86O, 2013A&A...557A..12Z, 2023Univ....9..113O, 2023MNRAS.525.3262B, 2022ApJ...940..169D, 2024MNRAS.tmp.1527D}.\\

To achieve the above-mentioned goals, in this paper, we analyse a set of well-sampled afterglow LCs from 200 GRBs detected by \swift-UVOT between 2005 and 2018. We examine observed features in GRB afterglow LCs, leveraging combined observations from \swift-UVOT, XRT, and BAT to enhance our understanding of the underlying physical processes. This study is the first to combine optical and X-ray data for such a large set of \swift-detected GRBs.

The structure of this paper is as follows: The data reduction process for the \swift-UVOT, along with a description of the final sample and the subsequent data analysis, is detailed in \S\ref{sec:sample}. The results are provided in \S\ref{sec:results}. In \S\ref{sec:discussion}, we discuss the various features observed by \swift-UVOT and explore their possible origins. Finally, the summary of the paper is presented in \S\ref{sec:summary}.

\section{The data reduction process and the sample} \label{sec:sample}

\begin{figure}
\includegraphics[width=\columnwidth]{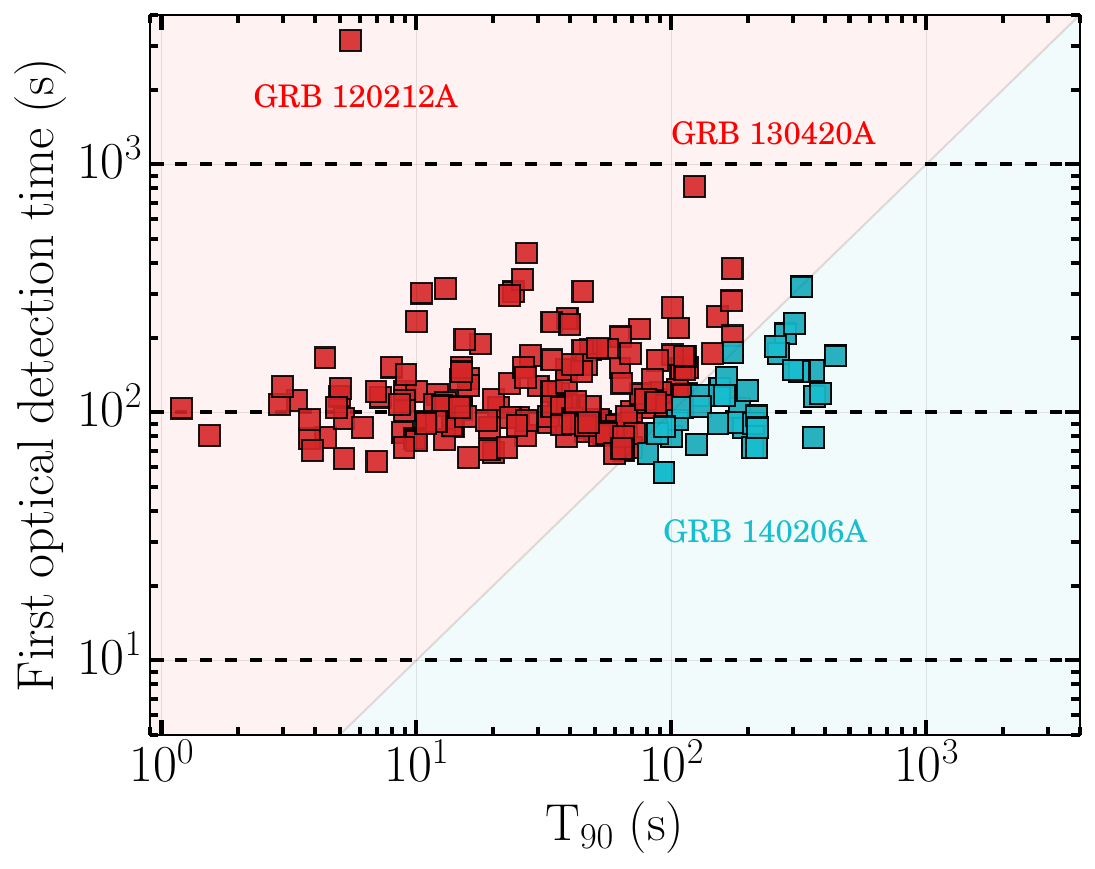}
\caption{Time of the first optical detections plotted versus their \tninty. Red squares represent GRBs with the first optical detection occurring after \tninty, while cyan squares indicate those detected before \tninty ends. The regions corresponding to \tninty less than or greater than the first optical detection are shown with the red and cyan shaded regions and separated by a thin line corresponding to \tninty = first optical detection. Additionally, three horizontal dashed lines are drawn at 10, 100, and 1000\,s. About 20\% of the GRBs were observed by UVOT while the $\gamma$-ray prompt emission was still going on.}
\label{fig:fod}
\end{figure}

\subsection{Advantage of \swift-UVOT over ground-based telescopes}
The use of space-based observation, as exemplified by the \swift satellite, offers several advantages. By avoiding the distorting effects of Earth's atmosphere, space-based telescopes can (but not always, depending on telescope size and detector quality) capture sharper, more detailed images and observe a broad electromagnetic spectrum, including optical/NIR, ultraviolet, X-rays, and $\gamma$-rays. Unlike ground-based telescopes, which are limited by atmospheric windows, longitude-latitude constraints, day-night cycles, and weather conditions, a space-based telescope can provide continuous, uninterrupted observations (although limited by the orbital constraints), resulting in multi-wavelength coverage with excellent spectral and temporal resolution. Additionally, ground-based observations from various instruments often suffer from calibration issues. \swift observations, on the other hand, are well-calibrated across all wavelengths, ensuring more reliable and accurate data. Therefore, unlike earlier studies that may have relied on observations made by several telescopes, our dataset benefits from the continuous and high temporal resolution observations of \swift-UVOT. Harnessing the precision of \swift-UVOT allows us to capture optical observations from seconds to several days after the burst, enabling in-depth exploration of GRB afterglows from the earliest to late phases.

\subsection{\swift-UVOT Data Reduction Process}
The data reduction process for \swift-UVOT is given in the \swift-UVOT data analysis guide\footnote{\url{https://www.swift.ac.uk/analysis/uvot/}} and \cite{2009MNRAS.395..490O}. Within a few seconds of the \swift-BAT trigger, \swift-XRT and UVOT immediately slew toward the burst and make the earliest possible observations of its decaying afterglow phase. UVOT utilises a more advanced observation technique than typical CCD observation. It involves photon counting with the help of photomultiplier tubes, unlike a CCD detector that accumulates photons during the exposure and provides counts at the end of the exposure in terms of analog-to-digital units. UVOT has two modes of observation. Event mode: where the detecting instrument records the temporal and spatial information of each photon. Image mode: following a predefined observation sequence, UVOT automatically switches to accumulation mode with increasing exposure time, allowing for detailed LC construction. UVOT obtains high cadence photometry in a set of seven filters (\textit{white}, \textit{v}, \textit{b}, \textit{u}, \textit{uvw1}, \textit{uvm2}, and \textit{uvw2}; \citealt{2005SSRv..120...95R, 2008MNRAS.383..627P}) and depending on the observed brightness of the GRBs by \swift-BAT, UVOT can perform spectroscopic observation, leveraging its grisms covering a wide wavelength range (1700-6500\,\AA).\\ 

As an initial step of data reduction, image astrometry was performed by cross-correlating the stars in the UVOT image, extracted in every 10\,s from the event file, with the USNO-B1 catalogue \citep{2003AJ....125..984M}. To perform photometry and to calculate the source count rate, an aperture of $5''$ was selected, while for faint sources, a $3''$ aperture was selected. For the background selection, a circular aperture of $20''$ was selected near the source; for details of the data reduction, please refer to \cite{2009MNRAS.395..490O}. Finally, to increase the signal-to-noise ratio, the LCs corresponding to different filters were normalised to \textit{v}-filter. The normalization of the LCs is only possible if there is no colour evolution during the afterglow phase. Information regarding the colour evolution of multi-band optical LCs from \swift-UVOT can be found in \cite{2009MNRAS.395..490O, 2024MNRAS.tmp.1527D}. From this process, we have around 200 count rate LCs of \swift-UVOT detected GRBs. 

\subsection{Data Analysis} \label{sec:data_analysis}

\begin{figure*}
\includegraphics[width=2\columnwidth]{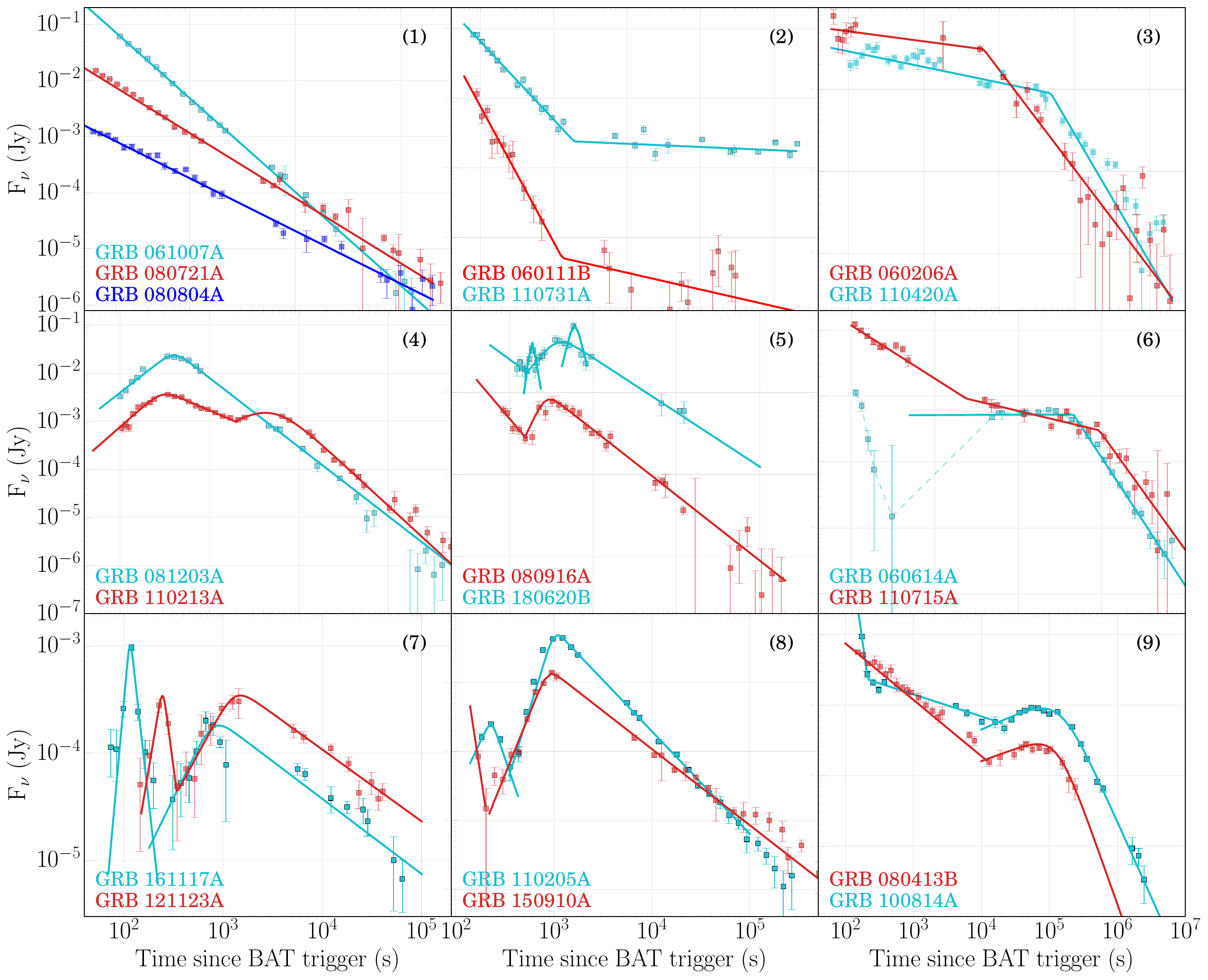}
\caption{Panel (1) to (9) represents the diverse range of distinct features such as breaks, bumps, flares, plateaus, and late re-brightening observed in \textit{v}-band LCs by the \swift-UVOT. The coloured square represents the observed UVOT data, and the corresponding fitted models are shown with solid coloured lines.}
\label{fig:opt_feat}
\end{figure*}

\subsubsection{\swift-UVOT and XRT LC fitting}
As discussed above, the data reduction process yielded 200-count rate LCs normalised to \textit{v}-filter. Utilising the zero point for \textit{v}-filter from \cite{2008MNRAS.383..627P}, we convert the count rate LC to the flux density LC for each GRB. The motivation for using flux density LCs over count rate LCs is that the flux density is more physically meaningful than the originally observed count rate LC. It allows for easier comparisons with theoretical models and enables direct comparison with the observations across UVOT, XRT, and BAT data. Further, flux density data can be used to construct spectral energy distributions across different wavelengths. To fit the flux density \swift-UVOT LCs in \textit{v}-filter, we have used the PL model with 0, 1, 2, and 3 breaks. Further, to fit the observed bump or flare, we used a smoothly joined broken PL \citep{2007AA...469L..13M}, in addition to the PL or broken PL models discussed above. These PL models with 0, 1, 2, and 3 breaks are consistent with those used to fit the XRT LCs available in the \swift-XRT catalogue \citep{eva09}. The shape of the model is illustrated in Figure 9 of \cite{eva09} and is also described in Fig. \ref{fig:fit_models} and equation \ref{eqn:eqn} in the appendix of this paper. To determine the best-fit model, we use the F-test, and an additional break is introduced only if the null hypothesis probability (p) is less than 0.0027, see Fig. \ref{fig:p_values}. For \swift-XRT and BAT, we retrieved the XRT and BAT flux density LCs at 10\,keV for each GRB from the burst analyser page \citep{eva07, eva09}. We applied a similar fitting procedure for XRT LCs as used for the UVOT LCs. When fitting the XRT LCs, overlapping flares are removed, information about which is given on the burst analyser webpage. We have not fitted the BAT observation due to its spiky nature, and use it as it is whenever required. We have utilised a Python-based \sw{emcee} \citep{2013PASP..125..306F} technique for LC fitting with 10000 iteration steps, where the first 500 steps were discarded as initial burn-in. Uncertainty estimates are provided at the 1$\sigma$ confidence level unless otherwise stated. The mathematical forms and the physical shape of these models are given in section \ref{eqn:eqn}, and Figures \ref{fig:fit_models} of the appendix, along with \swift-UVOT and XRT observations. In Fig. \ref{fig:fit_models}, the PL models with 0, 1, 2, and 3 breaks and a smoothly joined broken PL model are denoted as Pow0, Pow1, Pow2, Pow3, and SBPL, respectively. Temporal indices obtained from the fitting optical and X-ray LCs are shown in Fig. \ref{fig:indices}, \ref{fig:breaks}, and \ref{fig:box} and also listed in Table \ref{tab:pow0}, \ref{tab:pow1}, \ref{tab:onset}, \ref{tab:pow2}, \ref{tab:xpow0}, \ref{tab:xpow1}, \ref{tab:xpow2}, and \ref{tab:xpow3}.

\subsubsection{\swift-XRT spectral analysis}
Although this analysis is primarily based on the temporal analysis of \swift-UVOT and XRT observations. For completeness, we have retrieved the X-ray spectra for each GRB from the \swift burst analyser repository \citep{eva07, eva09}. The XRT operates in two observation modes: Windowed Timing (WT) for fast readout during bright afterglow phases and Photon Counting (PC) for 2D imaging of faint afterglow emission. Whenever available, we retrieved XRT spectra at three different epochs (not all GRBs have WT mode observations). The Late-time spectra, the spectra corresponding to the WT mode, and the spectra corresponding to the PC mode. We utilised an absorption \sw{Power-law} function with multiplicative absorption components \sw{PHABS} and \sw{ZPHABS} to model the \swift-XRT spectra in 0.3 - 10\,keV in \sw{XSPEC} \citep{1996ASPC..101...17A, 1999ascl.soft10005A}. For the detailed spectral analysis, please refer to \cite{2023ApJ...942...34R, 2024ApJ...971..163R}. We have extracted the late-time spectra a few thousand seconds after the burst trigger, corresponding to the phase when the afterglow becomes featureless, i.e., after the initial steep decay, flares, or plateau phase. The purpose of extracting late-time spectra is to constrain the host's neutral hydrogen column density (NH$_{z}$). The absorption column density for the Milky Way (NH$_{Gal}$) was taken from the repository \citep{eva07, eva09}. For redshift-known GRBs, we have used the given value of $z$ for spectral fitting, otherwise, a \(z=2\) is used for redshift-unknown GRBs \cite{2022JApA...43...82G}. The motivation for this choice is from the analysis results of \cite{2023Univ....9..113O}, where the redshift distribution of 477 GRBs observed by the \swift-UVOT yields an average redshift of \(z=2\). Similar results were obtained from the redshift distribution of GRBs detected by the \swift-XRT, as reported by \cite{2019ApJS..245....1T}. Although \(z=2\) is typical for long GRBs, this assumption is reasonable for our sample as it predominantly consists of long GRBs, with only two short GRBs (GRB 090426A and GRB 131004A; \citealt{2009A&A...507L..45A, pandey_2019}) included, both of which have known redshifts. Therefore, to calculate the distance-dependent properties for redshift unknown GRBs, we assume \(z=2\).

\subsubsection{Spectral energy distribution}
Finally, we created an optical-to-X-ray spectral energy distribution (SED) of 200 GRBs in our sample. Since the early WT observation is predominantly contaminated by the tail emission of the prompt phase, we utilised the spectrum obtained from the PC mode for constructing the optical-to-X-ray SED. Corresponding to the epoch of the X-ray spectrum, we used the optical flux in the \textit{v}-band. For Galactic extinction correction in the UVOT \textit{v}-band \citep{2023MNRAS.525.2701Y}, we employed the reddening map of \cite{2011ApJ...737..103S}. We have not applied the host extinction, and host extinction is assumed to be negligible. The optical-to-X-ray SED was then fitted using a PL function. A detailed description of the SED fitting process can be found in \cite{2007MNRAS.377..273S, 2010MNRAS.401.2773S}. The resulting spectral indices are listed in Tables \ref{tab:pow0} to \ref{tab:pow2}.

\subsection{The Sample}
After the data reduction and analysis/fitting process in the previous sections, our sample comprises 200 \swift-UVOT-detected GRBs. Each LC contains at least 5 optical data points sufficient to fit a PL function. The optical observations were further combined with corresponding XRT and BAT observations.
Based on traditional GRB classification using \tninty duration \citep{1993ApJ...413L.101K}, our sample consists of 2 short GRBs (\tninty $<$ 2\,s, GRB 090426A and GRB 131004A; \citealt{2009A&A...507L..45A, pandey_2019}), one ultra-long GRB (ULGRB with \tninty$>$1000\,s, GRB 111209A; \citealt{2013ApJ...766...30G, 2024ApJ...971..163R}), and the rest are long GRBs i.e \tninty $>$ 2\,s. Out of 200 GRBs, redshifts are available for 145, while for 55, there is no information. For 75 GRBs (37.5\%) in our sample, started observation in less than 100\,s. The earliest optical observation exists for GRB 140206A at around 57\,s after the burst trigger, and GRB 120212A has the most delayed optical detection at 3166\,s. The distribution of the first optical detections of GRBs along with their \tninty duration is shown in Fig. \ref{fig:fod}.

\begin{figure}
\includegraphics[width=\columnwidth]{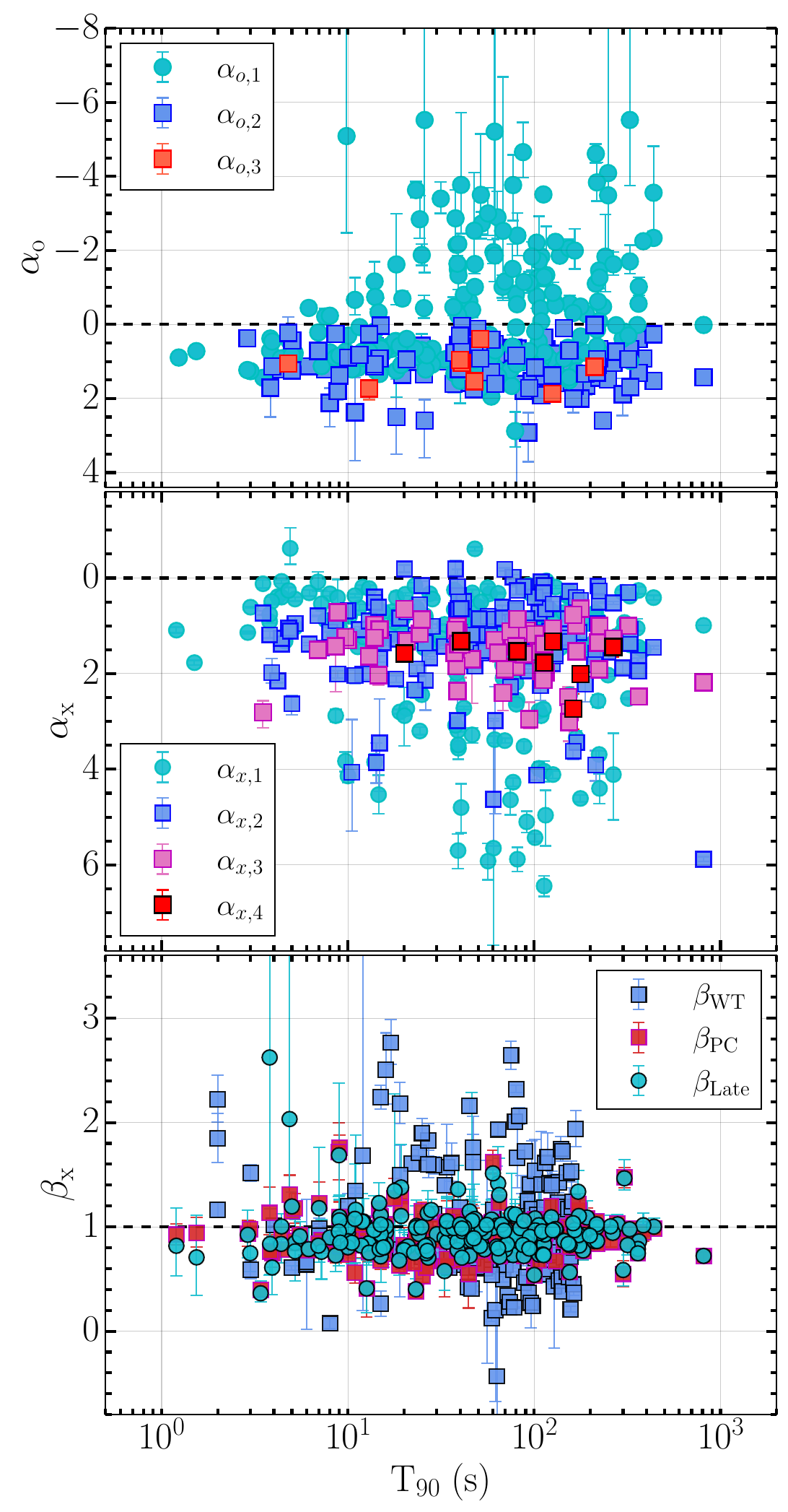}
\caption{The upper panel represents the decay indices $\alpha_{o,1}$, $\alpha_{o,2}$, and $\alpha_{o,3}$ obtained from the fitting of UVOT \textit{v}-band LCs with a smoothly joined broken PL model and PL models with 0-2 breaks. The decay indices are plotted against their \tninty duration, and a black dashed horizontal line is plotted at zero to distinguish between the rise or decay behaviour. Similarly, the middle panel represents the decay indices $\alpha_{x1}$, $\alpha_{x2}$, $\alpha_{x3}$, and $\alpha_{x4}$ of XRT LCs obtained from the fitting of PL with 0-3 breaks. The lower panel represents the spectral indices obtained from the fitting of the XRT spectrum observed in WT mode ($\beta_{x,WT}$), PC mode ($\beta_{x,PC}$) and late-time spectra always in PC mode ($\beta_{x,Late}$) in the energy range 0.3-10\,keV.}
\label{fig:indices}
\end{figure}

\begin{figure}
\includegraphics[width=\columnwidth]{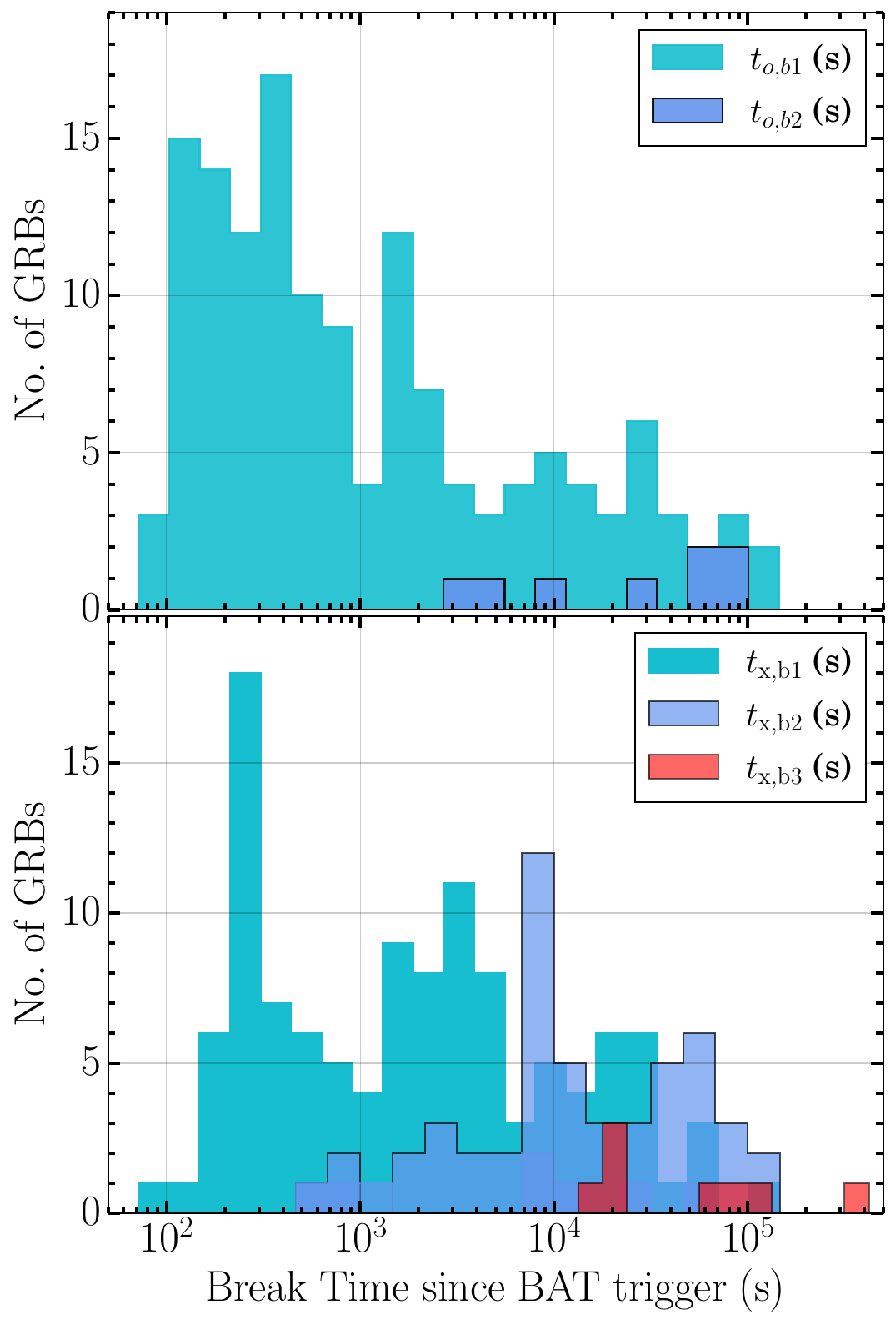}
\caption{The upper panel represents the distribution of the first ($t_{o,b1}$) and second break ($t_{o,b2}$) in the optical LCs in \textit{v}-band obtained from the fitting of PL models with 1 and 2 breaks. Similarly, the lower panel represents the distribution of breaks $t_{x,b1}$, $t_{x,b2}$, and $t_{x,b3}$ in the \swift-XRT LCs at 10 keV obtained from the fitting of PL models with 1, 2, and 3 breaks.}
\label{fig:breaks}
\end{figure}

\begin{figure}
\centering
\includegraphics[width=\columnwidth]{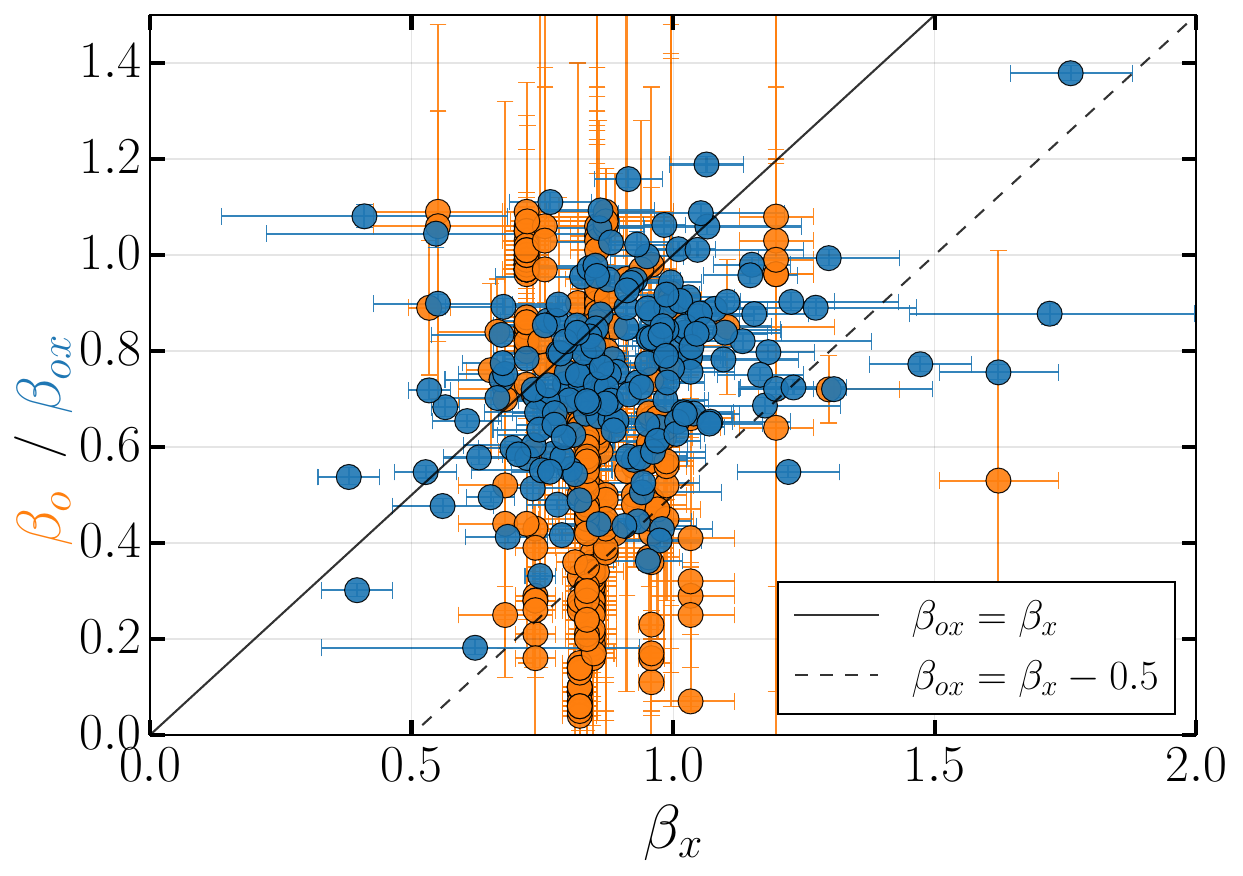}
\caption{On the X-axis, the spectral indices \(\beta_{x}\), obtained from the fitting of X-ray spectra in 0.3-10\,keV for the PC mode, are shown for all 200 GRBs in our sample. On the Y-axis, the \(\beta_{ox}\), obtained from the fitting of optical to X-ray SED of all 200 GRBs in our sample. The \(\beta_{ox}\) vs \(\beta_{x}\) plots is shown with the blue filled circles. The solid line represents the \(\beta_{ox}\) = \(\beta_{x}\) and the dashed line represents the $\beta_{ox}$ = \(\beta_{x}\) - 0.5. Optical spectral indices \(\beta_{o}\), as shown with orange colour, are taken from \citet{Dainotti_2022, 2024MNRAS.tmp.1527D}.}
\label{fig:box}
\end{figure}

\section{Results} \label{sec:results}
This section presents the results obtained from fitting the \swift-UVOT and XRT LCs and spectra in the previous sections.

\subsection{Afterglow features observed by \swift-UVOT}
Our sample includes at least nine distinct types of optical LCs. Two to three examples of each are shown in Fig. \ref{fig:opt_feat}. Although the observed features align with the synthetic features proposed by \cite{2012ApJ...758...27L}, each type exhibits unique characteristics in the arrangement of flares, bumps, and plateaus. The observed features are:

\begin{itemize}
\setlength\itemsep{0.5em}
\item In the first panel, the LCs display only a simple PL decay throughout the afterglow phase, with varying slopes, as shown for three specific GRBs: GRB 061007A \citep{2009MNRAS.395..490O, 2009ApJ...702..489R}, GRB 080721A \citep{Starling_2009}, and GRB 080804A. This behaviour is observed in approximately 76 LCs within our sample.

\item The second panel shows early optical LCs of GRB 060111B \citep{2009ApJ...702..489R} and GRB 110731A featuring an initial normal or steep decay phase followed by a plateau phase. In the introduction, we defined the early steep decay and late jet break as having $\alpha > 3$ and $\alpha > 1.5$, respectively. This LC pattern is common in the X-ray wavelength range but rare in the optical. In X-ray also, the plateau typically lasts up to 10$^3$ to 10$^4$\,s, and rarely extends to 10$^5$ \citep{2020MNRAS.492.2847B}. We only identified two clear examples of such LCs that are shown in panel 2 of Fig. \ref{fig:opt_feat}.

\item The third panel depicts LCs of GRB 060206A \citep{2009MNRAS.395..490O} and GRB 110420A, where a plateau phase ($0 < \alpha < 0.7$) is followed by a normal decay ($0.7 < \alpha < 1.5$). This pattern is observed across both optical and X-ray wavelengths. This is a common pattern found in X-ray wavelength, as 36/200 X-ray LCs consist of a plateau followed by a normal decay. It appears in 17 out of 200 LCs in our optical sample, indicating that while it's uncommon in optical, it is not entirely rare.

\item The fourth panel highlights LCs with a smooth bump during the early afterglow phase. GRB 081203A \citep{2009ApJ...697L..18K} displays a single smooth bump, while, in the case of GRB 110213A \citep{2013ApJ...774...13L}, an early smooth bump is followed by another re-brightening event. In our sample, we have 80 optical LCs that have at least a bump in the observed afterglow LCs.

\item The fifth panel presents an example of the early optical observations of GRB 080916A and 180620B, where a decay phase with slopes similar to the normal decay precedes the observed smooth bump peaks around 200\,s. The origin of the observed smooth bump can be any of the above-discussed cases. In some cases, both the PL decay and the early smooth bump in the LCs overlapped with the flares. We have observed this rare feature in the optical LCs of six bursts (GRB 080916A, GRB 130122A, GRB 140206A, GRB 140512A, GRB 161214B, GRB 180620B).

\item The sixth panel highlights the LCs of GRB 060614A \citep{2013ApJ...774...13L} and GRB 110715A \citep{2020ApJ...894L..22W}, both featuring a long plateau phase lasting up to $\sim$ 10$^{5}$\,s, followed by a steep decay ($\alpha_{o} \sim 2$) in the case of GRB 060614A and ($\alpha_{o} \sim 1.76$) in case GRB 110715A. The optical LC of GRB 060614A also includes an early steep decay ($\alpha_{o} = 4.49_{-0.50}^{+0.59}$) not connected to the plateau. A detailed discussion of these LCs is provided in section \ref{sec:internal_plateau}.

\item The LCs in the seventh panel display the double-bump features observed in GRB 121123A and GRB 161117A. The rise ($\alpha_{r} < -3$) and decay ($\alpha_{d} > 3$) indices of the first bump are too steep to be explained by external forward shock models. The shape of the second bump resembles the LCs presented in the fourth panel. Similarly, in the eighth panel, the observed LCs for GRB 110205A \citep{2011ApJ...743..154C} and GRB 150910A \citep{2020ApJ...896....4X} are similar to the LCs shown in the seventh panel, but the early flares have a relatively lower peak magnitude. For GRB 110205A, a complete flare is visible, whereas for GRB 150910A, only the decaying part of the flare is observed. We have identified 21 instances of this type of LC in our sample.

\item Finally, in the ninth panel, the LCs of GRB 080413B \citep{2011A&A...526A.113F} and GRB 100814A \citep{2014A&A...562A..29N} are shown, where an early plateau or normal decay phase, culminating in a late re-brightening bump, which is then followed by a normal decay phase. In our sample, LCs with a bump observed at very late times are classified as late re-brightening events, as discussed further in section \ref{sec:LRB}.
\end{itemize}

\subsection{Temporal and spectral Evolution}
Temporal and spectral indices obtained from fitting afterglow LCs and spectra for GRBs included in our sample are shown in Fig. \ref{fig:indices}, \ref{fig:breaks}, and \ref{fig:box} and also listed in Table \ref{tab:pow0}, \ref{tab:pow1}, \ref{tab:onset}, \ref{tab:pow2}, \ref{tab:xpow0}, \ref{tab:xpow1}, \ref{tab:xpow2}, and \ref{tab:xpow3}. The optical LC observed by \swift-UVOT generally reaches its sensitivity limit within 1-5 days \citep{2005SSRv..120...95R}, based on the source brightness. Any bumps observed within this time frame are inconsistent with typical expectations for supernovae. Generally, supernovae bump rises later than this time scale and peaks around 10-20 days after the burst trigger \citep{2014A&A...568A..19C, 2022NewA...9701889K, 2024arXiv240913391R}. Therefore, we discard the possibility of any supernova component in the observed LCs included in our sample.

\subsubsection{Temporal indices in optical LCs}
The comparison between the decay indices of optical LCs from the upper panel of Fig. \ref{fig:indices} revealed a clear distinction.

We identified that the 76/200 (38\%) GRBs are such that the observed optical LCs have shown only PL decay throughout the UVOT detection limit. All these LCs are fitted by a simple PL, and the observed decay indices of 67 LCs lie within the prediction of the forward shock model ($0.7 < \alpha < 1.5$), and 8 have shallow decay, and 1 has steep decay.

As shown in the upper panel of Fig. \ref{fig:indices}, 80/200 GRBs have an initial optical index ($\alpha_{o,1}$) that lies above the dashed line at zero, out of which 8 GRBs have shown two or more bumps. This means that around 40\% of the optical LCs rise at the beginning of the afterglow emission phase, which is consistent with the earlier results of \cite{2013ApJ...774...13L}, where they found a bump in the 64/146 optical LCs and 11 of them consist of two bumps. The upper panel of Fig. \ref{fig:indices} indicates that the probability of getting a rising slope is higher for GRBs with longer \tninty durations. These results are also consistent with the findings of \cite{2019MNRAS.488.2855P}.

Further, in our sample, we have 44/200 (22\%) optical LCs with at least one break, 36 LCs have one break, and 8 have two breaks. Out of which 44 LCs, 22 have shown shallow decay or plateau phase ($\alpha < 0.5$). Therefore, in total, there are 30 (15\%, i.e., 8 from simple PL, 19 from one break PL, and 3 from two break PL) optical LCs in our sample, where the shallow decay phase was observed. The rest of the LCs have initial or late decay indices that lie within the prediction of a forward shock model. 

21 GRBs have shown flares or steep decay at the start of the LC, before any bump, plateau, or normal decay is observed. These flares and steep decay in optical LCs do not always align with the steep decay and flares in X-ray LCs, discussed in more detail in section \ref{sec:RS}.

A distribution of GRBs within different sub-groups is shown in Figure \ref{fig:pie}. LCs with multiple features, such as a PL followed by a bump or a PL followed by a plateau, will be grouped under bumps and plateaus. For example, GRB 080413B and 100814A, where a normal decay phase is followed by the late re-brightening. Additionally, in GRBs such as GRB 060607A, GRB 060729A, GRB 070318A, GRB 100901A, GRB 110213A, and GRB 151027A, both the onset and rebrightening bump are observed. In the case of GRBs 060526A and 111209A, the plateau phase is followed by the late re-brightening. The 21 GRBs with an early flare followed by a PL, plateau, or bump are also included in their respective groups.

\subsubsection{Temporal indices in X-ray LCs}
Many of the X-ray LCs, on the other hand, show a steep decay in the early times. A possible reason for this is that, during the early steep decay phase in the X-ray LC, the high latitude prompt emission in X-rays dominates over the external afterglow emission at the early phase. A test for identifying early steep decay as tail emission is that the decay index should be close to \( \alpha_{x} \sim 2 + \beta_{x} \) \citep{2000ApJ...541L..51K}. However, observing high \( \alpha_{x} \) values is not uncommon in XRT LCs, with previous analyses indicating values up to \( \alpha_{x} \sim 10 \) are often observed \citep{2007AdSpR..40.1186Z}. One possible explanation is that some early X-ray LCs contain early flares, with the decay of these flares overlapping with the initial steep decay. Consequently, the very steep decay observed in the early X-ray LC likely results from a combined effect of internal shock emission and abrupt cessation of the central engine.\\

The early steep decay phase in X-ray LCs is generally followed by a plateau or normal decay phase. Unlike the optical LCs, early rises are rare in the X-ray LCs, only 3/200 X-ray LCs display an early rise. Further, following normal or shallow decay phases, a late steep decay phase is also observed in several X-ray LCs, most probably due to the geometrical jet break effect. Further, the plateau detection rate (70/200 $\sim$ 35\%) is much higher in X-ray compared to 15\% in the optical, calculated from our sample, consistent with the previous analysis of \cite{2007ApJ...670..565L, 2024ApJ...971..163R}. In the XRT LCS, 61/200 (30.5\%) are consistent with the simple PL, 79/200 (39.5\%) LCs exhibit a 1-break PL model, 52/200 (26\%) show two breaks, and 8/200 (4\%) show three breaks. The distribution of the F-test statistic, showing the distribution of the best-fit model, is presented in Fig. \ref{fig:p_values}. 

The comparison above shows that approximately 70\% of the XRT LCs exhibit one or more breaks, whereas only 22\% of optical LCs display this behaviour. Note that 156 optical LCs have no breaks, 76 exhibit only a single PL, and 80 LCs exhibit one or more bumps, again followed by a simple PL. Hence, more than one break is more common in X-rays than in optical LCs, as shown in the middle panel of Fig. \ref{fig:indices} and in Fig. \ref{fig:breaks}. Further, consistent with the analysis of \cite{2012ApJ...758...27L}, we obtained a comparatively smaller fraction of the plateau and flare in the optical LCs compared to the XRT LCs. Additionally, we have found that 40\% of optical LCs show at least a bump, which is consistent with the finding of \cite{2013ApJ...774...13L}, where 43\% of optical LCs showed the same features. Both our analysis and \cite{2013ApJ...774...13L} discard the bump in the early X-ray LCs due to the onset of afterglow. However, some late X-ray LCs show a simultaneous late re-brightening bump with optical LCs. 

This indicates that even though the eight-component synthetic optical LC by \cite{2012ApJ...758...27L} can be more complex than canonical 5 component X-ray LCs. However, our analysis results reveal that for a single GRB, the observed optical LC is simpler to model than the X-ray LC. In other words, X-ray LCs are more likely to experience multiple transitions (i.e., steep decay $\rightarrow$ normal $\leftrightarrow$ plateau phase $\rightarrow$ jet break) in their decay phases compared to optical LCs. This is expected as the X-ray emissions from external shock are predominantly contaminated by internal shock and energy ejection mechanisms.

\subsubsection{X-ray spectral indices}
\begin{figure*}
\centering
\includegraphics[width=1.8\columnwidth]{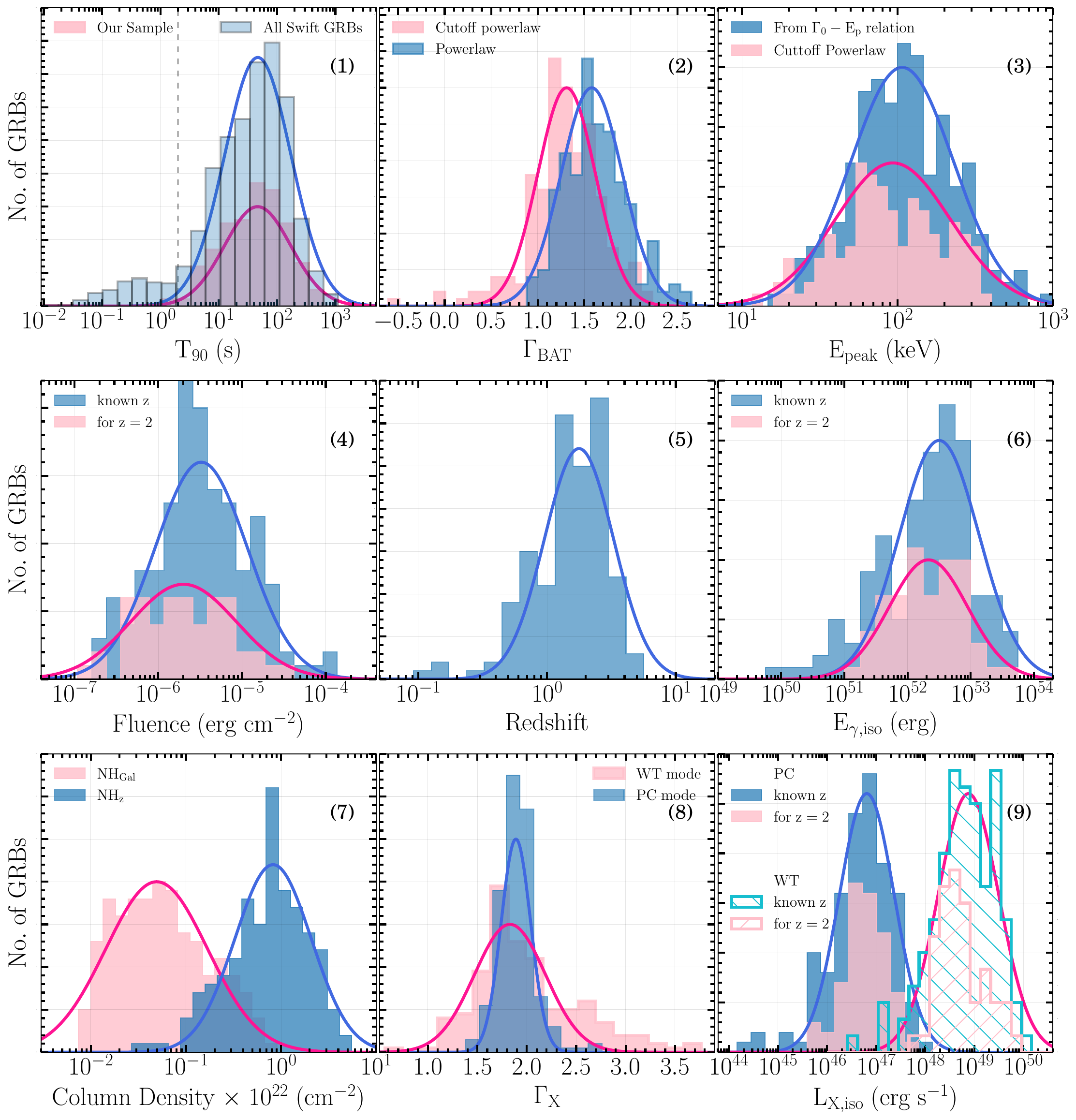}
\caption{Represents the distribution of prompt emission and X-ray afterglow properties of GRBs included in our sample. Panel 1 shows the distribution of \tninty for all GRBs detected by \swift (light blue) and for 200 GRBs in our sample, depicted in pink. Panels 2 and 3 display the distributions of spectral fit parameters, $\Gamma$ and \Ep. Here, $\Gamma$ obtained from the PL and cutoff PL models are shown in light blue and pink, respectively. \Ep from the cutoff PL is illustrated in pink, while that derived from the $\Gamma$-\Ep relation is shown in blue. Panels 4, 5, and 6 present the distributions for fluence, redshift, and \( E_{\gamma, \rm iso} \), with redshift-known GRBs in blue and redshift-unknown GRBs (assuming $z=2$) in pink. Panels 7, 8, and 9 illustrate parameters from X-ray spectral fitting. Panel 7 shows the neutral hydrogen column density contributions from the Milky Way (pink) and the host galaxy (light blue). Panel 8 represents the photon indices from WT spectra (pink), and from PC spectra in light blue. In panel 9, X-ray luminosity is depicted. The distributions are fitted with Gaussian functions, shown as red or blue lines overlaid on each distribution, with legends indicating the colour coding and corresponding fitted models.}
\label{fig:col_prop}
\end{figure*}

Our analysis mainly focuses on the temporal properties of GRBs. However, for completeness, we have also conducted a spectral analysis of the GRBs in our sample. As discussed in section \ref{sec:data_analysis}, whenever available, we have retrieved the X-ray spectra for each GRB at three different epochs (corresponding to WT mode, PC mode and Late time spectra) from the \swift burst analyser repository \citep{eva07, eva09}. The distribution of the column density for host galaxies obtained for the fitting of late-time spectra, along with the column density for the Milky Way, is shown in panel (7) of Fig. \ref{fig:col_prop}. We have obtained the mean value for column density for the host galaxy NH$_{z}$ = (0.91 $\pm$ 0.87) $\times$ 10$^{22}$ cm$^{-2}$ and Milky Way NH$_{Gal}$ = (0.13 $\pm$ 0.12) $\times$ 10$^{22}$ cm$^{-2}$ from the distributions of column densities in Fig. \ref{fig:col_prop}. The spectral indices obtained from the fitting of XRT spectra at epochs corresponding to late time, PC, and WT mode are shown in the lower panel of Fig. \ref{fig:indices} and panel (8) of Fig. \ref{fig:col_prop}. By fitting a Gaussian function to the distribution of spectral photon indices ($\Gamma_{x}$ = 1+$\beta_{x}$) for X-ray in the energy range 0.3-10\,keV, we obtained the mean values $\beta_{x, Late}$ = 0.89 $\pm$ 0.14, $\beta_{x,PC}$ = 0.89 $\pm$ 0.15, and $\beta_{x,WT}$ = 0.83 $\pm$ 0.36, respectively, for three spectra at a late time, PC, and WT mode. Finally, the flux obtained from the XRT spectral fitting is used to constrain the X-ray luminosity (L$_{\rm X,iso}$), as shown in panel (9) of Fig. \ref{fig:col_prop}. We obtained mean values of log(L$_{\rm X,iso}$/erg s$^{-1}$) = 46.81 $\pm$ 0.56 and 48.88 $\pm$ 0.62 for PC and WT mode, respectively. The spread in the distribution of $\Gamma_{WT}$ and a higher value of L$_{\rm X,iso}$ for WT mode indicates the presence of the prompt emission tail characterised by a steep decay phase generally observed in WT mode spectra.

\subsubsection{Optical to X-ray spectral energy distribution}
In this section, we present the results of the fitting of optical-to-X-ray SED performed in section \ref{sec:data_analysis}. The distribution of spectral index $\beta_{ox}$ obtained from the fitting of optical to X-ray SED vs the spectral index $\beta_{x}$ obtained from the fitting of X-ray spectra data in PC mode is shown in Fig. \ref{fig:box}. The spectral indices ($\beta_{o}$) for the optical spectra are taken from \cite{2024MNRAS.tmp.1527D} and are shown in the background, along with the $\beta_{ox}$ and $\beta_x$.

The synchrotron emission is characterised by three characteristic frequencies: the absorption frequency ($\nu_{a}$), typical synchrotron emission frequency ($\nu_{m}$), and cooling frequency ($\nu_{c}$). $\nu_{a}$ is important in the radio regime and does not affect the spectrum at optical and X-ray wavelengths. The spectral shape of the synchrotron spectrum at optical and X-ray wavelengths is defined by the position of characteristic frequencies $\nu_{m}$ and $\nu_{c}$. Therefore, the temporal and spectral indices in the afterglow emission are related to each other via some relations known as the closure relations \citep{Sari_1998}. These closure relations help us to determine the spectral regime from which the emission is coming. Based on the spectral regime and the obtained values of spectral indices and temporal indices, we can constrain the electron energy distribution index ($p$) and the type of ambient medium surrounding the burst. By fitting a Gaussian function to the distribution of $p$, we obtained a mean value of \( p = 2.47 \pm 0.37 \) for our sample. The resulting values for $p$ and the surrounding medium are listed in Tables \ref{tab:pow0} to \ref{tab:pow2}.

It is assumed that for $\beta_{ox}$ = $\beta_{x}$, both optical and X-ray emission are coming from the same spectral regime. For $\beta_{ox}$ $<$ $\beta_{x}$ - 0.5, the cooling frequency must lie within the X-ray and optical band. In Fig. \ref{fig:box}, we have shown the spectral indices $\beta_{ox}$ and $\beta_{x}$. The obtained $\beta_{ox}$ either very close to $\beta_{x}$ or lies in the range $\beta_{x}$-0.5 $<$ $\beta_{ox}$ $<$ $\beta_{x}$ within the error bars, as shown by the solid and dashed lines in \ref{fig:box}. This indicates that, for most of the GRBs in our sample, both the optical and X-ray emission lie in the same spectral regime, and for other GRBs, synchrotron cooling $\nu_{c}$ may lie between the two wavelengths. Further, in our sample, all GRBs have bright optical emission, and the optical emission mostly satisfies the closure relations with the observed X-ray emission.

\subsection{Collective properties of GRBs in our sample}
For completeness, we have also calculated the fluence and energy along with the temporal and spectral properties of the prompt and afterglow phases for each GRB included in our sample, using our analysis from previous sections. As we discussed in the data analysis section, we have not fitted the LCs and spectra for the BAT. Therefore, the spectral and temporal properties and the observed flux in the \swift-BAT energy range are taken from the BAT-GRB-Catalogue\footnote{\url{https://swift.gsfc.nasa.gov/results/batgrbcat/}} \citep{2016ApJ...829....7L}. The distribution of prompt emission and X-ray afterglow properties are shown in Fig. \ref{fig:col_prop}. As shown in panels 1-9 of Fig. \ref{fig:col_prop}, the fluence, redshift, \( E_{\gamma, \rm iso} \), and \( L_{X, \rm iso} \) obtained by assuming $z=2$ for redshift unknown GRBs perfectly align with the distribution of redshift known GRBs. The \tninty distribution of GRBs in our sample aligned well with the \tninty distribution of all \swift-detected GRBs, as shown in panel (1) of Fig. \ref{fig:col_prop}. Fitting of the Gaussian function to the log-normal distribution of \tninty in our sample and \tninty distribution of all \swift-detected long GRBs results in the same mean value for both the sample, log(\tninty/s) = 1.67 $\pm$ 0.57 \citep[see also][]{2024ApJ...971..163R}, the error bars are given at 1$\sigma$ level. \swift-BAT spectra in the energy range 15-350\,keV are commonly fitted with two empirical functions, a single PL and a PL with an exponential cutoff, i.e., \sw{Cutoff Power-law}. The mean values of the BAT photon index obtained from fitting the \sw{Power-law} model and \sw{Cutoff Power-law model}, respectively, are $\Gamma$ = 1.58 $\pm$ 0.33 and $\Gamma$ = 1.31 $\pm$ 0.31. The lower mean value of the photon index for the \sw{Cutoff Power-law} model suggests that this model generally predicts a harder spectrum compared to the \sw{Power-law} model. The distribution of the spectral peak energy (\Ep) of the \sw{Cutoff Power-law} model results in a mean value of log(\Ep/keV) = 1.97 $\pm$ 0.37. Since BAT is a soft energy instrument, it can only constrain \Ep between 15-150\,keV in a coded mask or 15-350\,keV energy range without a coded mask. Therefore, we utilise another method to constrain \Ep using equation (3) of \cite{2007ApJ...670..565L}. In this case, we get an as-expected higher mean value of log(\Ep/keV) = 2.03 $\pm$ 0.33. Further, the distribution of fluence observed in the 15-150\,keV energy range results in the mean value of log(Fluence/erg cm$^{-2}$) = -5.49 $\pm$ 0.55 and the mean isotropic energy release in $\gamma$ rays calculated from observed fluence is log(E$_{\rm \gamma, iso}$/erg) = 52.50 $\pm$ 0.63, calculated for 145 GRBs with known redshift shown in panel 5. For GRBs with unknown redshift, we have obtained similar mean values, log(Fluence/erg cm$^{-2}$) = -5.70 $\pm$ 0.63 and log(E$_{\rm \gamma, iso}$/erg) = 52.33 $\pm$ 0.62 by considering $z$ = 2. The distributions of temporal and spectral properties such as \( T_{90} \), \( \Gamma \), \( E_{\rm p} \), and the energy \( E_{\gamma, \rm iso} \), do not display any outliers. This indicates that our sample includes typical short and long GRBs with average temporal and spectral properties, as well as total energy values similar to those generally observed in short and long GRBs.

\begin{figure*}
\includegraphics[width=2.2\columnwidth]{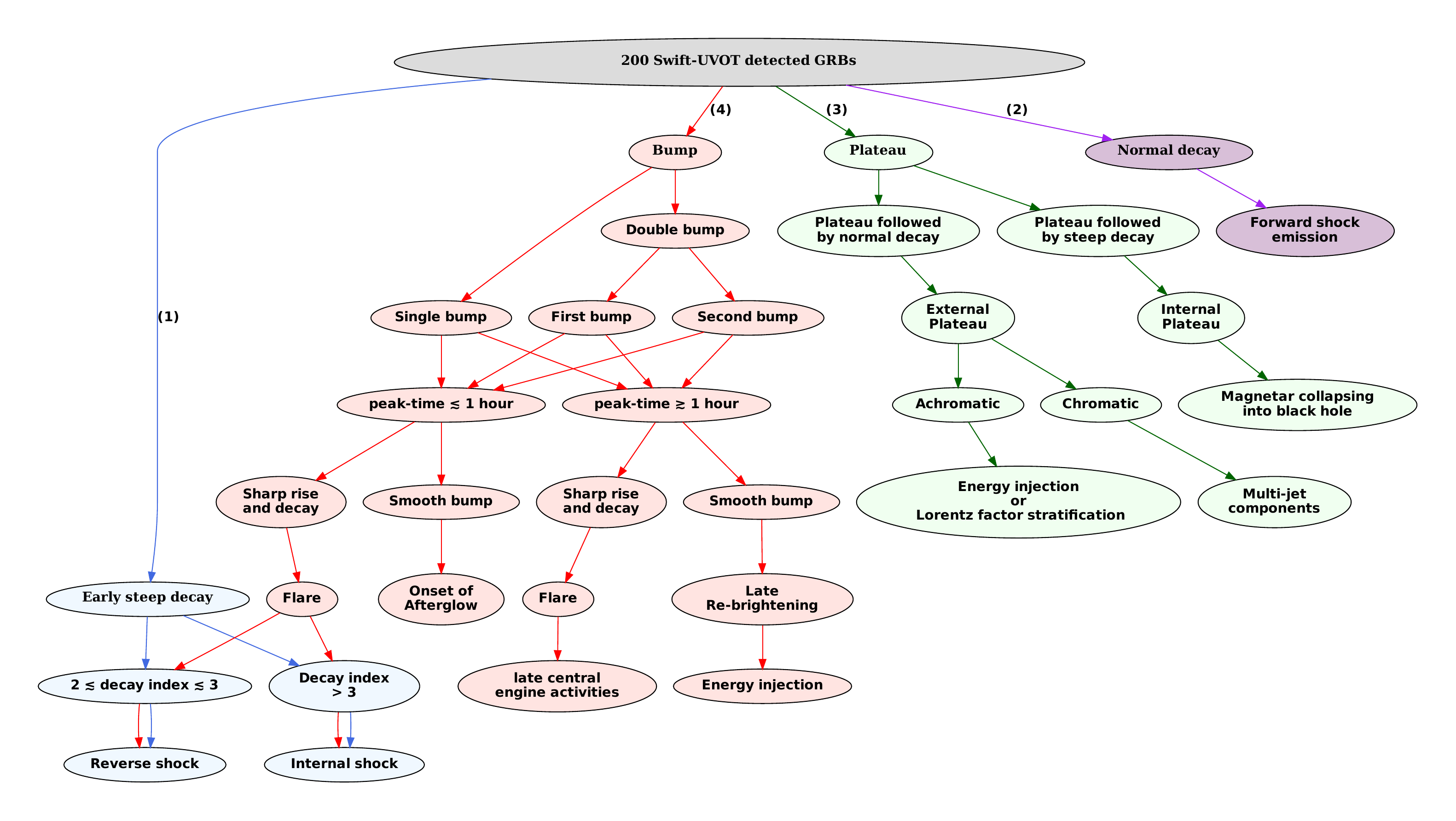}
\caption{Represents the flow chart of various characteristic features observed in \swift-UVOT \textit{v}-band LCs and their possible origin. The flow chart consists of four main branches corresponding to the observed optical LCs exhibiting a very initial flare or steep decay (1, blue branch), a simple PL behaviour throughout (2, purple branch), a plateau phase (3, green branch), and a bump in the optical LC (2, red branch).}
\label{fig:features}
\end{figure*}

\section{Discussion} \label{sec:discussion}
Early optical observations by \swift-UVOT have revealed several notable features. To guide the sequence of the present analysis, a flowchart about these observed features is shown in Fig. \ref{fig:features}. In the flow chart, we have shown four different branches, labelled as (1), (2), (3), and (4). A brief discussion of each branch is given below.

\begin{itemize}
\setlength\itemsep{0.5em}
\item The first branch, as shown with blue colour, represents the GRBs with optical LC consisting of a very early flare or steep decay preceding the early smooth bump, normal decay, or plateau phase. Early flare in the optical LCs generally occurs within 1000\,s. Flares due to internal shocks are often shorter, likely within \tninty duration of the burst, while reverse shock flares may extend longer than \tninty duration of the burst \citep{2015AdAst2015E..13G}. Although not shown in the flowchart, the first branch should be followed by one of the (2), (3), or (4) branches. The LCs with very early steep decay are shown in Table \ref{tab:ESD2}, Table \ref{tab:ESD1}, and Fig. \ref{fig:RS2}. We have constrained the origin of very early decay by comparing the observed value of decay indices with the theoretical predictions of decay indices of either reverse shock or internal shock, a detailed discussion of this comparison is provided in section \ref{sec:RS}.

\item The second branch, as shown with purple colour, represents the GRB with optical LC showing only simple PL decay throughout the afterglow phase, regardless of the behaviours of X-ray LC, as shown in Fig. \ref{fig:pow0_plots}. GRBs with optical LCs decay following a simple PL with decay index $\sim$ 1 are consistent with the forward shock in the external medium \citep{Sari_1998}. Therefore, the first branch ends with the remark of forward shock emission. PL decay due to external forward shock persists up to 10$^4$ to 10$^6$\,s. Typically followed by a jet break in the afterglow LCs, which steepens the rate of decay due to sideways expansion \citep{1999Sci...283.2069C, 1999Natur.398..389K}. If there are deviations from the normal decay phase, we require additional components to explain the observed afterglow LCs, a detailed explanation of which is given in section \ref{sec:SPL}.

\item The third branch is shown with green colour representing the GRBs with at least a plateau phase ($\alpha \leq 0.5 $) in the observed optical LC, as shown in Fig. \ref{fig:plateau_GRBs0} and \ref{fig:plateau_GRBs1}. The observed plateau phase usually starts within or after a few hundred seconds of the trigger and can last until 10$^4$–10$^5$\,s post-trigger \citep{2020MNRAS.492.2847B}. The observed shallow decay or plateau phase can not be explained through the standard external forward shock model and requires additional components. There are two types of plateaus observed in GRBs. A plateau followed by a normal decay phase is known as an external plateau. On the other hand, internal plateaus are followed by a steep decay phase \citep{2007ApJ...665..599T, 2007ApJ...670..565L, 2020ApJ...896...42Z}. A detailed discussion is provided in section \ref{sec:plateau}.

\item The fourth branch, as shown with the red colour, represents the GRBs with at least one smooth bump in the optical LC. Multi-band LCs of some of these GRBs are shown in Fig. \ref{fig:sbpl_plots} and \ref{fig:LRB_plots}. Depending on the shape, peak time, correlation with $\gamma$/X-ray LCs, and chromaticity of the observed bump in different UVOT bands, these can be further explained due to the onset of afterglow, off-axis jet, the passage of synchrotron frequency, late brightening, etc. The bump due to the onset of afterglow in the optical LCs often appears within the first hour of the burst trigger, based on the deceleration time of the blast wave, which depends on the total energy of the fireball, the initial Lorentz factor, and the ambient density. The late re-brightening bump appears much later in the afterglow, usually after one hour or more, and can last up to a few days after the initial burst \citep{2010ApJ...725.2209L, 2013ApJ...774...13L}. A detailed discussion of which is given in section \ref{sec:onset} and section \ref{sec:LRB}.
\end{itemize}

\subsection{Origin of very early flares: internal shock or external reverse shock} \label{sec:RS}
Flares refer to sharp rises and decays in the afterglow LC, with a characteristic width ($\Delta$t) at full width at half maximum (FWHM), where $\Delta t/t_{p} \ll 1$, and $t_{p}$ represents the peak time of the flare. In this section, we try to constrain the origin of early flares by relating them to the theoretical predictions of external reverse or internal shock. Reverse shock emission starts very early, sometimes before the end of the prompt emission phase, characterised by a sharp rise and decay (flares) in the early optical/NIR LC. A reverse shock travelling backwards through the fireball ejecta causes it to decelerate. Theoretically, there are two possible cases of an early reverse shock: relativistic (thick shell) or non-relativistic (thin shell) \citep{1995ApJ...455L.143S, 1999ApJ...513..669K}. In the thin shell case, the deceleration time of the fireball is longer than the GRB duration, i.e., when the observed peak time of the flare $t_{p}$ $>$ \tninty, and in the thick shell when $t_{p}$ $<$ \tninty. The shape of the optical LCs from the reverse shock depends on the profile of the medium surrounding the burst. Two types of surrounding medium are possible: an ISM-like medium where density remains constant, or a WIND medium where density is generally inversely proportional to the square of the distance from the central engine \citep{2000ApJ...536..195C}. In the intermediate case, a stratified medium is considered where the density profile is given by n(r) $\propto$ r$^{-k}$, where 0 $<$ k $<$ 2. In the ISM-like medium, reverse shock emission is expected to rise with slope $\alpha_{r}$ $\sim$ -5 and $\alpha_{r}$ $\sim$ -1/2, respectively, for thin and thick shell case and decrease with the decay index $\alpha_{r}$ $\sim$ 2. For wind-like media, the rise slope $\alpha_{r}$ $\sim$ -1/2 for both thin and thick shell cases, and the decay slope is $\alpha_{r}$ $\sim$ 3 \citep{2003ApJ...582L..75K, 2015AdAst2015E..13G}. Further, reverse shock emission dominates at lower frequencies (radio, NIR/optical), and the observed features must be uncorrelated with the corresponding X-ray or $\gamma$-ray LCs \citep{2000ApJ...545..807K, 2003ApJ...582L..75K, 2015AdAst2015E..13G, 2021MNRAS.505.4086G}. While in some GRBs, the decay indices of very early flares are very steep and inconsistent with the prediction of external reverse shock, and also, the observed optical flares coincide with the flares in $\gamma$/X-ray LCs. In these GRBs, the early steep decay is assumed to have a different origin, such as internal shock emission. Flares caused by reverse or internal shocks are observed very early in the optical afterglow, with most peaking below 200\,s, as shown in Tables \ref{tab:ESD2} and \ref{tab:ESD1}. Capturing these flares requires the observing instrument to slew rapidly toward the burst, a challenge often beyond the capabilities of ground-based instruments. However, there are exceptions, the robotic telescopes like Burst Observer and Optical Transient Exploring System (BOOTES; \citealt{Castro_1996, Castro_1999, 2004AN....325..679C}), Telescopes a Action Rapide pour les Objets Transitoires (TAROT; \citealt{2009AJ....137.4100K}), Mobile Astronomical System of TElescope Robots (MASTER; \citealt{2010AdAst2010E..30L}), Corrector de Optica Activa y de Tilts al Limite de dIfraccion (COATLI; \citealt{2016SPIE.9908E..5OW}), Reionization and Transients InfraRed (RATIR; \citealt{2012SPIE.8446E..10B}), Robotic Optical Transient Search Experiment (ROTSE-III; \citealt{2009ApJ...702..489R}), etc., with their rapid slew capabilities are able to detect such early features. Therefore, it is not always possible for detecting instruments to detect the complete flare or even the decaying part of the flares caused by reverse shock emission. However, the unprecedented slewing capabilities of \swift-UVOT have allowed us to observe these flares from a handful of GRBs. In our sample, 21 GRBs exhibit signatures of early flares in their early optical LCs. Out of which, 15 GRBs show only early decay slopes, while 6 GRBs display complete flares, as shown in the 8th panel of Fig. \ref{fig:opt_feat} and in Fig. \ref{fig:RS2}. The behaviour of the rest of the optical LCs is not determined by the presence of these initial steep decays or flares. There are three methods to identify the reverse shock origin of the observed optical LC \citep{2015AdAst2015E..13G}:

\begin{table}
\centering
\begin{tabular}{|c|c|c|c|c|}\hline
~~~ GRB ~~~ & $\alpha_{r}$ & ~~~~ $t_{p}(s)$ ~~~~ & ~~~ $\alpha_{d}$ ~~~ & ~~~~~~~ Origin ~~~~~~~ \\ \hline
060111B$^{1}$ & - & $<89$ & 1.95 & Reverse shock \\
060607A$^{2}$ & -2.21 & 201 & 1.57 & Reverse shock \\
060729A$^{1}$ & - & $<151$ & 2.00 & Reverse shock \\
080319B$^{3}$ & - & $<307$ & 1.89 & Reverse shock \\
080607A$^{4}$ & & $<105$ & 2.88 & Reverse shock \\
080714A & -2.31 & 130 & 2.42 & Reverse shock\\ 
100316B & - & $<78$ & 3.31 & Reverse shock \\
110205A$^{5}$ & -3.49 & 227.79 & 3.59 & Reverse shock\\
121011A$^{6}$ & - & $<119$ & 1.67 & Reverse shock \\ 
121024A$^{7}$ & - & $<173$ & 2.29 & Reverse shock \\ 
151027A$^{8}$ & - & $<105$ & 1.98 & Reverse shock \\
160417A & - & $<146$ & 2.84 & Reverse shock \\ \hline
\end{tabular}
\caption{GRBs with very early optical LC decay indices steeper than 1.5, i.e., the limit considered for the external forward shock. The observed flares are not correlated with XRT and BAT emission and possibly have reverse shock signatures. The observed indices are approximately obtained from fitting a few early data points; therefore, errors are not given. For GRBs where a clear \( t_p \) is not constrained, the first optical detection serves as an upper limit for \( t_p \). (1) \citealt{2009ApJ...702..489R}, (2) \citealt{2009MNRAS.395..490O}, (3) \citealt{2008Natur.455..183R}, (4) \citealt{2011AJ....141...36P}, (5) \citealt{2011ApJ...743..154C}, (6) \citealt{2016RAA....16...12X}, (7) \citealt{2016A&A...589A..37V}, (8) \citealt{2017A&A...598A..23N}.}
\label{tab:ESD2}
\end{table}

\begin{table}
\centering
\begin{tabular}{|c|c|c|c|c|}\hline
~~~ GRB ~~~ & $\alpha_{r}$ & ~~~~ $t_{p}$(s) ~~~~ & ~~~ $\alpha_{d}$ ~~~ & ~~~~~~~ Origin ~~~~~~~ \\ \hline
061121A$^{1}$ & $<$-10 & 80 & 6.00 & Internal shock \\
080310A$^{2}$ & - & $<116$ & 3.70 & Internal shock \\
100814A$^{3}$ & - & $<175$ & 2.14 & Internal shock \\
110119A & - & $<72$ & 7.00 & Internal shock \\
120327A$^{4}$ & - & $<151$ & 7.01 & Internal shock \\
121123A & -4.89 & 286.87 & 6.12 & Internal shock\\
150910A$^{5}$ & - & $<169$ & 9.34 & Internal shock \\
161117A & -8.50 & 120 & 8.53 & Internal shock \\
161214B & - & $<89$ & 1.43 & Internal shock \\ \hline
\end{tabular}
\caption{GRBs with very early optical LC decay indices steeper than the limit considered for the external forward shock and reverse. Also, the observed flare is consistent with the flare in XRT and BAT emission, hence showing an internal shock signature. The observed indices are approximately obtained from fitting a few early data points; therefore, errors are not given. For GRBs where a clear \( t_p \) is not constrained, the first optical detection serves as an upper limit for \( t_p \). (1) \citealt{2009MNRAS.395..490O}, (2) \citealt{2012MNRAS.421.2692L}, (3) \citealt{2014A&A...562A..29N}, (4) \citealt{2017A&A...607A..29M}, (5) \citealt{2020ApJ...896....4X}}.
\label{tab:ESD1}
\end{table}

\begin{itemize}
\setlength\itemsep{0.5em}
\item Correlation between multi-band peaks: The observed peak in the optical LC must be uncorrelated with the prompt $\gamma$/X-ray emission. If the optical flares are uncorrelated with the prompt $\gamma$-ray emission, this indicates that the optical and $\gamma$-ray emissions likely originate from different mechanisms \citep{1999Natur.398..400A}. Conversely, if the observed optical flare coincides with the $\gamma$/X-ray flare, it suggests a common origin for both the optical and high-energy emissions \citep{2005Natur.435..178V, 2006ApJ...638L..71B, 2008Natur.455..183R}.

\item Decay indices: As we have discussed earlier, the rise of the early optical/NIR LC due to reverse shock depends on whether the emission lies in a thick shell or a thin shell case. However, in both cases, the decay index $\alpha_{d} \sim 2$, indicates the reverse shock emission in the ISM-like medium and the $\alpha_{d} \sim 3$ in the wind-like medium \citep{2000ApJ...545..807K, 2003ApJ...582L..75K, 2015AdAst2015E..13G}.

\item Multi-band observations: The reverse shock emission dominates in the low-frequency range, i.e., in optical and NIR band \citep{2003ApJ...582L..75K}. By plotting the early multi-wavelength afterglow LC (including optical, X-ray, and others), we can identify the presence of distinct multiple components. Generally, early X-ray LCs are dominated by either internal shock or forward shock emission with energy injection from the central engine. If there is a reverse shock component, it must be prominent in optical/NIR wavelengths.
\end{itemize}

\subsubsection{Reverse Shock origin of early flare or steep decay}
In our analysis, we found that for 12 GRBs (GRB 060111B, GRB 060607A, GRB 060729A \citep{2009ApJ...702..489R}, GRB 080319B \citep{2009A&A...504...45P}, GRB 080607A, GRB 080714A, GRB 100316B, GRB 110205A, GRB 121011A, GRB 121024A, GRB 151027A \citep{2017A&A...598A..23N}, and GRB 160417A), the decay indices derived from the early steep decay phase fall within the range \(2 < \alpha_{d} < 3\), see Fig. \ref{fig:RS_GRBs}, \ref{fig:RS2}, and Table \ref{tab:ESD2}. Additionally, the observed peak in the optical LCs of these bursts shows no correlation with their corresponding X-ray and $\gamma$-ray LCs. This lack of correlation suggests that the observed optical emission in these GRBs have different origins and is consistent with the predictions of reverse shock emission occurring in either an ISM or wind-like medium \citep{2000ApJ...545..807K, 2003ApJ...582L..75K, 2015AdAst2015E..13G, 2010ApJ...714..799P}. Detailed discussions about the observed optical and X-ray LCs of individual GRBs are given in section \ref{sec:RS_signature}.

\subsubsection{Early steep decay or flare due to internal shock}
In 9 GRBs (GRB 061121A \citep{2009MNRAS.395..490O}, GRB 080310A \citep{2012MNRAS.421.2692L}, GRB 100814A \citep{2014A&A...562A..29N}, GRB 110119A, GRB 120327A, GRB 121123A, GRB 150910A, GRB 161117A, and 161214B) given in Table \ref{tab:ESD1}, the observed decay of early flare is too sharp to be considered with the external reverse or forward shock origin. At the same time, the observed steep decay coincides with the early flares observed in XRT and BAT data. The observed flare in the multiple bands indicates their common origin and is inconsistent with expectations for a reverse shock. Therefore, in the above-discussed GRBs, the origin of the observed flare can be internal shock within the relativistic jet \citep{2005Natur.435..178V}. The details of the observed LCs of each burst are given in the section \ref{sec:IS_signature}.

\subsubsection{Late flare overlapping the normal decay or plateau phase}
In addition to the above-discussed cases, the optical and X-ray LCs of several GRBs (GRB 060526A \cite{2009MNRAS.395..490O}, GRB 100901A \citep{2013ApJ...774...13L}, GRB 110420A, GRB 180620B, etc.) consist of flares overlapping the normal decay, smooth bump or plateau phase. The simultaneous sharp peak observed in the UVOT, XRT, and BAT LCs of these GRBs indicates the common origin of the observed flashes \citep{2005Sci...309.1833B, 2006ApJ...646..351L}. For example, the sharp peak overlapping on the plateau phase around 250\,s in the case of GRB 060526A, see panel 1 of Fig. \ref{fig:LRB_plots} and at $\sim 400$\,s on the onset bump in the case of GRB 100901A, see panel 5 of Fig. \ref{fig:RS_GRBs}. The observed flares overlapping the onset bump, plateau, or normal decay phase suggest the simultaneous existence of two independent emission components. The observed erratic flashes might be from the internal shock due to late central engine activity overlapping the emission from the external forward shock \citep{2005Sci...309.1833B}.

\subsection{Simple Power-law decay} \label{sec:SPL}

\begin{figure*}
\centering
\includegraphics[width=2\columnwidth]{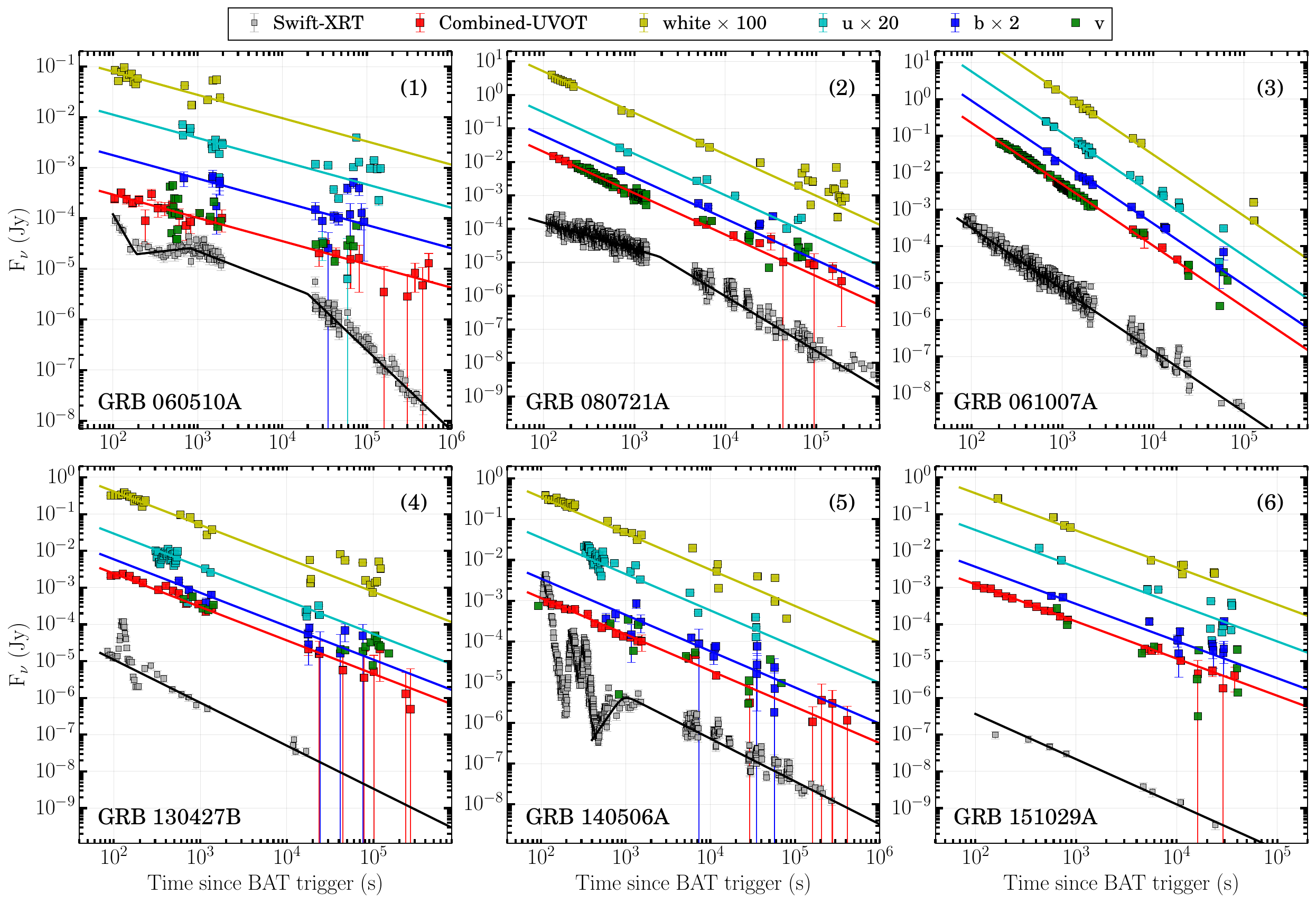}
\caption{Panels (1)-(6) represent multiband \swift-UVOT LCs of GRBs, which show only a PL decay throughout, regardless of the behaviour of XRT LCs at 10\,keV, i.e., X-ray LCs have different behaviours in each panel. The LCs in the different optical bands are scaled for better visualisation, as shown in the legends.}
\label{fig:pow0_plots}
\end{figure*}

In our analysis of the \swift-UVOT \textit{v}-band LCs, 76/200 LCs decay following a simple PL within the observation range of \swift-UVOT. Examples of this type of LCs in multiple UVOT bands are shown in Fig. \ref{fig:pow0_plots}. All these 76 \textit{v}-band LCs were fitted by only a single PL model, but the decay index $\alpha_{o}$ obtained from the fitting of LCs ranged from 0.2 to 2.88, with a mean value of $\alpha_{o, avg} = 0.92 \pm 0.37$. Theoretically, the afterglow LCs decaying with a PL index $\alpha_{o} \sim 1$ is consistent with the emission from the external forward shock \citep{Sari_1998}. Further, we have also found that 61/200 (30.5\%) X-ray LCs show PL decay, ignoring the early overlapping flares. Our fitting results revealed that, for \swift-XRT LC, the decay index ($\alpha_{x}$) of the PL model ranges from 0.61 to 1.77, with a mean value of $\alpha_{x, avg} = 1.20 \pm 0.28$.\\

\begin{figure}
\centering
\includegraphics[width=\columnwidth]{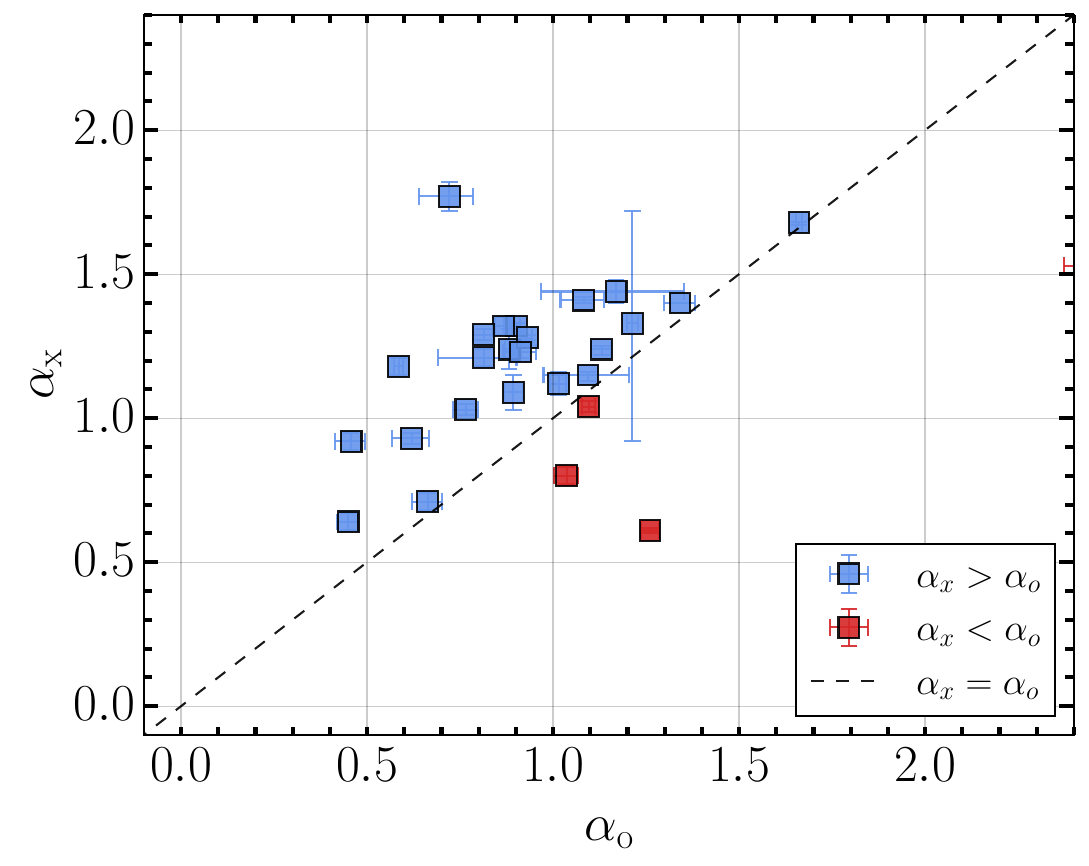}
\caption{{Represents the decay indices ($\rm \alpha_{x}~vs~\alpha_{o}$) obtained from the fitting of optical and X-ray LCs, where both show the simple PL decay. The black dashed line represents the $\rm \alpha_{x} = \alpha_{o}$.}}
\label{fig:A1L500}
\end{figure}

When we compared the optical LCs with X-ray LCs from Table \ref{tab:pow0} and \ref{tab:xpow0}, respectively, some dissimilarities between the two wavelengths were observed. Corresponding to these 76 optical LCs, only 27 GRBs have X-ray LCs with similar PL decay behaviour, i.e., for the rest of the GRBs, either optical has PL, but X-ray has one or more breaks or vice versa. A comparison between multi-band optical LCs and corresponding X-ray LCs at 10 keV is shown in Fig. \ref{fig:pow0_plots}, and the decay indices of optical and X-ray LCs are shown in Fig. \ref{fig:A1L500}. Among these 27 GRBs presented in Fig. \ref{fig:A1L500}, 24 exhibit a steeper decay index in the X-ray LCs compared to the optical ones ($\alpha_{x} > \alpha_{o}$), while the remaining 3 show the opposite trend, with $\alpha_{x} \lesssim \alpha_{o}$. The potential reason for this is that X-ray and optical bands often lie on different parts of the synchrotron spectrum. If the X-ray emission lies above the cooling frequency, the decay index follows \(\alpha_{x} = \frac{3p-2}{4}\). In an ISM-like medium, the X-ray LC exhibits a steeper decay than the optical, where \(\alpha_{o} = \frac{3p-3}{4}\). Conversely, in a wind-like medium, the X-ray LC decays more slowly than the optical, \(\alpha_{o} = \frac{3p-1}{4}\) \citep{Sari_1998}.\\

As we discussed, for more than 30 GRBs, where the optical LC exhibited only a PL decay, the X-ray LC displayed two or more components. Six examples of X-ray and multi-band optical LCs are shown in Fig. \ref{fig:pow0_plots}. There could be several reasons for the chromaticity observed between the X-ray and optical bands, i.e., the following four we describe below:

--- In the first panel of Fig. \ref{fig:pow0_plots}, a rise is observed in the early X-ray LC of GRB 060510A, though the complete X-ray LC is canonical. While the optical LC exhibits a smooth PL decay, it suggests different emission mechanisms or physical processes acting in the two bands. Further, the PL decay slope in the optical LCs is shallower than that of the normal decay phase, indicating the energy injection from the central engine. The observed chromatic behaviour in the afterglow LCs suggests that the optical and X-ray emission originate from different regions of the GRB outflow. One possible explanation for this behaviour is the two-component jet model, initially proposed by \cite{2003ApJ...591.1075K}, and further utilised by several authors to explain the observed chromaticity in the afterglow LCs \citep{2005NCimC..28..439P, 2007MNRAS.380..270O}. A two-component jet, consisting of a narrow/fast and a wide/slow jet with a common axis, can explain the observed chromatic behaviour \citep{2003ApJ...591.1075K, 2007MNRAS.380..270O}. The deceleration of the narrow jet likely occurs earlier, contributing predominantly to the X-ray emission \citep{2010ApJ...709L.146D}. The parameters obtained from the fitting of early X-ray LC are also consistent with the onset of forward shock in the external medium, more details are given in section \ref{sec:onset}. Meanwhile, the optical emission, arising from the wider jet, may not show a corresponding bump if the deceleration signature is masked by prolonged energy injection. Alternatively, if the wider jet component has not decelerated throughout the UVOT observation, the optical emission might have originated from a mechanism different from the standard external forward shock. We have two examples of such LCs, GRB 060510A and GRB 151114A, see panels 1 and 6 of Fig. \ref{fig:plateau_GRBs0}.

--- In the second panel of Fig. \ref{fig:pow0_plots}, the observed break in the X-ray LCs corresponding to the simple PL decay optical LCs can be due to the cooling frequency $\nu_{c}$ crossing the X-ray band, which may have caused a break in the X-ray LCs \citep{Sari_1998}. The difference between temporal and spectral indices of the X-ray LC before and after the break should be $\Delta \alpha \sim \pm$ 1/4 and $\Delta \beta \sim \pm$ 1/2 \citep{Sari_1998}. This is different from what we have found for the X-ray LCs and spectra of GRB 080721A $\alpha_{x,1} = 0.80$ and $\alpha_{x,2} = 1.63$ and $\beta_{x1}$= 0.7, $\beta_{x2}$ = 0.8, before and after the break time $t_{b,x}$=1890\,s, similar to those observed by \cite{Starling_2009}. Therefore, in this case, the passage of synchrotron frequency can not account for the break in X-ray LCs. Since the spectral index remains unchanged across the break, the nature of the break observed in X-ray LCs could be purely hydrodynamic. Therefore, the X-ray and optical emissions arise from different origins. Once again, a two-jet scenario emerges as a plausible explanation, with the narrow jet component likely exhibiting an early break around 1890\,s. However, the optical emission, originating from the outflow with a lower degree of collimation, does not exhibit a break at the same time \citep{2007MNRAS.380..270O, 2010ApJ...709L.146D}.

--- In the 3rd and 6th panels, the multi-band LCs of GRB 061007A and GRB 151029A, following a PL with the decay index within the error bar, are consistent with a similar origin. For both the GRBs, the emissions are consistent with the external forward shock model. For GRB 061007A the wind-like medium favoured with \(\alpha_{o} \sim \alpha_{x} \sim 1.66 \sim \frac{3p-1}{4}\) and in case of GRB 151029A the ISM-like medium is favoured with \(\alpha_{o} \sim \alpha_{x} \sim 1.01 \sim \frac{3p-3}{4}\) in the same spectral regime \citep{Sari_1998, 2022MNRAS.511.1694G}.

--- In the 4th and 5th panels, the multiple flares observed in the X-ray LC are not present on the optical LC, which instead shows a smooth decay. This high variability in the X-ray LC can not be explained by the external forward or reverse shock model, which dominates in the low energy (optical-NIR) emission \citep{2015AdAst2015E..13G}. Therefore, the flares observed in X-ray LCs suggest continued central engine activity and are likely attributed to internal shocks, a similar origin to that of the prompt emission \citep{2005Sci...309.1833B}. While the flares in the high-energy band are prominently distinguishable, flares in the low-energy band tend to exhibit lower significance and can often be obscured by background noise \citep{2013ApJ...774....2S}. This phenomenon is evident in Fig. \ref{fig:pow0_plots}. For instance, in GRB 130427B, the X-ray LC flare corresponds to only a minor deviation in the optical LC, whereas in GRB 140506A \citep{2024A&A...684A.164K}, the optical LCs remained unaffected despite the flare in the X-ray LC.\\

LCs with a temporal decay index in the range (\(0.7 < \alpha < 1.5\)) are widely observed in afterglows of GRBs and are well-described by the standard external forward shock model, depending on the detailed microphysics, such as the electron energy distribution and the ambient medium profile \cite{Sari_1998}. This model is supported by various observational studies and has been successfully used to interpret late-time (after an hour but before the jet break transition ($>$ 10$^4$\,s) into steep decay phase) afterglow LCs in almost all GRBs \citep{2006ApJ...637..889Z, 2009ApJ...707..328L, 2009MNRAS.395..490O}. However, some discrepancies in the optical LC were also observed, where the decay indices for eight were too shallow compared to predictions from the external forward shock model. The optical and X-ray LCs of these eight GRBs are shown in Fig. \ref{fig:plateau_GRBs0}. Moreover, in the UVOT observation range, the decay index for GRB 080607A ($\alpha_{o}$ $\sim$ 2.88) is too steep to be considered as a forward shock in the external medium, as shown in panel 4 of Fig. \ref{fig:RS2}. The origin of the observed discrepancies in the optical LCs given above is discussed in the sections \ref{sec:plateau} along with GRB having plateaus.\\

\subsection{Origin of plateau in \swift-UVOT LC} \label{sec:plateau}
The shallow decay or plateau phase refers to a period in the afterglow LC where the brightness decreases more slowly than expected from the standard fireball model \citep{Sari_1998}. Two types of plateaus are generally observed in GRB afterglows, external plateaus and the internal plateau \citep{2001ApJ...552L..35Z, 2007ApJ...665..599T, 2007ApJ...670..565L, 2007ApJ...658L..75G}. In both cases, to maintain the energy of the blast wave, the central engine must provide continuous energy at least till break time $t_{b}$ \citep{2006ApJ...642..389N, 2006ApJ...642..354Z, 2018ApJ...869..155S}. In this study, we have 30/200 GRBs that have a part of optical LCs showing a shallow decay phase. Eight optical LCs are such that throughout the UVOT observations, LCs show a shallow decay phase as shown in Fig. \ref{fig:plateau_GRBs0}. Whereas, for LCs have a plateau and one or more breaks are shown in Fig. \ref{fig:plateau_GRBs1}.\\

\begin{figure}
\centering
\includegraphics[width=\columnwidth]{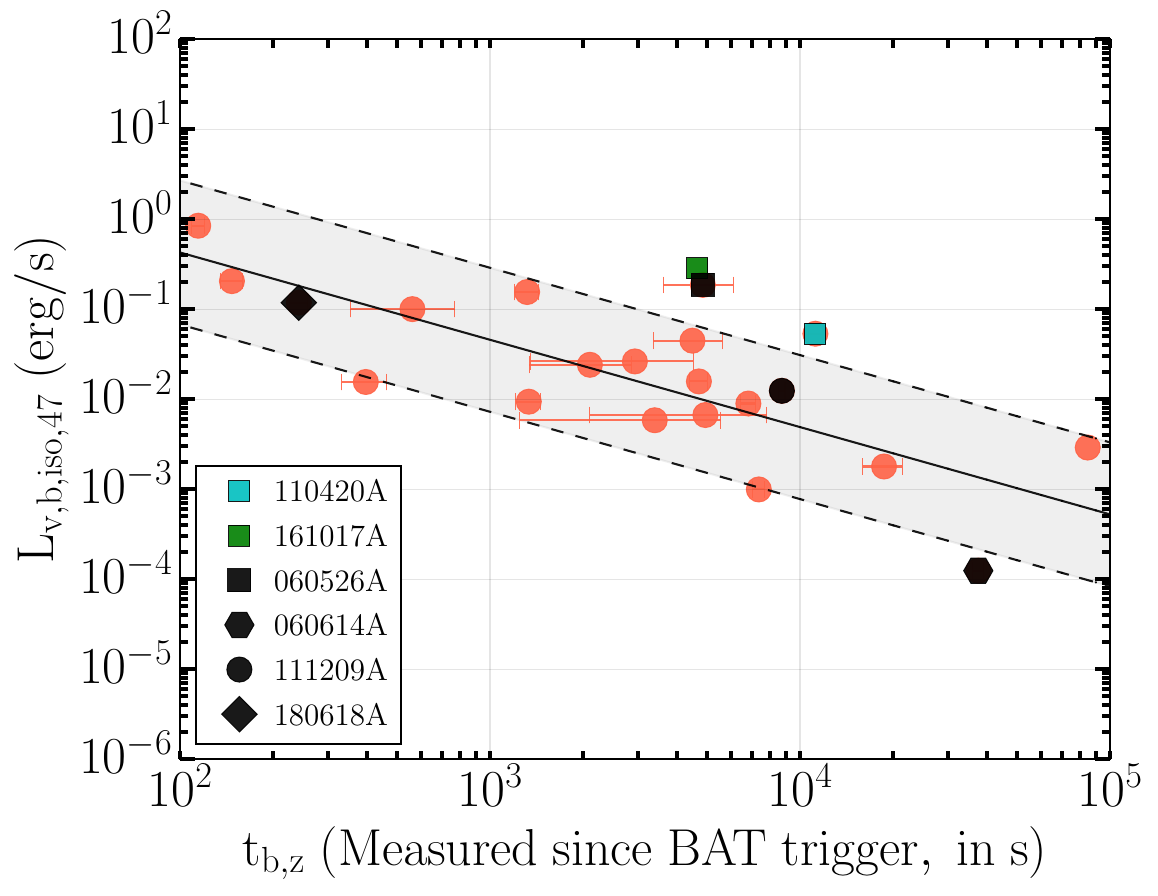}
\caption{Represent the correlation between the isotropic equivalent luminosity (L$_{v, b, iso, 47}$ = L$_{v, b, iso}$ $\times$ 10$^{-47}$) calculated at break time following the plateau phase and rest frame break time ($t_{b,z}$) for GRBs with at least a plateau in the \textit{v}-band LC (red filled circle). A linear fit between log($t_{b,z}$)-log(L$_{v, b, iso, 47}$) is shown with a black solid line, and the corresponding 1$\sigma$ error of the fit is shown with black dashed lines. The coloured squares are outliers to the fit. The GRBs shown with black markers possibly have internal plateaus in optical LCs.}
\label{fig:LPBTB}
\end{figure}

{Earlier studies have shown that a tight anti-correlation exists between X-ray luminosity at the break time and the break time observed in the rest frame \citep{2008MNRAS.391L..79D, 2010ApJ...722L.215D}. We have searched for a similar correlation for the plateau observed in the optical LCs. We have plotted the optical luminosity at the break time (L$_{v,b, iso, 47}$ = L$_{v, b, iso}$ $\times$ 10$^{-47}$) vs rest frame break time $t_{b,z}$ in Fig. \ref{fig:LPBTB}. In our analysis of the optical LCs, a strong correlation is found between log(L$_{v,b,iso, 47}$) and log($t_{b,z}$). The linear relation and Spearman correlation coefficients (r), along with the corresponding p-values, are given below:}

\begin{equation} \label{eqn:LPBTB}
log(L_{v,b,iso,47}) = (1.57 \pm 0.80) + (-0.97 \pm 0.23) log(t_{b,z})
\end{equation}

r = 0.70 and p $\sim$ 10$^{-3}$\\

\cite{2018ApJ...869..155S} claim that the luminosity-break time relation obtained by \cite{2008MNRAS.391L..79D, 2010ApJ...722L.215D} points toward the physical properties of the central engine of the burst. When a millisecond magnetar loses its rotational energy due to spinning down, it generates a Poynting-flux outflow or an electron-positron wind. In this way, the magnetar central engine injects energy into the surrounding medium to maintain the energy output of internal/external shock. \cite{2019ApJS..245....1T} shown that a magnetar central engine might carry a fixed energy reservoir, and an anti-correlation between L$_{v,b,iso}$ and $t_{b,z}$ is expected.

In Fig. \ref{fig:LPBTB}, we have shown the correlation between L$_{b, iso}$ and $t_{b,z}$ favour the energy injection model. {The four bursts, GRB 060526A ($z=5.11$), GRB 060614A ($z=0.125$), GRB 110420A (unknown $z$, considered at 2), and GRB 161017A ($z$ = 2.013), are outliers to the given correlation within 1$\sigma$. GRB 110420A has been excluded from the fitting due to its unknown redshift; however, it could align with the correlation if located at \( z \sim 4 \).} There could be several other reasons responsible for the observed deviation in these GRBs. The observed high luminosity in these cases might be attributed to the overlapping flares observed during the plateau phase. Therefore, the observed flux might be significantly affected by contamination from internal shock emissions. Another possibility is that the observed plateau in these GRBs might have a different origin than the energy injection from a magnetar central engine. An accreting black hole could be the potential central engine for these GRBs.\\

Besides the above-discussed magnetar wind model, there are several possible reasons for the shallow decay phase. Below, we have discussed them for each GRB in our sample.

\subsubsection{Origin of the plateaus followed by a steep decay} \label{sec:internal_plateau}
The plateau phase followed by a steep decay is referred to as the internal plateau \citep{2007ApJ...665..599T, 2007ApJ...670..565L}. In our sample, four GRBs (GRB 060526A, GRB 060614A, GRB 111209A, and GRB 180618A) appear to exhibit a plateau phase that is followed by a decay steeper than what is considered for the normal decay phase. In these four bursts, a similar late steep decay following the plateau was also observed by \cite{2009MNRAS.395..490O, 2007ApJ...670..565L, 2013ApJ...766...30G, 2022MNRAS.516.1584S}. Similarly, among the X-ray LCs, nine GRBs (GRB 050730A, GRB 060607A, GRB 120404A, GRB 111209A, GRB 130408A, GRB 130609B, GRB 170317A, GRB 180224A, and GRB 180620A) are such that the plateau phase in observed \swift-XRT LCs is followed by a steep decay with decay indices $>$ 2. Now, the question arises: do these GRBs exhibit internal plateaus in the \swift-XRT LCs? Only GRB 111209A shows indications of an internal plateau in both optical and X-ray LCs, while the others demonstrate distinct behaviour. Internal plateaus may be caused by the two-stage formation of a magnetar and, subsequently, black hole, \citep{2020ApJ...896...42Z}. This scenario is also favoured by ultra-long GRB (\tninty $>$ 1000\,s; \citealt{2013ApJ...766...30G, 2024ApJ...971..163R}), GRB 111209A, where a magnetar central engine is also required to explain their extended prompt emission duration.\\

If a magnetar is the origin of the internal plateau, then GRBs with internal plateaus must satisfy the correlation given in equation \ref{eqn:LPBTB}. Two GRBs with internal plateaus, GRB 060526A and GRB 060614A, are outliers to the given correlation, while the other two, GRB 111209A and GRB 180618A, aligned with the correlation given in equation \ref{eqn:LPBTB}, as shown in Fig. \ref{fig:LPBTB}. Therefore, our analysis supports a magnetar central engine for GRB 111209A and GRB 180618A, while a black hole central engine is favoured for GRB 060526A and GRB 060614A.

Further, if the internal plateau is caused solely by the abrupt cessation of emission from a magnetar central engine, we would expect an achromatic break in both the optical and X-ray LCs. However, in most of the bursts, the observed X-ray and optical LCs have different behaviour. There are several possible explanations for the absence of such achromatic breaks. One possibility is that the optical and X-ray emissions originate from different regions. For instance, the X-ray LC could exhibit an internal plateau dominated by internal emission \citep{2007ApJ...665..599T}, while the optical emission is primarily governed by external processes occurring far from the central engine. Another possibility is that the observed steep decay following the plateau phase may be due to a {jet break in the multi-component or structured jet} \citep{2010ApJ...709L.146D}. For some GRBs, (such as GRB 060607A, and GRB 111209A in X-rays, and GRB 060526A and GRB 060614A in optical, see Tables \ref{tab:xpow1}, \ref{tab:xpow2}, and \ref{tab:xpow3}) the break occurs after $10^{4}$\,s. This timing aligns with when the late jet break typically dominates \citep{1999ApJ...519L..17S}, suggesting that the LCs could transition directly from the plateau phase to the jet break phase. However, the decay index following the jet break can be as steep as the value of the electron spectral index, i.e. $\alpha \sim p \sim 2.5$ \citep{1999ApJ...519L..17S}. The observation of a very steep decay index for GRB 060607A ($\alpha_{x} \sim 4.12$) indeed requires an internal origin.

Details of the observed optical and X-ray LCs of four GRBs with the signature of internal plateaus are given in section \ref{sec:internal_plateau2}.

\subsubsection{Origin of the external plateau with achromatic breaks: Energy injection from a long active central engine}

The plateau phase, followed by a normal decay phase, is known as the external plateau. One possible explanation for the shallow decay phase is continuous energy injection from the long-lived central engine \citep{2001ApJ...552L..35Z, 2002ApJ...566..712Z}. \cite{Laskar_2015} have shown that the energy injection follows E $\propto$ t$^{m}$, where $m$ = (s-1)(3-k)/(7+s-2k), considering the given value of s=10, parameter related to the energy injection rate, we have the index $m$ $\sim$ 0.7 for the ISM medium k=0 and $m$ $\sim$ 1.2 for wind medium (k=2). The decay index of the LC follows as $\alpha = {m-1}$. Thus, depending on the medium, the energy injection scenario results in either a slightly rising LC or a shallow decay phase. Further, for a plateau in the afterglow LC to be consistent with the energy injection model, the following two conditions must be satisfied \citep{2006ApJ...642..389N, 2006ApJ...642..354Z}.

\begin{itemize}
\item {X-ray ($\beta_{x1}$,$\beta_{x2}$) or optical ($\beta_{o1}$,$\beta_{o2}$) spectral indices, before and after the break time, must be within the prediction of synchrotron emission in the same spectral regime, i.e \(\beta_{x1}-\beta_{x2} < 0.5\) and \(\beta_{o1}-\beta_{o2} < 0.5\), which indicates the break is not due to crossing of characteristic synchrotron frequency from the particular band.}
\item There should be an achromatic break in both optical and X-ray LCs, and the plateau must be followed by a normal decay phase.
\end{itemize}

When we compared the observed plateau in optical LCs with corresponding X-ray LCs in Fig. \ref{fig:plateau_GRBs0} and \ref{fig:plateau_GRBs1}, we found that only twelve GRBs (GRB 060526A, GRB 060614A, GRB 081126A, GRB 100316B, GRB 110319A, GRB 111209A, GRB 111228A, GRB 130609B, GRB 141004A, GRB 160117B, GRB 161219B, and GRB 170705A) have plateau phases observed in both X-ray and optical LCs \citep{2009MNRAS.395..490O, 2012ApJ...758...27L, 2016ApJ...817..152X, 2018ApJ...862...94L}. For the rest of the GRBs during the plateau phase observed in optical LCs, the X-ray LCs show either a steep decay, bump or normal decay phase.\\

Among the above-discussed GRBs that show a plateau in both X-ray and optical LCs, the observed achromatic breaks can primarily be attributed to hydrodynamic effects \citep{2006ApJ...642..354Z}. The energy injection from the central engine can sufficiently explain their origins. This energy injection prevents the LCs from decaying rapidly, as would be expected in a standard afterglow model, resulting instead in prolonged, stable emission before the eventual decay \citep{2001ApJ...552L..35Z, 2002ApJ...566..712Z}. To confirm this, for these twelve GRBs, we have obtained the spectral indices ($\beta_{x1}$) during the plateau phase and ($\beta_{x2}$) during the normal decay phase after the break, in the energy range 0.3-10\,keV. The observed spectral indices are shown in Table \ref{tab:xray_spec}. The observed spectral indices do not show significant deviation, which implies that the X-ray spectra did not evolve before and after the break in the X-ray LC. Further, in the last column of Table \ref{tab:xray_spec}, whenever available, we have provided the optical spectral indices from \cite{Dainotti_2022, 2024MNRAS.tmp.1527D}. The optical spectral indices are also consistent with the calculated spectral indices for X-rays in the 0.3 to 10 keV band. Our analysis suggests that the observed achromatic break in optical to X-ray wavelength is not due to the passage of the characteristic frequency. The possible origin of these achromatic breaks in the X-ray and optical LCs is due to fireball hydrodynamics, where the energy injection from the central engine to the external medium helps to maintain the constraint emission during the plateau phase.

\begin{table}
\centering
\begin{tabular}{|c|c|c|c|} \hline
GRB name & $\beta_{x1}$ & $\beta_{x2}$ & $\beta_{o}$\\ \hline
GRB 060526A & 0.98$_{-0.12}^{+0.10}$ & 1.00$_{-0.15}^{+0.15}$ & 1.10 $\pm$ 0.08\\
GRB 060614A & 0.72$_{-0.02}^{+0.02}$ & 0.88$_{-0.05}^{+0.05}$ & 0.86 $\pm$ 0.03\\ 
GRB 060708A & 1.08$_{-0.06}^{+0.06}$ & 1.07$_{-0.07}^{+0.07}$ & 0.88 $\pm$	0.05\\ 
GRB 100316B & 1.10$_{-0.63}^{+0.21}$ & 0.98$_{-0.21}^{+0.24}$ & 1.14 $\pm$	0.14\\ 
GRB 111209A & 0.41$_{-0.01}^{+0.01}$ & 0.65$_{-0.04}^{+0.04}$ & 0.84 $\pm$	0.10\\ 
GRB 111228A & 1.10$_{-0.03}^{+0.04}$ & 0.99$_{-0.05}^{+0.05}$ & 0.78 $\pm$	0.05\\ 
GRB 110319A & 1.02$_{-0.13}^{+0.13}$ & 1.17$_{-0.08}^{+0.08}$ & - \\
GRB 130609B & 0.94$_{-0.02}^{+0.02}$ & 0.95$_{-0.06}^{+0.06}$ & - \\ 
GRB 141004A & 0.93$_{-0.13}^{+0.13}$ & 0.61$_{-0.13}^{+0.14}$ & - \\ 
GRB 160117B & 0.65$_{-0.06}^{+0.06}$ & 0.83$_{-0.11}^{+0.11}$ & 0.53 $\pm$	0.48\\
GRB 161219B & 0.87$_{-0.05}^{+0.05}$ & 0.80$_{-0.07}^{+0.07}$ & 0.85 $\pm$	0.09\\ 
GRB 170607A & 0.92$_{-0.02}^{+0.02}$ & 0.98$_{-0.04}^{+0.04}$ & 0.90 $\pm$	0.04\\ 
GRB 170705A & 0.83$_{-0.02}^{+0.02}$ & 0.88$_{-0.04}^{+0.04}$ & 0.77 $\pm$	0.15\\ \hline
\end{tabular}
\caption{Represents the spectral indices obtained from the fitting of X-ray spectra for GRB with an achromatic plateau in the X-ray and optical LCs. The spectra are extracted at two epochs in the energy range 0.3-10\,keV separated by the observed break in the X-ray LCs following the plateau phase. The Available values of optical spectral indices $\beta_{o}$ during the plateau phase are taken from \citet{Dainotti_2022, 2024MNRAS.tmp.1527D}.}
\label{tab:xray_spec}
\end{table}

Details on the origin of individual LCs are given in the section \ref{sec:external_plateau1} of the appendix.

\subsubsection{Origin of the external plateau with chromatic breaks: Multiple jet components}
In our sample, we have eighteen GRBs (GRB 060111B, GRB 060510A, GRB 060708A, GRB 061021A, GRB 090929B, GRB 101219B, GRB 110420A, GRB 110731A, GRB 111225A, GRB 130418A, GRB 130725B, GRB 140907A, GRB 151114A, GRB 151215A, GRB 160425A, GRB 161017A, GRB 170607A and 180618A) where the optical LCs show a plateau, but the X-ray LC decay follows either a steep decay, bump or normal decay \citep{2009MNRAS.395..490O, 2012ApJ...758...27L, 2022MNRAS.516.1584S, 2018ApJ...861...48S, 2022MNRAS.516.1584S}. As discussed above, the chromatic features observed in optical and X-ray LC from these GRBs can not be considered from the energy injection from a long-lasting central engine.

A two-component jet system can be one of the possible explanations. A two-component jet has already been proposed in several GRBs to account for the observed chromatic behaviour in optical and X-ray LCs \citep{2005NCimC..28..439P, 2007MNRAS.380..270O, 2010ApJ...709L.146D}. In a two-component jet, a fast-narrow and slow-wider jet system, where the jets have a common axis, the less energetic wider jet in the field of view observer can produce the PL decaying afterglow emission \citep{2013ApJ...774...13L}. Meanwhile, the more energetic narrow jet, initially outside the field of view of the observer, gradually enters the observer's field of view and can supply the additional energy needed to sustain the plateau phase observed in the optical LCs \citep{1998ApJ...503..314P}.

Further, a break in a given energy band can also result from the passage of the synchrotron break frequency across that band. In this scenario, a spectral evolution is typically observed before and after the break \citep{Sari_1998}. If the observed break in the optical band is due to the passage of a synchrotron break frequency, the optical spectral indices before ($\beta_{o1}$) and after the break ($\beta_{o2}$) should change. If $\nu_c$ crosses the optical band, a spectral change of $\Delta \beta \sim 0.5$ is expected. On the other hand, if $\nu_m$ crosses the optical band, the spectral index changes from $\beta_{o1} = -\frac{1}{3}$ to $\beta_{o2} = \frac{p-1}{2}$ \citep{Sari_1998}. The detailed SED fitting of optical data alone is beyond the scope of this paper, which we might plan in future papers. Details of individual LCs for these GRBs with chromatic plateaus are given in the section \ref{sec:external_plateau2} of the appendix.

\subsection{GRBs with bumps in optical LC} \label{sec:onset}
\begin{figure*}
\centering
\includegraphics[width=2\columnwidth]{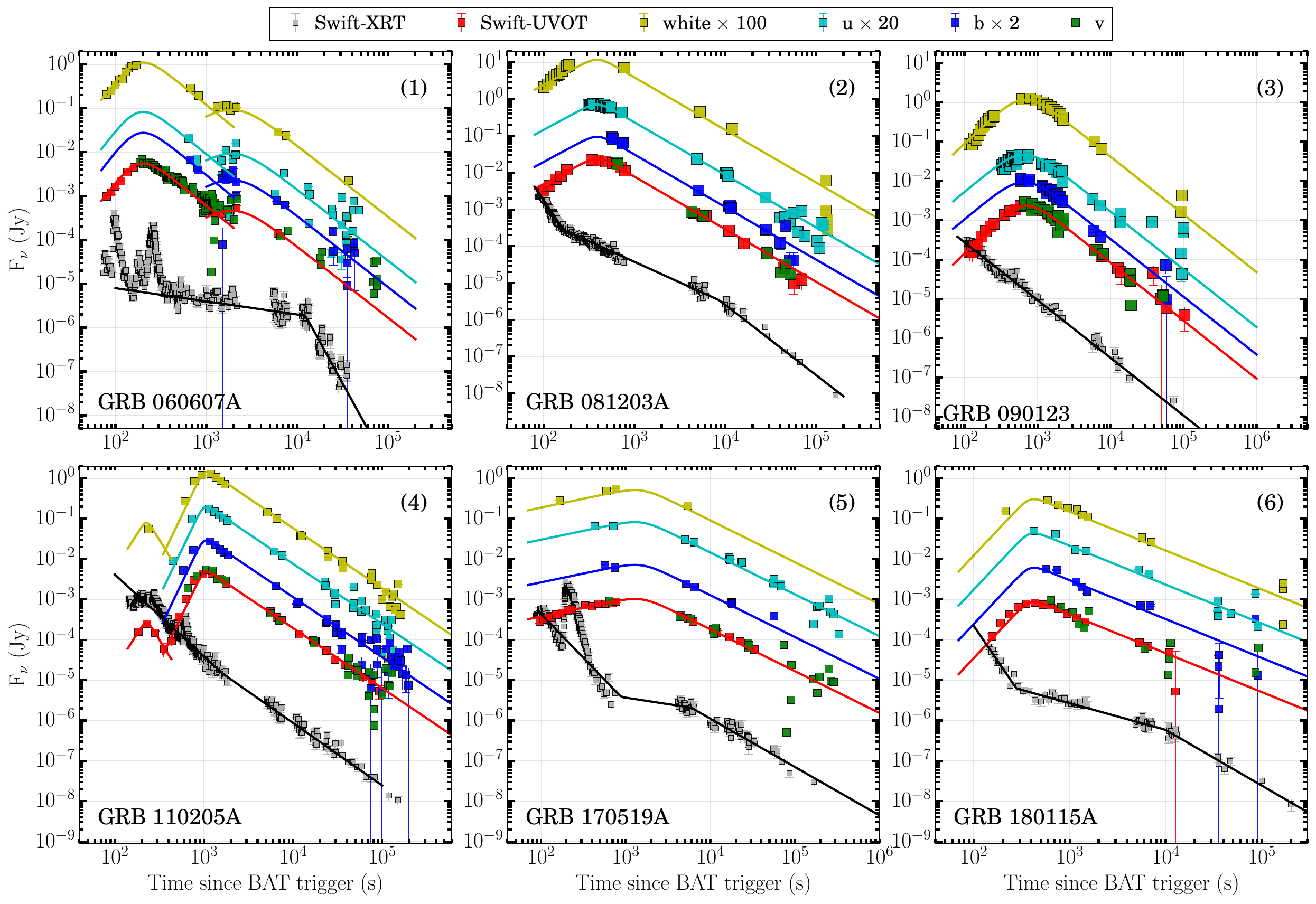}
\caption{Panels (1)-(6) represent multiband \swift-UVOT LCs of GRBs, which show a bump in the early optical LCs regardless of the behaviour of XRT LCs at 10\,keV, i.e., X-ray LCs have different behaviours in each panel. The UVOT LCs in the different optical bands are scaled for better visualization, as shown in the legends.}
\label{fig:sbpl_plots}
\end{figure*}

Around 80/200 GRBs in our sample have at least one smooth bump in the optical LC, as shown in panels 4, 7, and 8 of Fig. \ref{fig:opt_feat}. The multi-wavelength LCs of some GRBs with bumps are also shown in Fig. \ref{fig:sbpl_plots}. There are several reasons for bumps in the afterglow LC, such as the onset of afterglow, passage of synchrotron characteristic frequency, density fluctuation in the surrounding medium, late re-brightening originating due to energy injection from the central engine, or due to the structured outflow with different components coming to the view later, etc. \citep{2009MNRAS.395..490O, 2010ApJ...725.2209L}. The definition of an onset bump given in the literature is that the onset of the afterglow is characterised by a smooth bump in an early optical LC within one hour of GRB trigger {(Note: \cite{2009MNRAS.395..490O} have found that the onset bump peaks $\lesssim$ 1000\,s post trigger} and analysis of \cite{2013ApJ...774...13L} revealed peak time of onset bump up to 3000\,s. Therefore, we select the limit to $\sim$ 1 hour, which is also favoured by our analysis in the next section). The decay following the early smooth bump should be consistent with the external forward shock model, i.e., 0.7 $\leq$ $\alpha$ $\leq$ 1.5 \citep{2007AA...469L..13M, 2010ApJ...725.2209L, 2013ApJ...774...13L}. By studying the onset bump in 20 optical and 12 X-ray LCs, \cite{2010ApJ...725.2209L} found some correlation between the observed properties of the burst, such as parameters of the onset bump, Lorentz factor, and total energy release (E$_{\gamma, iso}$). Sometimes late ($\geq$ 1-hour) re-brightening also occurs, which has a similar shape as the onset of afterglow and follows some of the correlation as the onset bump but occurs at a much later time \citep{2013ApJ...774...13L}, some examples are shown in Fig. \ref{fig:LRB_plots}. Further, we have eight GRBs showing more than one bump in the optical LC (see Fig. \ref{fig:RS_GRBs}), where the first bump corresponds to either onset or flare, and the second bump might be either onset, re-brightening, or flare depending on its shape and peak time. In the following subsections, we will discuss these two types of processes in detail.

\subsubsection{\textbf{Early bump due to Onset of Afterglow}} 

As we have discussed in the introduction section, the fireball model predicts a smooth bump in the early afterglow LC, known as the onset of the afterglow \citep{1999ApJ...520..641S, 2007AA...469L..13M, 2010ApJ...725.2209L, 2013ApJ...774...13L}. To consider a bump as a signature of the onset of afterglow, we follow the same criteria as followed by \cite{2010ApJ...725.2209L} that decay indices following the bump must be shallower than 2. We impose an additional constraint that the peak time \( t_p \lesssim 1 \) hour. This choice is motivated by \cite{1999ApJ...520..641S}, where Equation 7 indicates that the deceleration time for a fireball is approximately 1 hour, assuming a bulk Lorentz factor \( \Gamma = 10 \) and isotropic-equivalent $\gamma$-ray energy \( E_{\gamma, \mathrm{iso}} = 10^{51} \) erg. Note that this deceleration time is sensitive to \( \Gamma \) and decreases as \( \Gamma \) increases. Consequently, the onset peak must occur within 1 hour. 

The smooth bump is generally fitted by a smoothly joined broken PL function (see eq. \ref{eqn:eqn}) which consists of four parameters: rising and decaying index $\alpha_{r}$ and $\alpha_{d}$, a break time $t_{b}$ which is further related to the peak time ($t_{p}$ = $t_{b}$$(\frac{-\alpha_{r}}{\alpha_{d}})^{\frac{1}{s(\alpha_{d}-\alpha_{r})}}$; \citealt{2007AA...469L..13M}) and a smoothness parameter s, which we kept fixed at 1 or 3, depending on the flatness or sharpness of the bump and select the one that provides the best fit. The fitting parameters of GRBs with onset signatures are listed in Table \ref{tab:onset}. In addition to these parameters, using the best fit model, we have also calculated the width (W) at the full width and half a maxima (FWHM), rise time ($t_{r}$), and decay time ($t_{d}$) around the FWHM in the observer's frames. Further, the early onset bump is considered the most reliable method to constrain the Lorentz factor of the bursts \citep{2007AA...469L..13M}. To calculate the Lorentz factor of the bursts in our sample, we have used the relation between $t_{\rm p}$, E$_{\gamma, iso}$, and $\Gamma_0$ by \cite{2007AA...469L..13M, 2018AA...609A.112G}.

In agreement with \cite{2010ApJ...725.2209L}, we have found significant correlations between the onset parameters. In Fig. \ref{fig:onset}, we have shown some of the correlations given by \cite{2010ApJ...725.2209L} and reconfirm them with a larger sample of GRBs in this study. The linear relationship between the parameters and the Spearman correlation coefficients (r), along with the corresponding p-values, are given in Table \ref{tab:onset_corr}.\\

Further, our sample also includes three GRBs (GRB 060510A, GRB 110213A, and GRB 151114A) that display smooth bumps in their early X-ray LCs. Among these, one GRB 110213A exhibits bumps in both optical and X-ray LCs \citep{2011ApJ...743..154C}. For two GRBs (GRB 060510A and GRB 151114A), the optical LCs show a plateau corresponding to the bump in the X-ray LCs.\\

\begin{table}
\centering
\scriptsize
\begin{tabular}{|c|c|c|c|c|c|}\hline
Y & X & \multicolumn{2}{c|}{log(Y) = m $\times$ log(X) + c} & r & p \\ 
 & & c &m & & \\ \hline
 $W$         & $t_{p,z}$              &  0.14 $\pm$ 0.20  &  1.08 $\pm$ 0.10 &0.85&  $<$ $10^{-4}$\\
$t_{r}$      & $t_{p,z}$              & -0.51 $\pm$ 0.20  &  1.02 $\pm$ 0.11 &0.84&  $<$ $10^{-4}$\\
$t_{d}$      & $t_{p,z}$              &  0.54 $\pm$ 0.17  &  0.97 $\pm$ 0.07 &0.90&  $<$ $10^{-4}$\\
$T_{90}$     & $T_{90}/t_{p,z}$       & -1.64 $\pm$ 0.18  &  0.78 $\pm$ 0.06 &0.53&  $<$ $10^{-4}$\\
$L_{p,v,47}$ & $t_{p,z}$              &  1.12 $\pm$ 0.38  & -0.67 $\pm$ 0.18 &-0.33&  $\sim$ 0.003\\ 
$\Gamma_{0}$ & $E_{\gamma,iso,52}$    &  2.33 $\pm$ 0.02  &  0.23 $\pm$ 0.02 &0.70&  $<$ $10^{-4}$ \\
$\Gamma_{0}$ & $t_{p,z}$              &  3.59 $\pm$ 0.05  & -0.53 $\pm$ 0.01 &-0.93&  $<$ $10^{-4}$ \\
$t_{p,z}$    & $E_{\gamma,iso,52}$    &  2.19 $\pm$ 0.05  & -0.48 $\pm$ 0.01 &-0.45&  $\sim$ $10^{-4}$ \\
$L_{p,v,47}$ & $E_{\gamma,iso,52}$    & -0.67 $\pm$ 0.01  &  0.67 $\pm$ 0.15 &0.48&  $<$ $10^{-4}$ \\
 \hline
\end{tabular}
\caption{Represents the linear relationship (log(Y) = m $\times$ log(X) + c), where the parameters corresponding to Y and X are given in Columns 1 and 2. Columns 3 and 4 consist of coefficients obtained from the linear fit between the burst properties such as total energy divided by 10$^{52}$ ($E_{\gamma,iso,52}$), Lorentz factor ($\Gamma_{0}$), \tninty and the parameters of the onset bump defined as peak time ($t_{p}$), width (W) at the full width and half a maxima (FWHM), rise time ($t_{r}$), decay time ($t_{d}$) around the FWHM, and the optical peak luminosity in \textit{v}-band divided by 10$^{47}$ ($L_{p,v,47}$). The Spearman correlation coefficients (r), along with the corresponding p-values, are given in the last two columns. The major correlations presented above show consistency in both trend and strength with the results from \citet{2010ApJ...725.2209L}. For example, the observed slope of the tight positive $\Gamma_{0}-E_{\gamma,iso}$ correlation is $0.23 \pm 0.02$ in this work, compared to $0.25 \pm 0.03$ in \citet{2010ApJ...725.2209L}, which is consistent within 1 sigma. Similarly, the slope of the tight anti-correlation between $\Gamma_{0}-t_{p}$ is $-0.53 \pm 0.01$ in this work, while \citet{2010ApJ...725.2209L} report $-0.59 \pm 0.05$, again consistent within 1 sigma. Additionally, other correlations, such as $t_{p,z}-E_{\gamma,iso}$, $W-t_{p,z}$, $t_r - t_{p,z}$, and $t_{d} - t_{p,z}$, are in good agreement and show similar tight correlations as those reported by \citet{2010ApJ...725.2209L}.}
\label{tab:onset_corr}
\end{table}

Besides the onset of afterglow, there are several other reasons that can cause an early rise in the afterglow LC. For completeness, we briefly discuss some of these reasons here. For more details, please refer to \cite{2009MNRAS.395..490O}.

\subsubsection{\textbf{Early bump due to Passage of synchrotron frequency}} As we discussed, the synchrotron spectrum is described by three characteristic frequencies $\nu_{a}$, $\nu_{m}$, and $\nu_{c}$. $\nu_{a}$ generally effects the radio emission and $\nu_{c}$ mostly lies above the optical band \citep{2013NewAR..57..141G}. The peak synchrotron frequency lies near the optical band and always decreases with time following $\nu_{m}$ $\propto$ t$^{-3/2}$. Passage of synchrotron frequency $\nu_{m}$ through the observing band can cause a bump in the afterglow LC. In this case, the LC will show chromatic behaviour and spectral indices change across the breaks. If the LC in all bands shows an achromatic bump, then this scenario can be discarded. A subset of 6 GRBs from the 80 GRBs with bumps is shown in Fig. \ref{fig:sbpl_plots}, as plotting all of them would unnecessarily increase the length of the paper. As we can see in Fig. \ref{fig:sbpl_plots}, the multi-band LCs of GRBs with smooth bumps do not show any evolution with the colour. \cite{2007AA...469L..13M} have also studied the onset bump in GRB 060418A and GRB 060607A, where multi-colour NIR /optical LCs did not show any evolution with colour. A similar result was also found in \cite{2009MNRAS.395..490O}. Therefore, we discard the passage of synchrotron frequency as the origin of the onset bump in GRBs.
 
\subsubsection{\textbf{Early bump due to Fluctuation in ambient medium density}} \label{sec:density_bump}
Long GRB environments are assumed to be highly dusty due to the pre-existing material surrounding the dying star \citep{2006ApJ...651..985R, 2024A&A...683A..55C}. Radiation pressure from the GRB can destroy this dust over time \citep{2000ApJ...543...56P}. Decreasing extinction can increase the flux rate and cause a bump in the early LC, accompanied by a change in colour \citep{2002A&A...396L...5L}. Initially, the emission would appear redder, with the peak shifting to bluer wavelengths later. However, as discussed above and shown in Fig. \ref{fig:sbpl_plots}, no colour evolution was observed in the afterglow LC of GRBs with early onset signature. Therefore, we discard this scenario as the cause of the bump in the early optical LC.

\subsubsection{\textbf{Early bump due to Off-axis jet model}} \cite{1998ApJ...503..314P} have explained these bumps due to the off-axis jet model. The shape of the afterglow LC is defined by whether the observer is looking within the opening angle of the jet $\theta_{jet} \propto \Gamma^{-1}$ or outside the opening angle. {As the jet decelerates and spreads laterally, the observed emission either increases or decays more slowly, depending on the observing angle $\theta_{obs}$ $>$ 2$\theta_{jet}$ or $\theta_{obs}$ $<$ 2$\theta_{jet}$, respectively, resulting either in a bump or in a shallower phase in the early optical LC \citep{2001_Granot, Granot_2002}.} This model explains the observed bump as a geometric effect rather than a direct change in the energy output. This scenario can be used to explain plateau, onset bump, and late brightening. {Furthermore, in a two-component jet system, a narrow jet propagating along a wider jet with a common axis} \citep{2005NCimC..28..439P}, the observer's viewing angle could be outside the narrow jet and inside the wider jet, which can also cause a chromatic bump {(a bump in optical but not in X-ray)} in the afterglow LCs of some GRBs in our sample. There is insufficient evidence to confirm or rule out the off-axis jet model in comparison to the onset of afterglow, as both scenarios can produce similar smooth bumps in the early afterglow LC.

\subsubsection*{\textbf{Why not all GRBs have Onset bump ?}}
The onset bump (deceleration peak) occurs when the jet starts to decelerate significantly due to interaction with the circumburst medium. Although a bump is theoretically expected in the optical and NIR afterglows of GRBs, there are still several possible reasons that can mask the onset bump. First, if the jet has an extremely high initial Lorentz factor, the deceleration radius becomes very small $R_{d} \propto \Gamma_{0}^{-2/3}$ \citep{1995ApJ...455L.143S, 1999ApJ...520..641S}. As a result, the deceleration peak occurs very early, potentially during the prompt emission phase, and may be missed by observing instruments. In our sample, as shown in Fig. \ref{fig:pow0_plots}, the optical and X-ray LCs of several GRBs have started before 100\,s, and still, we have not found any signature of onset bump in these GRBs. Therefore, deceleration may have occurred before this timescale. Further, in a low-density environment, jet deceleration occurs at later times, resulting in either a weaker onset bump or its appearance outside standard observational timeframes \citep{1999ApJ...520..641S}. Additionally, in some cases, the reverse shock emission can dominate the early afterglow LC, producing bright, sharp features that may obscure the onset bump signature \citep{2015AdAst2015E..13G}. Another possibility is that the observed bump in the optical LCs could also be explained by an off-axis jet \citep{1998ApJ...503..314P}. As discussed earlier, both the onset of the afterglow and an off-axis jet can produce similar features in early optical LCs. In the off-axis jet scenario, an onset bump is expected, whereas an on-axis view would result only in a single PL decay LC. This will also explain the observation of a bump in one band but not in another or vice versa.

\subsection{Late time Re-brightening} \label{sec:LRB}

\begin{figure*}
\centering
\includegraphics[width=2\columnwidth]{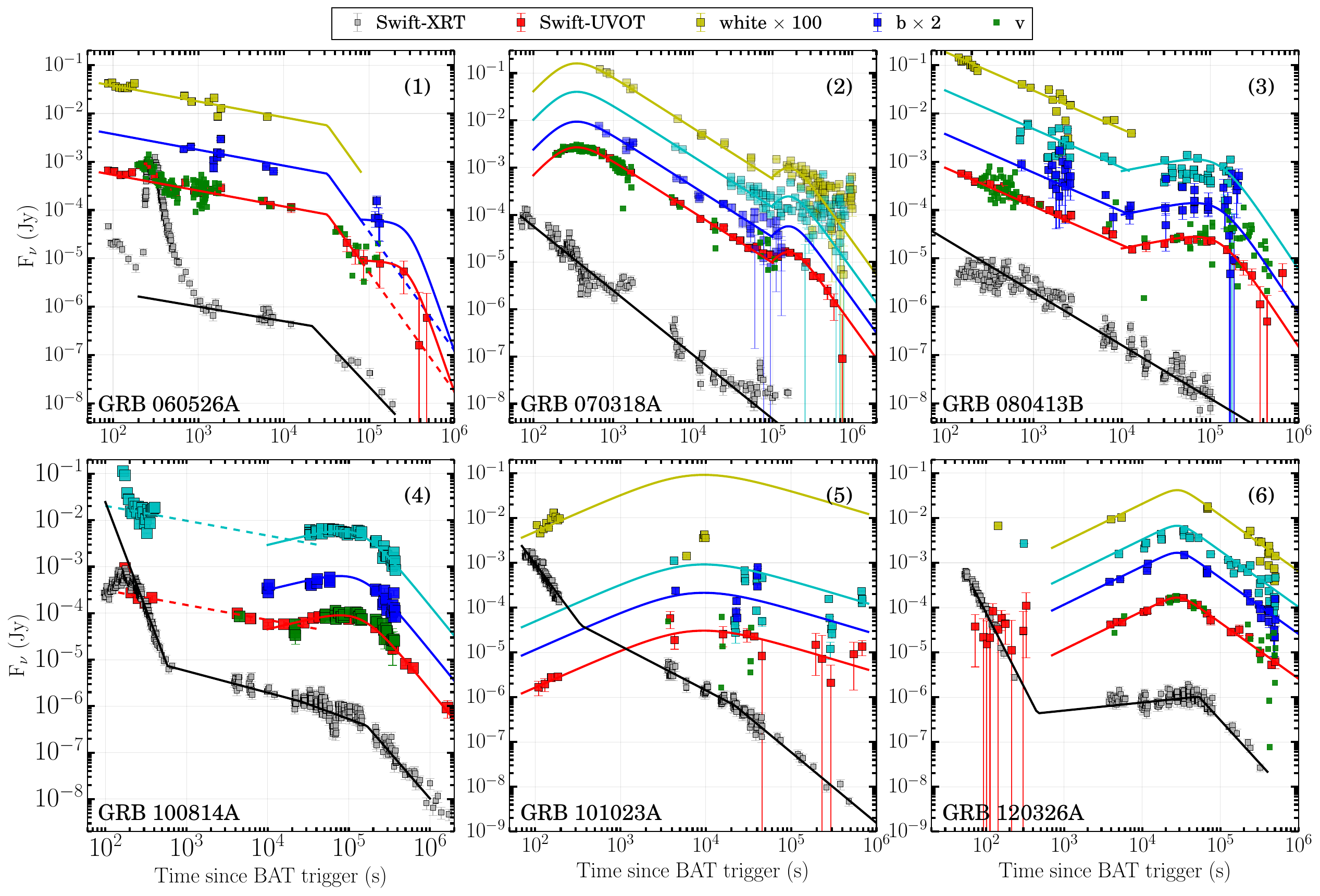}
\caption{Panels (1)-(6) represent the \swift-UVOT LCs in multiple filters (coloured squares) and XRT LC at 10 keV (black squares) for GRBs with the signature of late re-brightening. The fitted models to the multi-band UVOT LCs are plotted with the coloured lines as shown in the legend, and XRT LC at 10 keV is plotted with the black line.}
\label{fig:LRB_plots}
\end{figure*}

Late re-brightening bumps share a similar shape to the onset bump (i.e, smooth rise and decay) but occur much later ($>$ 1 hour; \citealt{2013ApJ...774...13L}) in the afterglow LCs. As we have discussed in section \ref{sec:onset}, GRB with a smooth bump in the early optical LC ($t_{p}$ $\lesssim$ 1 hour) is generally explained by the onset of forward shock in the external medium. The LCs with late bump $t_{p}$ $\gtrsim$ 1 hour do not satisfy all the correlations (see Fig. \ref{fig:onset}) studied by \cite{2010ApJ...725.2209L}; therefore, they may have a different origin. Eight GRBs in our sample have shown two consecutive bumps in the early optical LC, where the second bump in some GRBs exceeds this time limit, see Fig. \ref{fig:RS_GRBs}. In addition to the double bump GRBs, the optical LCs of some bursts show a normal decay or plateau phase in the early time, but have a bump or re-brightening at the late time. Some examples of this type of LC are shown in Fig. \ref{fig:LRB_plots}. In Fig. \ref{fig:onset}, we can see that some GRBs (GRB 060526A, GRB 060729A, GRB 070318A, GRB 080413B, GRB 080928A, GRB 100814A, GRB 100901A, GRB 101023A, GRB 110213A, GRB 120326A and GRB 151027A), with $t_{p} \gtrsim 1$ hour, are not consistent with all the correlation studied in section \ref{sec:onset}. In the cases of GRB 060526A, 060729A, 070318A, 100901A, 110213A, 120326A, and 151027A similar late rebrightening events were also observed by \cite{2013ApJ...774...13L, 2013A&A...557A..12Z, 2014ApJ...785..113H, 2017A&A...598A..23N}. This indicates that the observed bumps in the optical LC of these GRBs may not be due to the onset of the afterglow. There are several possible reasons for the origin of late re-brightening in the optical/NIR LCs, as discussed below.

\subsubsection{Re-brightening due to energy injection or refreshed shock}
Earlier studies have shown that, in most cases, late energy injection from the central engine causes a refreshed shock in the external medium, which is able to explain the late re-brightening observed in the afterglow LC of GRBs \citep{2002ApJ...566..712Z, 2014ApJ...785..113H, 2022MNRAS.513.2777K}. The energy injection scenario suggests that if the energy is supplied by the central engine, the observed late bump in the multi-band afterglow LC should be achromatic \citep{Laskar_2015}. In our sample, we have six GRBs (GRB 060729, GRB 070318A, GRB 100814A, GRB 100901, GRB 120326A, and GRB 151027A), such that corresponding to the late re-brightening in the optical LCs; the X-ray LCs also consist of a plateau or similar re-brightening, supporting the energy injection scenario.

As shown in Figures \ref{fig:LRB_plots} and \ref{fig:RS_GRBs}, the optical LCs of GRB 060729A, GRB 100814A, and GRB 100901A have a bump with the peak time lies within 10$^4$\,s $<$ $t_{p}$ $<$ 10$^5$\,s. Additionally, the X-ray LCs of three GRBs also have a long plateau extended up to 10$^4$-10$^5$\,s. The observed achromatic breaks in the optical and X-ray LCs of these three GRBs suggest that the central engine is active up to this timescale \citep{Laskar_2015}.

Further, both the X-ray and optical LCs of GRB 120326A have a late bump. Similarly, a simultaneous re-brightening at around 5\,ks is also observed in the optical and X-ray LCs of GRB 151027A. The observed peak times ($t_{p}$) of the late-re-brightening in X-ray and optical observations of both GRBs are within the error bars. The simultaneous observation of the late re-brightening bumps in the multi-wavelength LCs of both bursts suggests the energy injection model. GRB 120326A has one of the most delayed bumps, peaking around $t_{p}$ $\sim$ 30\,ks. \cite{2014ApJ...785..113H} also observed similar features and proposed a magnetar central engine for the burst lies in a stellar wind-like medium.

\subsubsection{Re-brightening due to density fluctuation}
A density bump or a sudden jump in the circumburst medium can enhance the energy dissipation rate, potentially leading to re-brightening in the afterglow LCs \citep{Berger_2000, 2002A&A...396L...5L}. Further, \cite{2002A&A...396L...5L, 2006ApJ...651..985R} have shown that a late-time bump caused by a density fluctuation can occur in the optical band when the optical frequencies lie between the synchrotron peak frequency \(\nu_m\) and the cooling frequency \(\nu_c\). This implies that if \(\nu_m\) initially resides within the optical band (and subsequently decreases with time, as expected; \citealt{Sari_1998}), multi-wavelength observations should reveal a temporal shift of the LC peak from blue to red wavelengths. Additionally, if X-ray emission lies above cooling frequency, the density fluctuation in the circumburst medium can not cause a bump in the X-ray LCs. Furthermore, the magnitude of re-brightening due to density fluctuation is very low for optical and NIR LCs as well \citep{Geng_2014}. {As we can see in Fig. \ref{fig:LRB_plots}, multi-band optical LCs do not show any colour evolution. However, the wavelength coverage and the quality of data are not sufficient to completely rule this out. Moreover, we have also observed simultaneous re-brightening in optical and X-ray bands. Our observation suggests that the Late re-brightening is not due to density fluctuation in the ambient medium.}

\subsubsection{Re-brightening due to Multiple jet components}
In some GRBs (GRB 080413B, GRB 080928A, GRB 101023A, and GRB 110213A), we found a late bump in the optical LC, but the corresponding X-ray LC was found to decay normally at the same time. There can be several possible reasons for the different behaviours of LCs at different wavelengths. One possibility is that the energy distribution within the jet is not isotropic, and one should account for a structured jet or two-component jet, i.e., a jet with angular energy variation \citep{2005NCimC..28..439P}. Another possible explanation is the residual activity of the central engine. The late ejected shells from the central engine could collide with slower shells emitted earlier, leading to a sudden release of energy that manifests as re-brightening in the afterglow \citep{2013A&A...560A..70G}. These interactions are typically more noticeable in the X-ray band, making X-ray flares more common compared to optical flares or late re-brightening.\\

To address chromatic late re-brightening, observed in the optical but not in X-ray, we assume a multiple-jet model consisting of at least two co-axial jet components: a narrow, highly energetic inner jet and a wider, less energetic outer jet surrounding it \citep{2005NCimC..28..439P}. If the observing angle ($\theta_{obs}$) is within the jet opening angle ($\theta_{jet}$) of the narrow jet component, a PL decay LC is expected \citep{Laskar_2015}. It is assumed that the prompt emissions and the entire PL decaying X-ray afterglow emissions come from the narrow jet component. The deceleration of the narrow jet component is also responsible for the early bump in the optical LCs, followed by a PL decay. Further, a wider jet, during its deceleration phase, can be responsible for the late re-brightening of the optical LCs \citep{2011A&A...526A.113F}. The observation of both early and late bumps in the optical LCs of some GRBs favour this two-component jet model. Discussion on individual GRBs with chromatic late re-brightening is given in section \ref{sec:cromatic_lrb}.

\section{Summary and Conclusion} \label{sec:summary}
This paper presents a statistical analysis of the morphological features observed in the optical afterglow LCs of GRBs, utilizing \swift-UVOT observations. Our analysis of \swift-UVOT data surpasses those observed from ground-based facilities in many ways: i.e., by leveraging the continuous, high temporal and multi-wavelength observation capabilities of UVOT, our sample consists of optical observations of GRBs from a few seconds to days after the trigger. \swift multi-band observations are well-calibrated and analysed using the same method, which ensures the reliability of the data and is free from the calibration challenges often faced by ground facilities. We have analysed 200 GRBs detected by \swift-UVOT, with well-sampled optical LCs in the \textit{v}-band observed between 2005 to 2018. To ensure comprehensive analysis, whenever available, we also included UVOT observation in other bands (\textit{white}, \textit{u}, \textit{b} etc.) along with \swift-XRT and BAT LCs.

The prompt emission properties (\tninty, $\Gamma$, \Ep, E$_{\gamma, iso}$, etc.) of these GRBs align with those of typical long and short GRBs. The redshift is available for 145 GRBs, however, for redshift unknown bursts, assuming $z$=2 yields a similar distribution of characteristic properties as those with known redshift, as shown in Fig. \ref{fig:col_prop}. Further, our analysis highlights the features observed in early optical LCs of these GRBs, with the earliest observation for GRB 140206A starting at 57\,s post-burst. In contrast, GRB 120212A had the most delayed optical observation.

We employed several PL models with 0, 1, 2, and 3 breaks to fit the observed data. Our findings indicate that 40\% of the optical LCs show an initial rise followed by normal decay or a plateau phase. Conversely, the early X-ray LC typically decays steeply at early times, likely due to early optical emissions being dominated by the external forward shock, whereas early X-ray emission is dominated by the tail of the prompt emission. Most of the optical LCs in our sample either display a bump followed by a PL (80/200 LCs) or a single PL decay (76/200 LCs) throughout. Only 36/200 GRBs have a break in optical LC, and more than one break is found in only 8/200 GRBs. Further, 21 LCs in our sample have shown very early steep decay preceding the bump, normal decay or plateau. Multiple breaks are less common in optical LCs compared to X-ray LCs, which often exhibit two or more breaks.\\

The unprecedented early observation capability of \swift has enabled the detection of several early flares in the afterglow LCs. In our sample, 21 GRBs exhibit this early steep decay. The observation of the complete flares is not always possible, and in most cases, only a steep decay is observed preceding the onset bump, normal decay or plateau phase. The decay indices of most of these flashes or steep decay phases are consistent with the reverse shock emission under the theoretical prediction of \cite{2015AdAst2015E..13G}, as described in section \ref{sec:RS}. Other flares with sharp decay indices and peaks coinciding with the corresponding X-ray and gamma-ray emission are consistent with internal shock emission.\\

In our sample, 76/200 GRBs exhibited only a PL component throughout the observation. Not all PL decay LCs are consistent with the forward shock in the external medium. Based on our analysis, eight GRBs with shallow require energy injection, and one is consistent with reverse shock origin. In our sample corresponding to 76 optical LCs, the X-ray LCs of 27 GRBs have similar PL decay behaviour. 24 X-ray LCs have decay indices slightly steeper than the decay in optical LCs. This reveals that during the normal decay phase, on average, the X-ray LC has a slightly steeper decay index than the optical LC. This suggests that, for these GRBs, the X-ray and optical emissions originate from different regimes of the synchrotron spectrum. The faster decay of the X-ray LC compared to the optical may be attributed to the cooling break frequency, $\nu_{c}$, lying between these two bands.\\

Based on the morphological study of GRB LCs in our sample, 30 GRBs exhibit a plateau in the early phase. In two GRBs (GRB 060526A, GRB 060614A), observed plateaus are likely originating from the internal shock, where the plateau in the optical LCs is followed by a steep decay phase. A hint of internal origin in the plateau phase is also observed in GRB 180618A and 111209A, where the decay indices following the plateau are steeper than the prediction of the forward shock model. Therefore, four GRBs in our sample have optical LCs that have the signature of a plateau originating from the internal shock. A possible origin of the internal plateau is a magnetar central engine collapsing into a black hole \cite{2020ApJ...896...42Z}. Another possibility is that the steep decay is due to a jet break occurring directly following the plateau phase. 

In the case of optical LCs with plateaus followed by a normal decay phase, our analysis revealed that energy injection from a long-active central engine to the external forward shock is responsible for the plateau phase. However, other mechanisms, such as multiple jet components, are also required to explain chromatic breaks observed in the optical and X-ray LCs of GRBs with plateaus. Further, the optical luminosity observed during the plateau is tightly correlated with the break time, indicative of a magnetar as their possible central engine.\\

Further, 80/200 GRBs in our sample exhibit one or more bumps in their optical LCs. In section \ref{sec:onset}, we studied several possible reasons for the smooth bumps observed in the early optical LCs. Most GRBs with early bumps in their optical LCs are consistent with the onset of afterglow in the external medium. We derived various correlations between the parameters of the onset bump, and our results are consistent with those of \cite{2010ApJ...725.2209L}.\\

Furthermore, for some GRBs in our sample, the observed optical bumps do not align with the correlations studied for the onset of afterglow. The smooth bumps observed in these GRBs appear at a later time ($>$ 1 hour), which we have defined as a limit between the onset of afterglow and late re-brightening, which is also supported by the theoretical and observation findings of \cite{1999ApJ...520..641S, 2009MNRAS.395..490O, 2013ApJ...774...13L}. These late bumps are consistent with a refreshed shock produced by late central engine activity or the two jet components, where the narrow jet comes to the observer's field of view at a later time, also the late deceleration of the wider jet can produce similar results.\\

In our sample, out of 200 UVOT LCs, 80 LCs begin with a rise, 76 show only a PL decay, and 40 LCs consist of one or more breaks. Therefore, our analysis results revealed that any subset of observed features in the optical and X-ray LCs of \swift-detected GRBs could be explained in terms of various models. For example, optical LCs with early bump and PL decay are consistent with the forward shock model in the external medium. The deviation of observed decay indices from the prediction of the forward shock model can be explained through the energy injection or off-axis jet model, which clearly indicates a diverse set of physical mechanisms or emissions involved during the afterglow phase. Further, 21 GRBs with early flares, with some of them having their origin with the relativistic jet through the internal shock, and some of them are consistent with the prediction of the external reverse shock. Therefore, the optical emission from the GRBs seems to originate from different outflow locations.

\section*{Acknowledgements}
SBP and AKR acknowledge support from DST grant no. DST/ICD/BRICS/Call-5/CoNMuTraMO/2023(G) for the present work. SBP also acknowledges discussions held with Mat Page, Sam Oates (MSSL-UCL) and the contemporary \swift team about UVOT data of GRB afterglows over the years. RG was sponsored by the National Aeronautics and Space Administration (NASA) through a contract with ORAU. The views and conclusions contained in this document are those of the authors and should not be interpreted as representing the official policies, either expressed or implied, of the National Aeronautics and Space Administration (NASA) or the U.S. Government. The U.S. Government is authorised to reproduce and distribute reprints for Government purposes, notwithstanding any copyright notation herein. AA acknowledges the Yushan Young Fellow Program by the Ministry of Education, Taiwan, for financial support (MOE-111-YSFMS-0008-001-P1). AJCT acknowledges support from the Spanish Ministry project PID2023-151905OB-I00 and Junta de Andalucia grant P20\_010168. This research has used data obtained through the HEASARC Online Service, provided by the NASA-GSFC, in support of NASA High Energy Astrophysics Programs.

\section*{Data Availability}
This work utilised the data of \swift-UVOT for 200 well-sampled LC. The X-ray and $\gamma$-ray data used in our analysis are taken from \swift-XRT, and BAT burst analyser page \citep{eva07, eva09}.

\bibliographystyle{mnras}
\bibliography{ref} 
\newpage
\FloatBarrier
\appendix
\parindent 0em
\section{Fitting models used in this paper} \label{eqn:eqn}
Various PL models used to fit optical and X-ray LCs are:

SBPL
\[
f(t)= A
\begin{cases}
\left[\left(\frac{t}{t_{\rm b}}\right)^{s \alpha_{r}}+ \left(\frac{t}{t_{\rm b}}\right)^{s \alpha_{d}}\right] ^{{1}/{s}}
 \end{cases}
\]
Pow0
\[
 f(t)= A
\begin{cases}
t^{-\alpha_1}, & \text{for~all}~t
\end{cases}
\]

Pow1
\[
 f(t)= A
\begin{cases}
 t^{-\alpha_1},& \text{if } t \leq t_{b1} \\
 t_{b1}^{(\alpha_2-\alpha_1)} ~ t^{-\alpha_2}, & \text{if } t_{b1} < t
\end{cases}
\]

Pow2
\[
 f(t)= A
\begin{cases}
 t^{-\alpha_1},& \text{if } t \leq t_{b1} \\
 t_{b1}^{(\alpha_2-\alpha_1)} ~ t^{-\alpha_2}, & \text{if } t_{b1} < t < t_{b2}\\
 t_{b1}^{(\alpha_2-\alpha_1)} ~ t_{b2}^{(\alpha_3-\alpha_2)} ~ t^{-\alpha_3}, & \text{if } t_{b2} < t
\end{cases}
\]

Pow3
\[
 f(t)= A
\begin{cases}
 t^{-\alpha_1},& \text{if } t \leq t_{b1} \\
 t_{b1}^{(\alpha_2-\alpha_1)} ~ t^{-\alpha_2}, & \text{if } t_{b1} < t < t_{b2}\\
 t_{b1}^{(\alpha_2-\alpha_1)} ~ t_{b2}^{(\alpha_3-\alpha_2)} ~ t^{-\alpha_3}, & \text{if } t_{b2} < t < t_{b3}\\
 t_{b1}^{(\alpha_2-\alpha_1)} ~ t_{b2}^{(\alpha_3-\alpha_2)} ~ t_{b3}^{(\alpha_4-\alpha_3)} ~ t^{-\alpha_4}, & \text{if } t_{b3} < t
\end{cases}
\]

\section{Some extra figures and Tables} \label{fig:figures}

\begin{figure}
\includegraphics[width=\columnwidth]{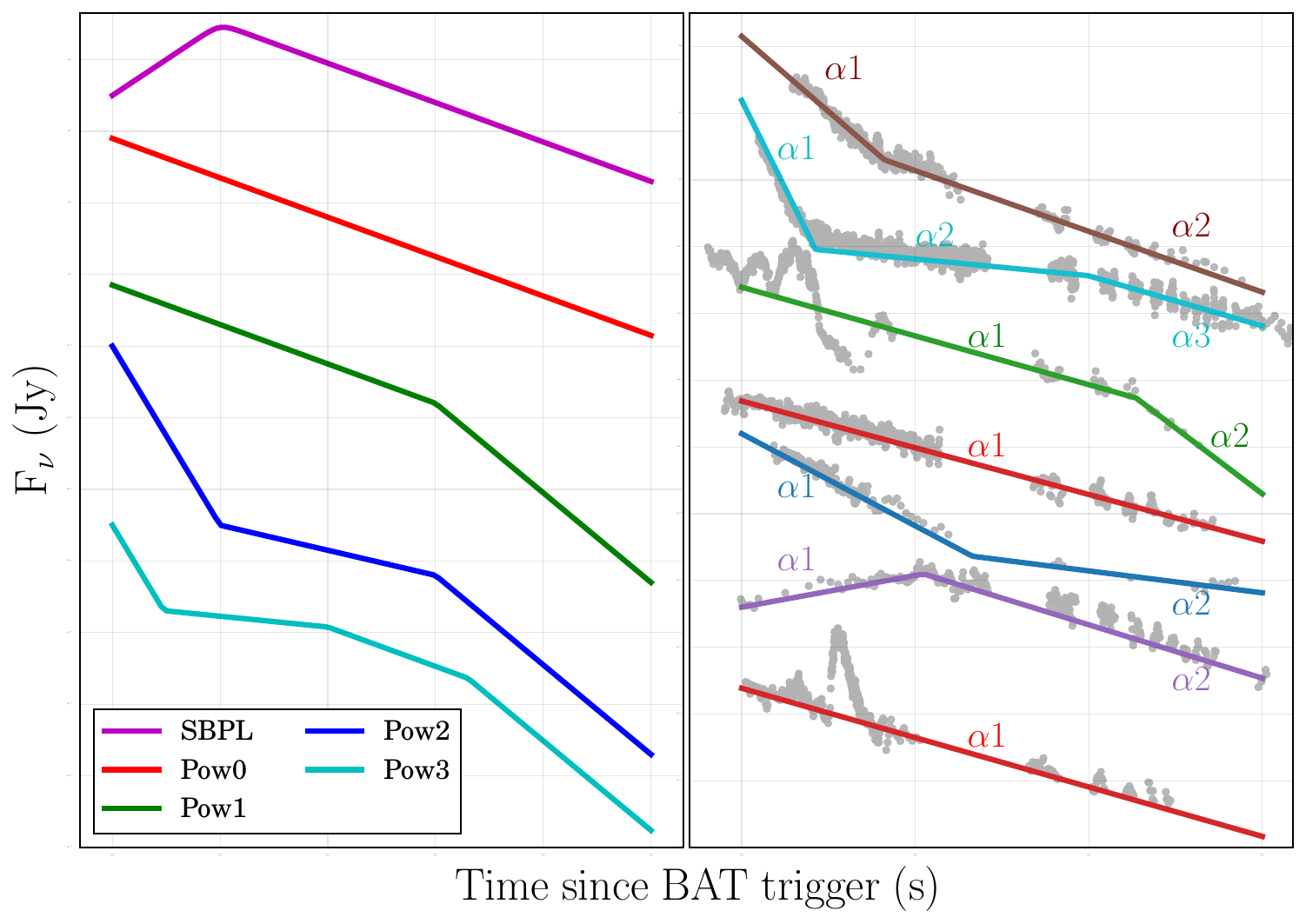}
\caption{Represents the temporal shape of the fitting model and the example of a few observed LCs by the \swift-UVOT and XRT. The PL models with 0, 1, 2, and 3 breaks and a smoothly joined broken PL model are denoted as Pow0, Pow1, Pow2, Pow3 and SBPL, respectively.}
\label{fig:fit_models}
\end{figure}

\begin{figure}
\includegraphics[width=\columnwidth]{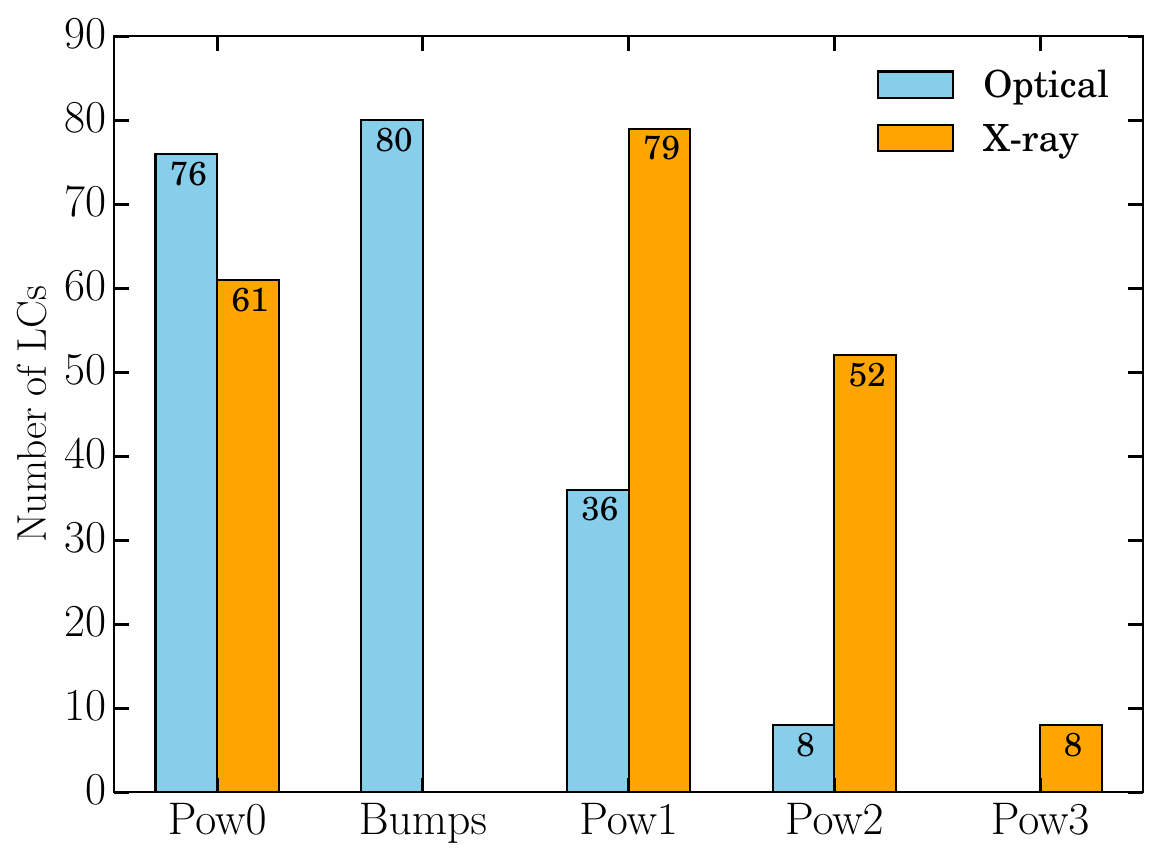}
\caption{Distribution of LC morphologies in the optical and X-ray afterglows of 200 GRBs in our sample. In the optical, 76/200 show a simple PL decay (Pow0), 80/200 exhibit at least a bump, 36/200 have a single break (Pow1), and 8/200 display two breaks (Pow2). In the X-ray, 61/200 show no breaks (Pow0), 79/200 have one break (Pow1), 52/200 have two breaks (Pow2), and 8/200 display three breaks (Pow3).}
\label{fig:pie}
\end{figure}

\begin{figure*}
\includegraphics[width=\columnwidth]{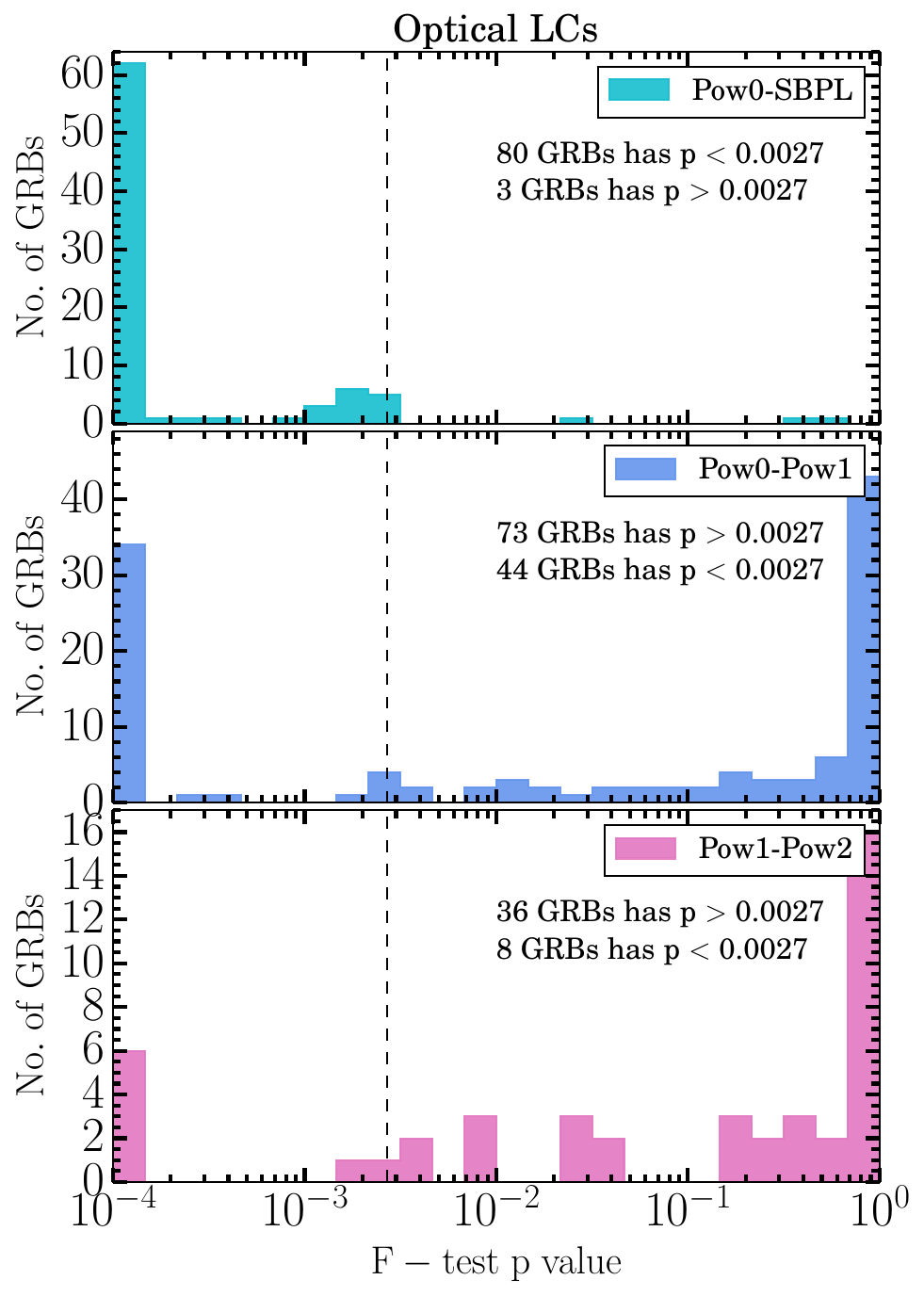}
\includegraphics[width=\columnwidth]{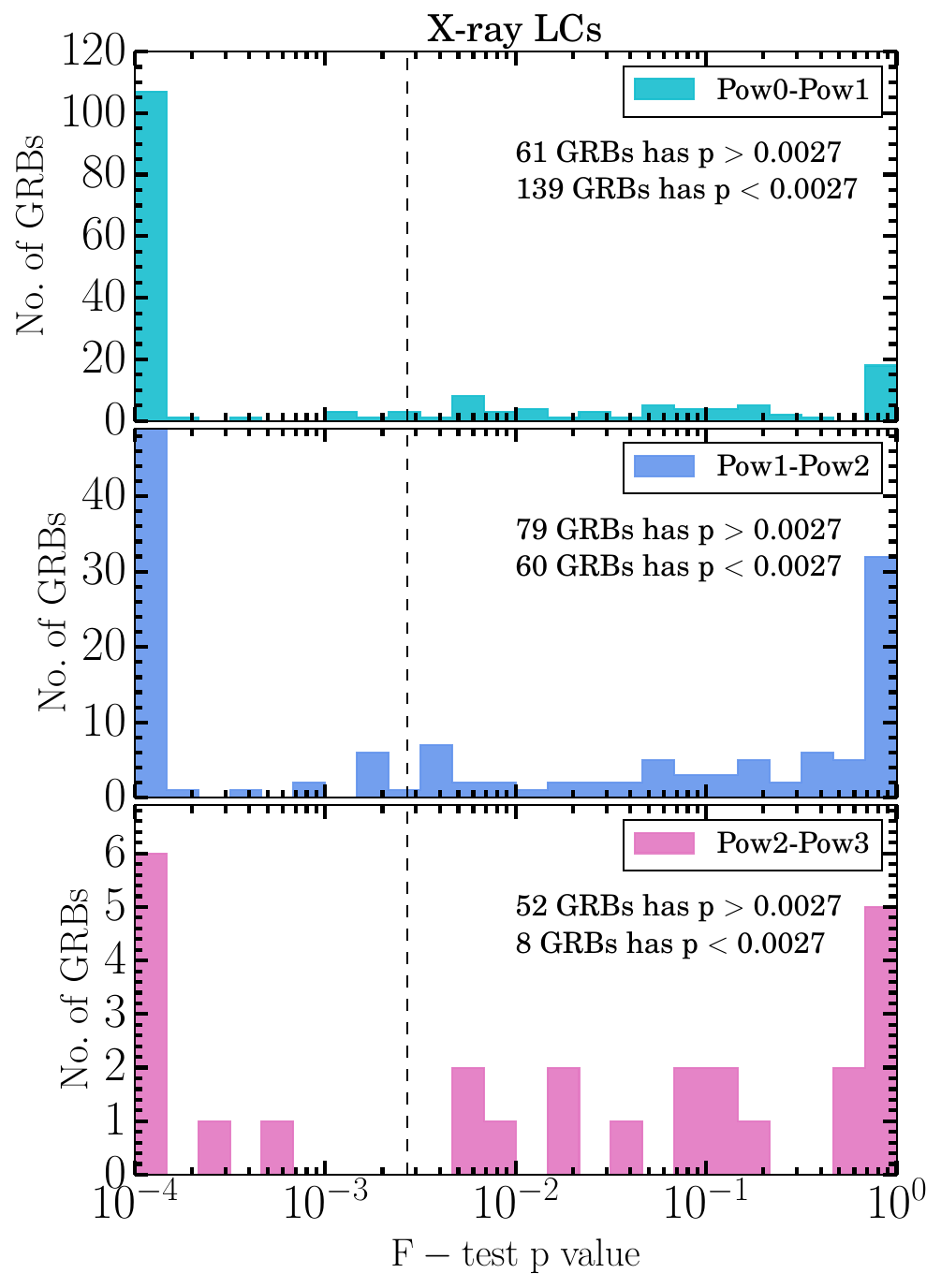}
\caption{The left and right panels show the distribution of the best-fit models for LCs observed by the \swift-UVOT and XRT, respectively. In both the optical and X-ray LCs, an additional break is introduced only if the null hypothesis probability \( p \) is less than 0.0027, as shown with the dashed line. Accordingly, in each panel, two PL models are compared, where the second PL model includes one additional break relative to the first.}
\label{fig:p_values}
\end{figure*}

\begin{figure*}
\centering
\includegraphics[width=1.6\columnwidth]{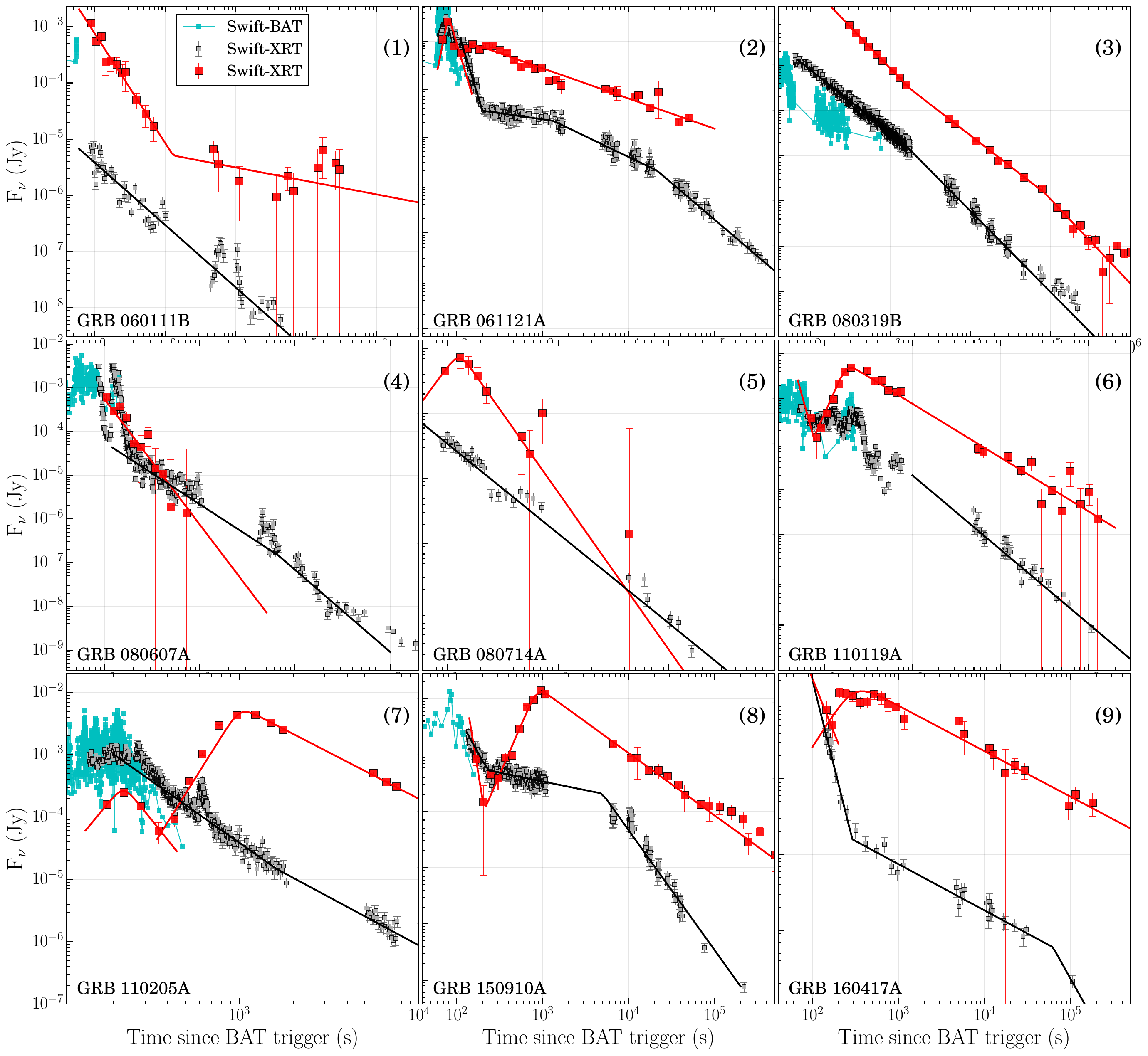}
\caption{Panels (1)-(9) represents multi-wavelength LC of few GRBs from Table \ref{tab:ESD2} \& \ref{tab:ESD1}, with the signature of reverse or internal shock in the \swift-UVOT \textit{v}-band LC. The UVOT, XRT, and BAT observations are shown in red, black, and cyan squares. The fitted models to the UVOT LC in \textit{v}-band and XRT LC at 10 keV are respectively shown with the solid red and black lines.}
\label{fig:RS2}
\end{figure*}

\begin{figure*}
\includegraphics[width=1.7\columnwidth]{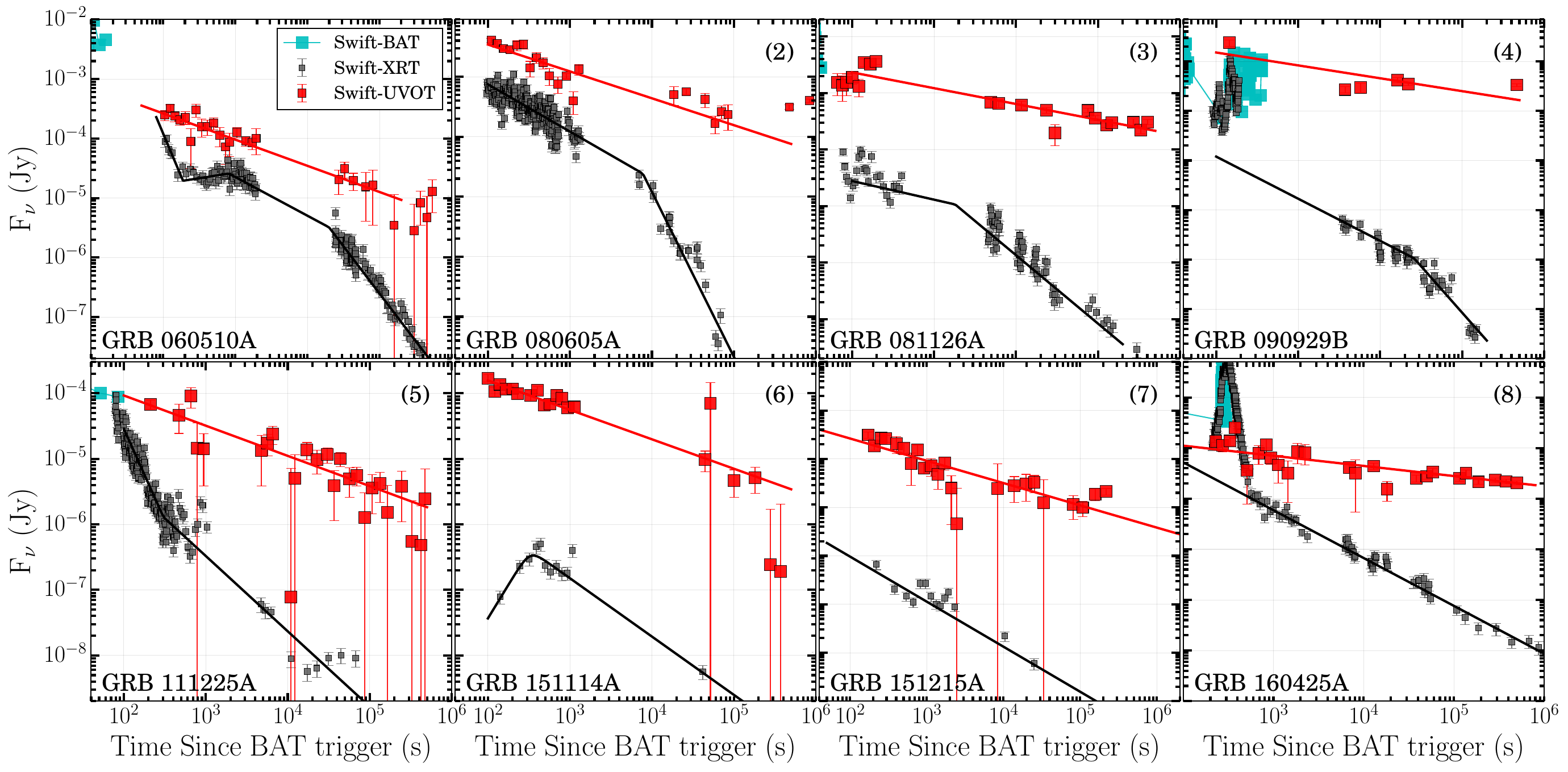}
\caption{Panels (1)-(8) represent the GRBs with a plateau in their optical LC without any breaks. The \textit{v}-band LCs observed by \swift-UVOT are denoted by red squares. Similarly, the XRT LC at 10\,keV is represented by a black square, while the BAT LCs are shown as cyan squares. The best-fit models for the X-ray and optical data are shown by solid red and black lines, respectively.}
\label{fig:plateau_GRBs0}
\end{figure*}

\begin{figure*}
\centering
\includegraphics[width=2\columnwidth]{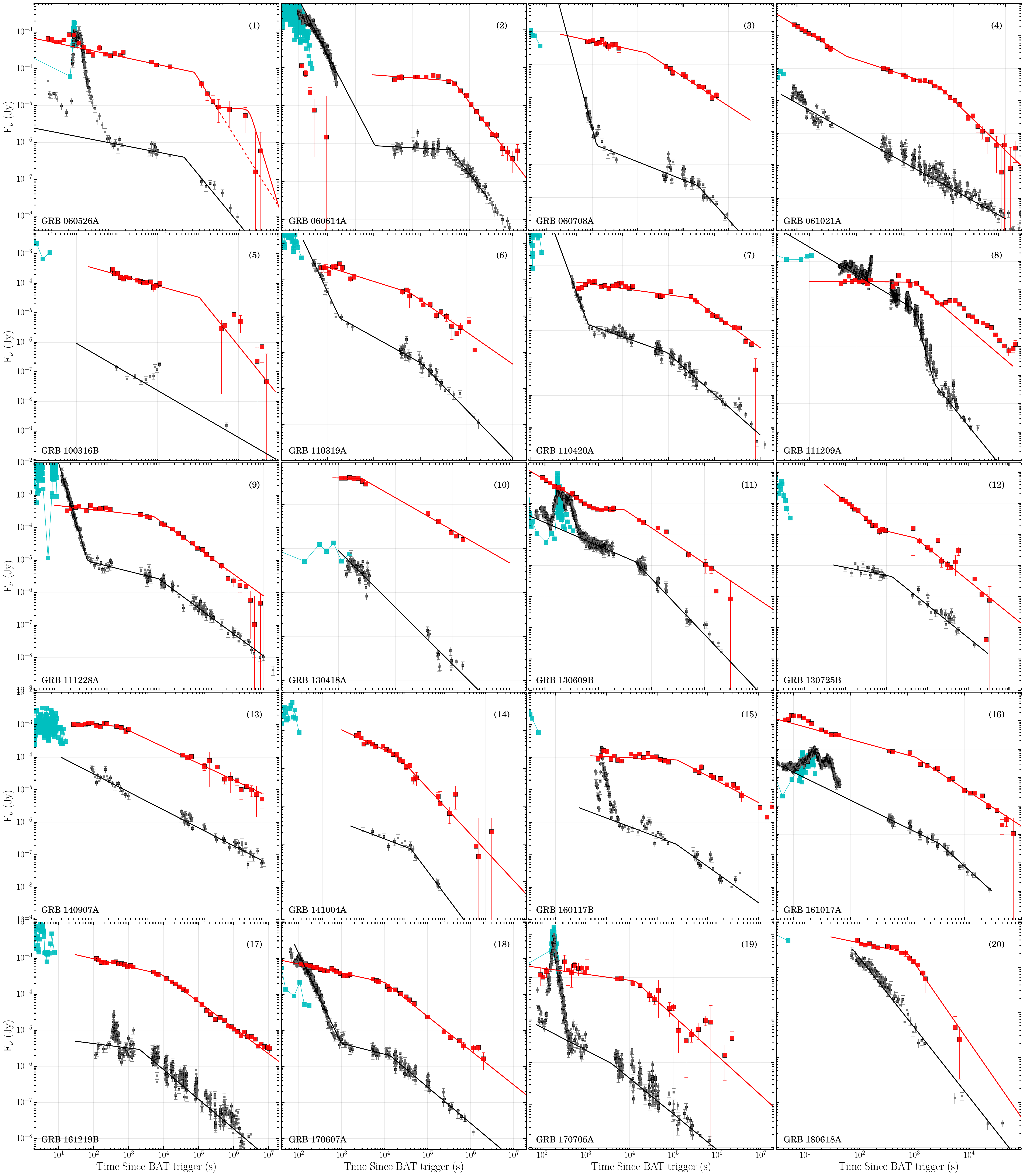}
\caption{Panels (1)-(20) represent the GRBs with at least one plateau and one of more breaks in their optical LC. The \textit{v}-band LCs observed by \swift-UVOT are denoted by red squares. Similarly, the X-ray LC at 10\,keV is represented by the black squares, while the BAT LCs are shown as cyan squares. The best-fit models for the X-ray and optical data are shown by solid red and black lines, respectively. The dotted line, if present, represents an extension of the model that would have fit the LCs in the absence of any deviation.}
\label{fig:plateau_GRBs1}
\end{figure*}

\begin{figure*}
\centering
\includegraphics[width=2\columnwidth]{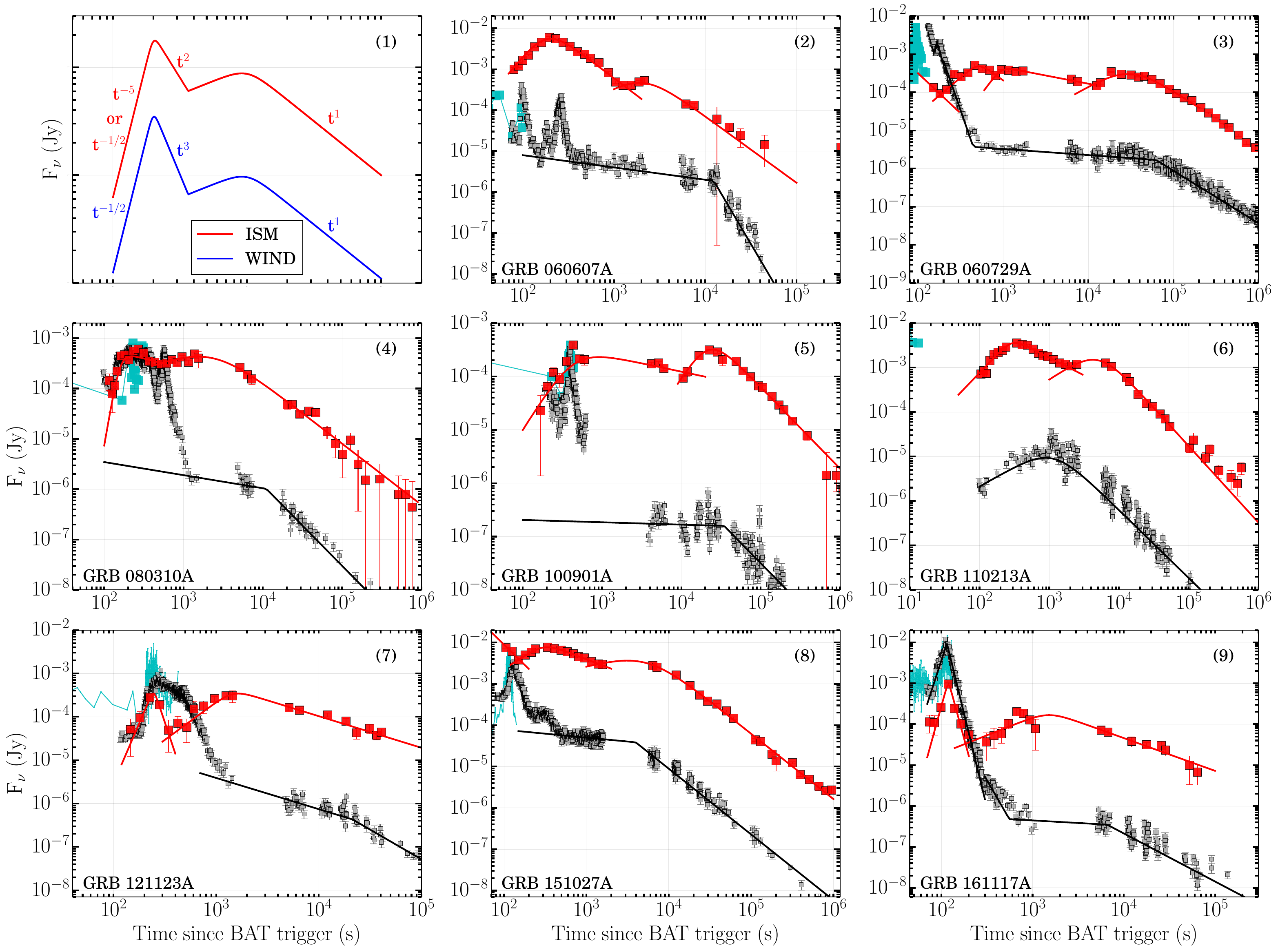}
\caption{Panel (1) represents the theoretical prediction of reverse shock emission from \citet{2015AdAst2015E..13G}. Panels (2)-(9) represent the GRBs with at least two bumps in the \textit{v}-filter LC. The UVOT, XRT, and BAT observations are shown in red, black, and cyan squares. The fitted models to the UVOT LC in \textit{v}-band and XRT LC at 10 keV are respectively shown with the solid red and black lines.}
\label{fig:RS_GRBs}
\end{figure*}

\begin{figure*}
\centering
\includegraphics[width=2\columnwidth]{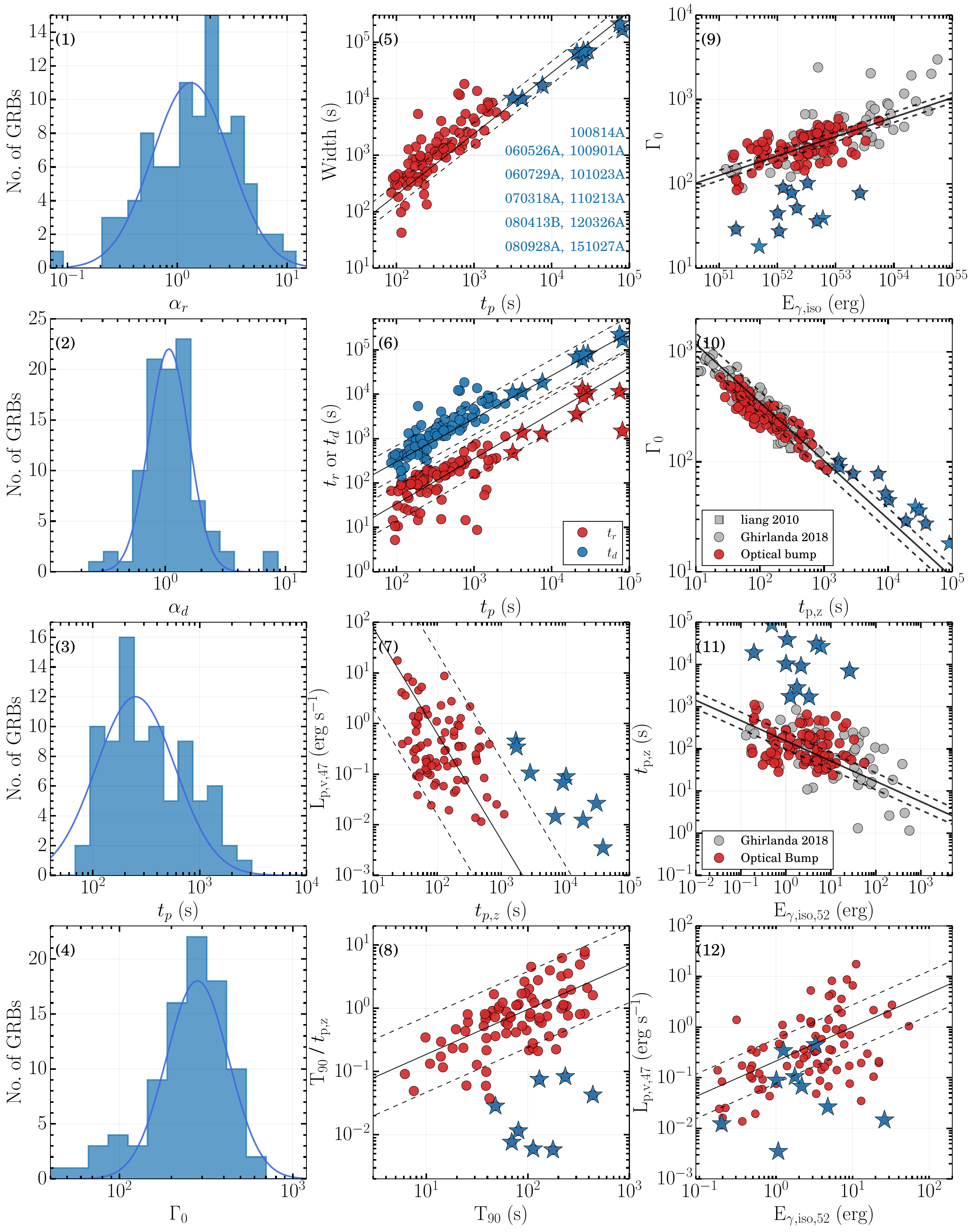}
\caption{Panels (1) to (12) represent the distribution and the correlation of parameters derived from the fitting of optical LCs (red circle and blue stars) with the signature of the onset bump. The observations taken from \citet{2018AA...609A.112G}, wherever used, are shown with grey circles in the background. Panels 1-3 present the distribution of parameters obtained from the fitting of UVOT LCs with a smoothly joined broken PL model. The distribution of the Lorentz factor derived using the $t_{p,z}$-$\Gamma_{0}$ relation is shown in panel (4). Panels (5) to (8) show the various correlations found between the peak time ($t_{p}$), width at FWHM, rise time ($t_{r}$), and decay time ($t_{d}$), in observer's frame. Panels (9) to (12) illustrate the correlations that exist between the bulk Lorentz factor $\Gamma_{0}$, $E_{\gamma,iso}$, $t_{p,z}$, and the observed peak luminosity ($L_{p,v}$) in the \textit{v}-band. The black solid lines represent the linear relationships between the parameters obtained by fitting a PL function, and the corresponding 2$\sigma$ significance levels of the fits are shown with dashed lines.}
\label{fig:onset}
\end{figure*}

\FloatBarrier
\onecolumn
\renewcommand{\arraystretch}{1.25}
\small
\begin{longtable}{|c|c|c|c|c|c|c|c|c|c|}
\caption{Lists the parameters obtained from the fitting of a simple PL model to the \swift-UVOT LCs in \textit{v}-band for GRBs in which the observed optical LCs are not showing any breaks or bumps. F-test and corresponding p-values are shown in columns 5 and 6. p-value $>$ 0.0027 indicates that adding one more break has no statistical significance. The parameters obtained from the fitting of X-ray spectra in 0.3-10\,keV, optical to X-ray SED and the results obtained from applying the closure relation are shown in the last two columns. Distance-dependent properties from redshift unknown GRBs are calculated by considering $z$ $\sim$ 2.}
\label{tab:pow0}\\ \hline

GRBs & T$_{90}$ (s) & z & $\alpha_{o,1}$ & F-Test & p-value  & $\beta_{x}$ & $\beta_{ox}$ & $p$ & Medium\\
\hline

\endfirsthead
\hline 
\multicolumn{10}{|c|}{{Continued on next page}} \\ \hline
\endfoot
\hline

GRBs & T$_{90}$ (s) & z & $\alpha_{o,1}$ & F-Test & p-value  & $\beta_{x}$ & $\beta_{ox}$ & $p$ & Medium\\
\hline

\endhead
\endlastfoot
050319A & 151.6 & 3.24 & $0.63_{-0.03}^{+0.03}$ & 0.95 & 0.4 & $1.01_{-0.05}^{+0.05}$ & $0.90_{-0.01}^{+0.01}$ & 2.01$\pm$0.10 & ISM\\
050712A & 51.4 & $\sim$2 & $0.62_{-0.05}^{+0.05}$ & 0.65 & 0.52 & $1.02_{-0.08}^{+0.08}$ & $0.91_{-0.02}^{+0.02}$ & 2.05$\pm$0.17 & ISM\\
050721A & 105 & $\sim$2 & $0.74_{-0.18}^{+0.43}$ & 1.22 & 0.32 & $0.67_{-0.11}^{+0.11}$ & $0.74_{-0.02}^{+0.02}$ & 2.34$\pm$0.21 & Wind\\
050726A & 46.5 & 0.165 & $0.78_{-0.34}^{+0.11}$ & -0.87 & 1 & $0.86_{-0.06}^{+0.06}$ & $0.71_{-0.01}^{+0.01}$ & 2.73$\pm$0.13 & ISM\\
050802A & 27.5 & 1.71 & $0.83_{-0.02}^{+0.02}$ & 2.87 & 0.07 & $0.81_{-0.04}^{+0.04}$ & $0.70_{-0.01}^{+0.01}$ & 2.63$\pm$0.08 & ISM\\
051109A & 37.2 & 2.35 & $0.79_{-0.02}^{+0.02}$ & 0.13 & 0.87 & $0.96_{-0.05}^{+0.05}$ & $0.61_{-0.01}^{+0.01}$ & 1.93$\pm$0.11 & ISM\\
060223A & 11.3 & 4.41 & $1.17_{-0.20}^{+0.18}$ & 0 & 1 & $0.84_{-0.14}^{+0.14}$ & $0.72_{-0.02}^{+0.02}$ & 2.69$\pm$0.28 & Wind\\
060502A & 28.4 & 1.51 & $0.66_{-0.05}^{+0.04}$ & -0.06 & 1 & $1.03_{-0.08}^{+0.08}$ & $0.67_{-0.01}^{+0.01}$ & 2.07$\pm$0.17 & ISM\\
\hline
\end{longtable}

\FloatBarrier
\small
\begin{longtable}{|c|c|c|c|c|c|c|c|c|c|c|c|c|}
\caption{Lists the parameters obtained from the fitting of a PL model with 1-break to the \swift-UVOT LCs in \textit{v}-band for GRBs in which the observed optical LCs consist of a break. F-test and corresponding p-values are shown in columns 7 and 8, p-value $<$ 0.0027 indicates that 1 break PL is preferred over a simple PL. The parameters obtained from the fitting of X-ray spectra in 0.3-10\,keV, optical to X-ray SED and the results obtained from applying the closure relation are shown in the last two columns. Distance-dependent properties from redshift unknown GRBs are calculated by considering $z$ $\sim$ 2.}
\label{tab:pow1}\\ \hline

GRBs & T$_{90}$ & z & $\alpha_{o,1}$ & $t_{o,b}$ & $\alpha_{o,2}$ & F-Test & p-value & $\beta_{x}$ & $\beta_{ox}$ & $p$ & Medium \\
 & (s) &  &  & (s) & & & &  &  &  &  \\
\hline

\endfirsthead
\hline 
\multicolumn{13}{|c|}{{Continued on next page}} \\ \hline
\endfoot
\hline

GRBs & T$_{90}$ & z & $\alpha_{o,1}$ & $t_{o,b}$ & $\alpha_{o,2}$ & F-Test & p-value & $\beta_{x}$ & $\beta_{ox}$ & $p$ & Medium \\
\hline

\endhead
\endlastfoot
050922C & 4.6 & 2.19 & 0.83$_{0.02}^{0.01}$ & 11482.91$_{2877.54}^{3058.78}$ & 1.24$_{0.06}^{0.08}$ & 18.05 & 1.00E-04 & $1.15_{-0.07}^{+0.07}$ & $0.98_{-0.01}^{+0.02}$ & 2.30$\pm$0.14 & ISM \\
060111B & 58.8 & $\sim$2 & $1.95_{-0.15}^{+0.18}$ & $1252.94_{-541.89}^{+873.83}$ & $0.42_{-0.25}^{+0.29}$ & 9.65 & 0.0024 & $1.04_{-0.11}^{+0.11}$ & $0.87_{-0.02}^{+0.02}$ & 2.09$\pm$0.21 & ISM\\
060512A & 8.4 & 0.44 & 0.84$_{0.02}^{0.02}$ & 21881.46$_{7118.5}^{9805.18}$ & 1.38$_{0.18}^{0.32}$ & 12.62 & 0.000284 & $1.76_{-0.11}^{+0.11}$ & $1.37_{-0.01}^{+0.02}$ & 3.52$\pm$0.23 & ISM \\
060526A & 298 & 5.11 & $0.39_{-0.02}^{+0.02}$ & $29692.05_{-7557.54}^{+5636.35}$ & $1.88_{-0.43}^{+0.57}$ & 23.75 &  $<10^{-4}$ & $0.85_{-0.08}^{+0.08}$ & $1.05_{-0.02}^{+0.03}$ & 2.71$\pm$0.17 & Wind\\
060614A & 109.1 & 0.125 & $-0.09_{-0.03}^{+0.03}$ & $42294.06_{-1762.24}^{+1951.54}$ & $1.91_{-0.06}^{+0.06}$ & 343.68 &  $<10^{-4}$ & $0.98_{-0.08}^{+0.08}$ & $0.69_{-0.01}^{+0.01}$ & 2.97$\pm$0.16 & Wind\\
060708A & 10 & 1.8 & $0.30_{-0.12}^{+0.18}$ & $1573.48_{-582.25}^{+6247.69}$ & $0.89_{-0.04}^{+0.09}$ & 35.86 &  $<10^{-4}$ & $1.06_{-0.07}^{+0.07}$ & $1.18_{-0.02}^{+0.02}$ & 2.12$\pm$0.14 & ISM\\
060912A & 5 & 0.937 & $1.07_{-0.03}^{+0.03}$ & $3086.45_{-529.59}^{+679.71}$ & $0.43_{-0.02}^{+0.02}$ & 136.09 &  $<10^{-4}$ & $0.69_{-0.24}^{+0.23}$ & $0.65_{-0.02}^{+0.02}$ & 2.38$\pm$0.48 & ISM\\
081008A & 187.8 & 1.97 & $0.76_{-0.01}^{+0.01}$ & $3579.38_{-399.40}^{+390.99}$ & $1.33_{-0.05}^{+0.05}$ & 81.33 &  $<10^{-4}$ & $0.91_{-0.07}^{+0.07}$ & $0.89_{-0.01}^{+0.02}$ & 2.82$\pm$0.15 & ISM\\
\hline
\end{longtable}
\FloatBarrier
\addtolength{\tabcolsep}{-1pt}
\small
\begin{longtable}{|c|c|p{0.6cm}|p{1.2cm}|c|c|c|c|c|c|c|c|c|}
\caption{Lists the parameters (rise index $\alpha_{r}$, decay index $\alpha_{d}$, break time $t_b$ and smoothness parameter s) obtained from the fitting of a smoothly joined broken PL model to the \swift-UVOT LCs in v-band for GRBs in which the observed optical LCs consist of at least a bump. F-test and corresponding p-values are shown in columns 8 and 9. The p-value $<$ 0.0027 indicates that the smoothly joined broken PL is preferred over a simple PL. The parameters obtained from the fitting of X-ray spectra in 0.3-10\,keV, optical to X-ray SED and the results obtained from applying the closure relation are shown in the last two columns. Distance-dependent properties from redshift unknown GRBs are calculated by considering $z$ $\sim$ 2.}
\label{tab:onset}\\ \hline

GRBs & T$_{90}$ & z & $\alpha_{r}$ & $t_{b}$ & $\alpha_{d}$ & s & F-Test & p-value & $\beta_{x}$ & $\beta_{ox}$ & $p$ & Medium\\ 
 & (s) & & &(s) & & & & & & & & \\ 
\hline
\endfirsthead
\hline 
\multicolumn{13}{|c|}{{Continued on next page}} \\ \hline
\endfoot
\hline

GRBs & T$_{90}$ & z & $\alpha_{r}$ & $t_{b}$ & $\alpha_{d}$ & s & F-Test& p-value & $\beta_{x}$ & $\beta_{ox}$ & $p$ & Medium\\ 
\hline

\endhead
\endlastfoot
050730A &155&3.97& $-0.56_{-0.22}^{+0.20}$ & $1608.96_{-448.26}^{+751.82}$ & $1.25_{-0.13}^{+0.16}$ &1&37.91&  $<10^{-4}$ & $0.56_{-0.02}^{+0.02}$ & $0.68_{-0.02}^{+0.02}$ & 2.12$\pm$0.05 & ISM\\
050801A &19.6& $\sim$2 & $-0.70_{-0.19}^{+0.18}$ & $174.51_{-17.45}^{+22.73}$ & $1.36_{-0.05}^{+0.06}$ &1&75.87&  $<10^{-4}$ & $0.82_{-0.16}^{+0.16}$ & $0.95_{-0.02}^{+0.02}$ & 2.65$\pm$0.33 & ISM\\
050820A &241&2.61& $-1.83_{-1.14}^{+0.95}$ & $594.11_{-214.43}^{+871.46}$ & $0.83_{-0.02}^{+0.02}$ &1&126.51&  $<10^{-4}$ & $0.88_{-0.03}^{+0.03}$ & $0.78_{-0.01}^{+0.01}$ & 2.76$\pm$0.06 & ISM\\
060206A &7.6&4.05& $-0.21_{-0.22}^{+0.14}$ & $2604.42_{-1125.41}^{+1683.81}$ & $1.21_{-0.07}^{+0.08}$ &1&78.39&  $<10^{-4}$ & $0.22_{-0.19}^{+0.20}$ & $0.90_{-0.02}^{+0.02}$ & 2.45$\pm$0.40 & ISM\\
060418A &109&1.49& $-1.83_{-0.14}^{+0.14}$ & $163.46_{-4.71}^{+5.24}$ & $1.18_{-0.01}^{+0.01}$ &1&583.93&  $<10^{-4}$ & $0.84_{-0.11}^{+0.10}$ & $0.97_{-0.02}^{+0.02}$ & 2.68$\pm$0.20 & ISM\\
060605A &79.8&3.8& $-0.81_{-0.24}^{+0.20}$ & $386.02_{-85.28}^{+136.78}$ & $1.02_{-0.08}^{+0.10}$ &1&168.72&  $<10^{-4}$ & $0.99_{-0.06}^{+0.06}$ & $0.80_{-0.02}^{+0.02}$ & 1.99$\pm$0.12 & Wind\\
060607A &103&3.08& $-2.21_{-0.10}^{+0.09}$ & $201.54_{-3.25}^{+3.47}$ & $1.57_{-0.03}^{+0.03}$ &3&126.33&  $<10^{-4}$ & $0.53_{-0.04}^{+0.04}$ & $0.71_{-0.02}^{+0.02}$ & 2.06$\pm$0.08 & Wind\\
060607A &103&3.08& $-0.75_{-0.39}^{+0.32}$ & $2109.29_{-283.17}^{+357.48}$ & $1.29_{-0.13}^{+0.16}$ &3&126.33&  $<10^{-4}$ & $0.53_{-0.04}^{+0.04}$ & $0.71_{-0.02}^{+0.02}$ & 2.06$\pm$0.08 & Wind\\
\hline

\end{longtable}

\begin{longtable}{|p{1cm}|p{0.45cm}|p{0.95cm}|p{1.6cm}|p{0.95cm}|p{1.8cm}|p{0.95cm}|c|c|c|c|c|c|}
\caption{Lists the parameters obtained from the fitting of a PL model with 2-break to the \swift-UVOT LCs in \textit{v}-band for GRBs in which the observed optical LCs consist of two breaks. F-test and corresponding p-values are shown in columns 8 and 9. The p-value $<$ 0.0027 indicates that the 2 break PL is preferred over the 2 break PL. The parameters obtained from the fitting of X-ray spectra in 0.3-10\,keV, optical to X-ray SED and the results obtained from applying the closure relation are shown in the last two columns. Distance-dependent properties from redshift unknown GRBs are calculated by considering $z$ $\sim$ 2.}
\label{tab:pow2}\\ \hline

GRBs & z & $\alpha_{o,1}$ & $t_{o,b1}$ & $\alpha_{o,2}$ & $t_{o,b2}$ & $\alpha_{o,3}$ & F-Test & p-value & $\beta_{x}$ & $\beta_{ox}$ & $p$ & Medium \\
 & & & (s) & & (s) & & & & & & & \\
\hline

\endfirsthead
\hline 
\multicolumn{5}{|c|}{{Continued on next page}} \\ \hline
\endfoot
\hline

GRBs & z & $\alpha_{1}$ & $t_{b1}$ & $\alpha_{2}$ & $t_{b2}$ & $\alpha_{3}$ & F-Test & p-value & $\beta_{x}$ & $\beta_{ox}$ & $p$ & Medium\\
\hline
\endhead
\endlastfoot
061021A & 0.35 & $0.97_{-0.05}^{+0.05}$ & 900.77$_{-199.24}^{+304.35}$ & $0.52_{-0.04}^{+0.06}$ & $52471.60_{-4916.40}^{+3845.19}$ & $1.54_{-0.09}^{+0.08}$ & 48.08 & $<10^{-4}$ & $0.87_{-0.04}^{+0.04}$ & $0.40_{-0.01}^{+0.01}$ & 1.80$\pm$0.08 & ISM\\
080319B & 0.94 & $1.84_{-0.01}^{+0.01}$ & 1721.23$_{-115.81}^{+123.58}$ & $1.38_{-0.02}^{+0.02}$ & 84149.9$_{-13079.1}^{+10733.1}$ & 1.87$_{-0.13}^{+0.16}$ & 8.15 & 0.00188 & $0.73_{-0.04}^{+0.04}$ & $0.70_{-0.01}^{+0.01}$ & 2.47$\pm$0.07 & ISM\\
110715A & $\sim$2 & $0.81_{-0.07}^{+0.08}$ & 2361.21$_{-803.42}^{+1461.71}$ & $0.31_{-0.04}^{+0.03}$ & 93163.1$_{-5728.83}^{+7985.84}$ & 1.76$_{-0.25}^{+0.34}$ & 19.87 & $<10^{-4}$ & $0.76_{-0.09}^{+0.13}$ & $0.86_{-0.01}^{+0.01}$ & 2.52$\pm$0.22 & Wind\\
130609B & 0.82 & $1.01_{-0.04}^{+0.04}$ & $867.45_{-102.37}^{+77.32}$ & $0.07_{-0.22}^{+0.18}$ & $3404.31_{-506.63}^{+1260.93}$ & $1.18_{-0.07}^{+0.06}$ & 48.96 & $<10^{-4}$ & $0.85_{-0.05}^{+0.05}$ & $0.59_{-0.01}^{+0.01}$ & 2.60$\pm$0.11 & Wind\\
130725B & $\sim$2 & $1.06_{-0.04}^{+0.04}$ & $719.14_{-108.41}^{+97.25}$ & $0.24_{-0.18}^{+0.47}$ & $4432.76_{-2777.05}^{+1986.83}$ & $1.05_{-0.13}^{+0.11}$ & 7.98 & 0.00248 & $0.74_{-0.07}^{+0.07}$ & $0.74_{-0.01}^{+0.02}$ & 2.48$\pm$0.16 & ISM\\
150403A & 2.06 & 1.32$_{-0.01}^{+0.01}$ & 1857.69$_{-194.53}^{+257.7}$ & 0.60$_{-0.06}^{+0.05}$ & 59465.41$_{-18542.45}^{+33284.66}$ & 1.04$_{-0.07}^{+0.09}$ & 17.38 & $<10^{-4}$ & $0.70_{-0.03}^{+0.03}$ & $0.58_{-0.01}^{+0.01}$ & 2.40$\pm$0.06 & ISM\\
170607A & 0.557 & 1.52$_{-0.23}^{+0.61}$ & 361.18$_{-57.53}^{+49.95}$ & 0.30$_{-0.02}^{+0.03}$ & 10969.5$_{-964.25}^{+3612.68}$ & 0.96$_{-0.02}^{+0.05}$ & 53.24 & $<10^{-4}$ & $0.97_{-0.04}^{+0.04}$ & $0.83_{-0.01}^{+0.01}$ & 1.95$\pm$0.09 & Wind\\
180314A & 1.445 & 1.72$_{-0.09}^{+0.09}$ & 406.11$_{-28.46}^{+40.49}$ & 0.92$_{-0.03}^{+0.03}$ & 32044.34$_{-7373.6}^{+9075.68}$ & 0.39$_{-0.06}^{+0.06}$ & 15.56 & 0.00005 & $0.77_{-0.07}^{+0.07}$ & $0.64_{-0.01}^{+0.01}$ & 2.54$\pm$0.14 & ISM\\
\hline
\end{longtable}

\begin{longtable}{|c|c|c|c||c|c|c|c||c|c|c|c|}
\caption{Fitted parameters of \swift-XRT light curves at 10\,keV for GRBs in our sample, using a simple power law model.}
\label{tab:xpow0}\\ \hline

GRB & $\alpha_{x,1}$ & F-Test & p-value & GRB & $\alpha_{x,1}$ & F-Test & p-value & GRB & $\alpha_{x,1}$ & F-Test & p-value\\ \hline
\endfirsthead
\hline 
\multicolumn{12}{|c|}{{Continued on next page}} \\ \hline
\endfoot
\hline
GRB & $\alpha_{x,1}$ & F-Test & p-value & GRB & $\alpha_{x,1}$ & F-Test & p-value & GRB & $\alpha_{x,1}$ & F-Test & p-value\\
\hline
\endhead
\endlastfoot
050712A & 0.93$_{-0.02}^{+0.02}$ &2.93&0.061& 081011A & 0.71$_{-0.03}^{+0.03}$ &0&1& 130605A & 0.80$_{-0.03}^{+0.03}$ &1.4&0.3\\
050721A & 1.59$_{-0.02}^{+0.02}$ &5.4&0.006& 090102A & 1.41$_{-0.01}^{+0.01}$ &5&0.009& 130610A & 1.31$_{-0.02}^{+0.02}$ &2.5& 0.096\\
050801A & 0.86$_{-0.04}^{+0.04}$ &0&1& 090123A & 1.48$_{-0.01}^{+0.01}$ &2&0.126& 131004A & 1.77$_{-0.05}^{+0.05}$ &0& 1\\
050820A & 1.31$_{-0.01}^{+0.01}$ &5.4&0.005& 090401B & 1.40$_{-0.01}^{+0.01}$ &1.9&0.14& 140129A & 0.61$_{-0.01}^{+0.01}$ &6.2& 0.003\\
050922C & 1.32$_{-0.01}^{+0.01}$ &3.4&0.039& 090426A & 1.09$_{-0.06}^{+0.06}$ &4.1&0.03& 140215A & 1.46$_{-0.01}^{+0.01}$ &0& 1\\
060111B & 1.12$_{-0.02}^{+0.02}$ &5.3&0.054& 100111A & 0.92$_{-0.02}^{+0.02}$ &3.3&0.05& 140506A & $1.04_{-0.02}^{+0.02}$ &3& 0.09\\
060206A & 0.88$_{-0.02}^{+0.02}$ &5.6&0.012& 100316B & 0.91$_{-0.03}^{+0.03}$ &5.7&0.028& 140907A & $0.91_{-0.02}^{+0.02}$ &5.1& 0.02\\
060223A & 1.44$_{-0.04}^{+0.04}$ &1.2&0.33& 100728B & 1.23$_{-0.02}^{+0.02}$ &2&0.15& 141020A & 1.29$_{-0.02}^{+0.02}$ &6.2& 0.007\\
\hline
\end{longtable}

\begin{longtable}{|c|c|c|c|c|c||c|c|c|c|c|c|}
\caption{Fitted parameters of \swift-XRT light curves at 10\,keV for GRBs in our sample, using a 1-break power law model. A few GRBs are also fitted by a smoothly joined broken power law, as shown in bold.}
\label{tab:xpow1} \\ \hline

GRB & $\alpha_{x,1}$ & $t_{x,b1}$ (s) & $\alpha_{x,2}$ & F-Test & p-value & GRB & $\alpha_{x,1}$ & $t_{x,b1}$ (s) & $\alpha_{x,2}$ & F-Test & p-value\\  \hline

\endfirsthead
\hline 
\multicolumn{12}{|c|}{{Continued on next page}} \\ \hline
\endfoot
\hline
GRB & $\alpha_{x,1}$ & $t_{x,b1}$ (s) & $\alpha_{x,2}$ & F-Test & p-value & GRB & $\alpha_{x,1}$ & $t_{x,b1}$ (s) & $\alpha_{x,2}$ & F-Test & p-value\\ 
\hline
\endhead
\endlastfoot
050726A & $0.31_{-0.07}^{+0.08}$ & $1463.08_{-160.36}^{+196.75}$ & $1.74_{-0.06}^{+0.06}$ & 40 & $<10^{-4}$ & 130408A & $0.39_{-0.11}^{+0.07}$ & $3294.26_{-453.62}^{+286.17}$ & $2.15_{-0.04}^{+0.04}$ & 43 & $<10^{-4}$\\
050802A & $0.33_{-0.04}^{+0.04}$ & $3915.90_{-232.57}^{+253.72}$ & $1.67_{-0.03}^{+0.03}$ & 118 & $<10^{-4}$ & 130420A & $0.65_{-0.03}^{+0.05}$ & $54962.80_{-11190.59}^{+124039.76}$ & $1.11_{-0.06}^{+0.16}$ & 7.7 & $<10^{-4}$\\ 
051109A & $0.66_{-0.05}^{+0.06}$ & $3524.67_{-858.40}^{+1534.96}$ & $1.14_{-0.02}^{+0.02}$ & 56 & $<10^{-4}$ & 130505A & $1.05_{-0.01}^{+0.01}$ & $20933.58_{-1678.88}^{+1942.28}$ & $1.48_{-0.02}^{+0.02}$ & 99 & $<10^{-4}$\\ 
060502A & $0.44_{-0.02}^{+0.02}$ & $25533.28_{-4818.24}^{+6058.21}$ & $1.01_{-0.04}^{+0.04}$ & 421 & $<10^{-4}$ & 130701A & $0.40_{-0.05}^{+0.05}$ & $452.21_{-47.74}^{+54.44}$ & $1.26_{-0.03}^{+0.03}$ & 26 & $<10^{-4}$\\ 
060526A & $0.3_{-0.06}^{+0.06}$ & $21554.04_{-2848.99}^{+2988.75}$ & $1.88_{-0.11}^{+0.12}$ & 20 & $<10^{-4}$ & 130725B & $0.26_{-0.06}^{+0.07}$ & $1354.98_{-224.09}^{+225.43}$ & $1.05_{-0.04}^{+0.04}$ & 19 & $<10^{-4}$\\ 
060607A & $0.36_{-0.01}^{+0.01}$ & $12712.80_{-265.46}^{+252.74}$ & $4.12_{-0.10}^{+0.10}$ & 200 & $<10^{-4}$ & 131002A & $5.92_{-0.51}^{+0.59}$ & $87.02_{-3.49}^{+3.42}$ & $0.9_{-0.02}^{+0.02}$ & 29 & $<10^{-4}$\\ 
060714A & $0.06_{-0.07}^{+0.11}$ & $2590.26_{-332.88}^{+668.36}$ & $1.16_{-0.03}^{+0.05}$ & 29 & $<10^{-4}$ & 131030A & $0.60_{-0.13}^{+0.07}$ & $690.00_{-108.67}^{+52.05}$ & $1.14_{-0.01}^{+0.01}$ & 19 & $<10^{-4}$\\ 
060804A & $0.41_{-0.12}^{+0.13}$ & $2146.97_{-498.27}^{+861.42}$ & $1.28_{-0.05}^{+0.05}$ & 34 & $<10^{-4}$ & 140102A & $1.25_{-0.01}^{+0.01}$ & $3800.63_{-1103.42}^{+1119.05}$ & $1.52_{-0.05}^{+0.05}$ & 10 & $<10^{-4}$\\
\end{longtable}
\begin{longtable}{|c|c|c|c|c|c|c|c|c|}
\caption{Fitted parameters of \swift-XRT light curves at 10\,keV for GRBs in our sample, using a 2-break power law model.}
\label{tab:xpow2} \\ \hline
GRB & $\alpha_{x,1}$ & t$_{x,b,1}$ & $\alpha_{x,2}$ & t$_{x,b,2}$ & $\alpha_{x,3}$ & F-Test & p-value\\  \hline
\endfirsthead
\hline 
\multicolumn{9}{|c|}{{Continued on next page}} \\ \hline
\endfoot
\hline
GRB & $\alpha_{x,1}$ & t$_{x,b,1}$ & $\alpha_{x,2}$ & t$_{x,b,2}$ & $\alpha_{x,3}$ & F-Test & p-value\\ 
\hline
\endhead
\endlastfoot
050319A & $0.37_{-0.04}^{+0.03}$ & $6418.49_{-1638.20}^{+786.96}$ & $0.85_{-0.06}^{+0.06}$ & $72804.00_{-6787.71}^{+18058.81}$ & $2.49_{-0.21}^{+0.47}$ & 32 & $<10^{-4}$\\ 
050525A & $0.41_{-0.38}^{+0.18}$ & $3546.27_{-1835.05}^{+2124.42}$ & $2.01_{-0.07}^{+0.08}$ & $101377.20_{-23326.66}^{+30486.14}$ & $0.72_{-0.20}^{+0.19}$ & 12 & 0.00018\\ 
050730A & $2.44_{-0.06}^{+0.06}$ & $294.98_{-3.11}^{+4.07}$ & $0.28_{-0.01}^{+0.01}$ & $9028.87_{-141.15}^{+145.59}$ & $2.81_{-0.03}^{+0.03}$ & 65 & $<10^{-4}$\\ 
060605A & $2.19_{-0.27}^{+0.36}$ & $238.25_{-32.14}^{+29.39}$ & $0.67_{-0.04}^{+0.04}$ & $8379.01_{-669.99}^{+667.85}$ & $1.86_{-0.08}^{+0.09}$ & 6.8 & 0.00202\\ 
060614A & $2.88_{-0.02}^{+0.02}$ & $2201.02_{-74.45}^{+78.7}$ & $0.08_{-0.05}^{+0.04}$ & $46982.4_{-2521.49}^{+1896.95}$ & $1.95_{-0.03}^{+0.04}$ & 34 & $<10^{-4}$\\ 
060708A & $4.14_{-0.13}^{+0.14}$ & $183.83_{-8.94}^{+9.51}$ & $0.68_{-0.05}^{+0.04}$ & $12542.07_{-3678.03}^{+2458.58}$ & $1.30_{-0.05}^{+0.05}$ & 106 & $<10^{-4}$\\ 
060729A & $6.44_{-0.21}^{+0.22}$ & $445.93_{-10.05}^{+10.21}$ & $0.16_{-0.02}^{+0.02}$ & $62458.68_{-2099.14}^{+2075.06}$ & $1.38_{-0.01}^{+0.01}$ & 144 & $<10^{-4}$\\ 
070420A & $4.27_{-0.09}^{+0.09}$ & $270.8_{-6.09}^{+7}$ & $-0.02_{-0.05}^{+0.05}$ & $2391.31_{-106.42}^{+125.13}$ & $1.43_{-0.02}^{+0.02}$ & 138 & $<10^{-4}$\\
\hline
\end{longtable}

\begin{longtable}{|c|c|c|c|c|c|c|c|c|c|}
\caption{Fitted parameters of \swift-XRT light curves at 10\,keV for GRBs in our sample, using a 3-break power law model.}
\label{tab:xpow3} \\ \hline

GRB & $\alpha_{x,1}$ & $t_{x,b,1}$ (s) & $\alpha_{x,2}$ & $t_{x,b,2}$ (s) & $\alpha_{x,3}$ & $t_{x,b,3}$ (s) & $\alpha_{x,4}$ & F-Test & p-value\\ \hline

\endfirsthead
\hline 
\multicolumn{8}{|c|}{{Continued on next page}} \\ \hline
\endfoot
\hline
GRB & $\alpha_{x,1}$ & $t_{x,b,1}$ (s) & $\alpha_{x,2}$ & $t_{x,b,2}$ (s) & $\alpha_{x,3}$ & $t_{x,b,3}$ (s) & $\alpha_{x,4}$ & F-Test & p-value\\ 
\hline
\endhead
\endlastfoot

061121A & $5.88_{-0.25}^{+0.26}$ & $198.87_{-4.77}^{+4.22}$ & $0.26_{-0.08}^{+0.09}$ & $1327.46_{-202.20}^{+954.59}$ & $0.86_{-0.03}^{+0.10}$ & $21346.77_{-2401.67}^{+3148.43}$ & $1.53_{-0.03}^{+0.03}$ & 23.0 & $<10^{-4}$\\
060510A & $2.88_{-0.59}^{+0.63}$ & $185.91_{-11.19}^{+14.5}$ & $-0.19_{-0.08}^{+0.07}$  & $811.42_{-97.75}^{+102.43}$ & $0.65_{-0.15}^{+0.03}$ & $19947.66_{-4984.26}^{+7636.01}$ & $1.58_{-0.19}^{+0.11}$ & 16.1 & $<10^{-4}$\\
090530A & $4.80_{-0.50}^{+0.53}$ & $148.34_{-8.35}^{+11.66}$ & $0.64_{-0.03}^{+0.02}$ & $57093.40_{-25281.17}^{+17399.18}$ & $1.39_{-0.51}^{+0.69}$ & $97541.88_{-43873.53}^{+70343.55}$ & $1.32_{-0.18}^{+0.16}$ & 10.8 & 0.00012\\ 
090618A & $4.04_{-0.21}^{+0.22}$ & $390.26_{-2.43}^{+1.90}$ & $0.82_{-0.14}^{+0.13}$ & $10102.47_{-197.01}^{+154.34}$ & $1.38_{-0.02}^{+0.02}$ & $14723.77_{-1503.62}^{+1208.38}$ & $1.47_{-0.01}^{+0.01}$ & 17.2 & $<10^{-4}$\\
100814A & $4.61_{-0.07}^{+0.07}$ & $582.98_{-14.69}^{+16.48}$ & $0.47_{-0.04}^{+0.03}$ & $25999.24_{-11285.73}^{+6758.91}$ & $0.64_{-0.04}^{+0.05}$ & $165691.65_{-10044.29}^{+10238.97}$ & $2.01_{-0.08}^{+0.08}$ & 7.7 & 0.00053\\ 
110102A & $4.11_{-0.86}^{+0.95}$ & $516.73_{-18.95}^{+24.11}$ & $0.52_{-0.03}^{+0.03}$ & $9944.08_{-1218.11}^{+1132.83}$ & $1.27_{-0.17}^{+0.04}$ & $70866.35_{-48960.76}^{+105264.17}$ & $1.44_{-0.07}^{+0.21}$ & 11.9 & $<10^{-4}$\\
161117A & $4.11_{-0.08}^{+0.08}$ & $516.34_{-14.76}^{+16.50}$ & $0.54_{-0.05}^{+0.04}$ & $8352.99_{-1899.08}^{+1436.87}$ & $1.06_{-0.06}^{+0.05}$ & $121506.49_{-56054.37}^{+53995.58}$ & $1.33_{-0.09}^{+0.12}$ & 24.0 & $<10^{-4}$\\
171209A & $1.30_{-0.04}^{+0.04}$ & $253.11_{-7.01}^{+5.11}$ & $3.62_{-0.18}^{+0.17}$ & $543.30_{-40.98}^{+43.36}$ & $0.77_{-0.07}^{+0.07}$ & $19062.16_{-1357.19}^{+1532.26}$ & $2.73_{-0.31}^{+0.43}$ & 9.0 & 0.00014\\
\hline
\end{longtable}

\FloatBarrier
\section{Individual GRB optical LCs showing deviations from simple PL Model}

\subsection{Details of LCs with the signature of early reverse shock} \label{sec:RS_signature}
Throughout the UVOT detection, the optical LC of GRB 080607A is too steep $\alpha_{o}$ $\sim$ 2.88 to be consistent with the external forward shock model. However, the X-ray LC of the burst is canonical, fitted with PL with 3 breaks, as shown in Fig. \ref{fig:RS2} and Table \ref{tab:xpow3}. One of the possible reasons for the steep decay in optical LC is that the early optical emission is coming from the tail of prompt emission from higher latitudes. The steep decay observed in optical LC is simultaneous with the steep decay phase of the X-ray LC, but the steep decay phase in X-ray LC lasts only up to 185\,s and is then transformed into the normal decay phase. Further, the early X-ray emission consists of flares that are not present in the optical LC, suggesting a different origin for optical emission. Another possibility is that jet breaks occur much earlier, causing a steep decay of the optical LC. However, if the effect is only geometric, then both X-ray and optical LCs should follow the same trend, but this scenario is also not supported by the canonical X-ray LC. Furthermore, reverse shock emission in the wind-like medium can also account for the steep decay ($\alpha_{o} ~ \sim ~ 3$) in the optical LC. The closure relation derived from the optical and X-ray spectral and temporal indices also favoured a wind-like medium surrounding the burst, see Table \ref{tab:pow0}. This conclusion is further supported by the afterglow analysis of \cite{2010ApJ...723L.218C} of the same burst. Our analysis suggests the reverse shock nature of the optical LC of GRB 080607A. \\

As shown in Fig. \ref{fig:RS_GRBs}, the optical LC of GRB 060607A consists of an early bump. The decay slope of the bump is $\alpha_{d} \sim 1.6$, which is slightly steeper than the prediction of forward shock and shallower than the prediction of reverse shock in the ISM-like medium \citep{2015AdAst2015E..13G}. At the same time, the early X-ray LC of the burst consists of several flares that overlap with BAT data. The observed optical bump is smooth and independent of the observed variability in the X-ray and $\gamma$-ray LCs. This indicates that the optical emission has a different origin (afterglow onset or reverse shock) from the early X-ray emission (internal shock).\\

Similarly, in the case of GRBs 060111B and GRB 080319B (see also, \citealt{2009ApJ...702..489R, 2008Natur.455..183R}), the early optical decay is steep ($\alpha_o$ $\sim$ 1.9), but the X-ray LCs of both the bursts follow a normal decay. After 1000\,s, the optical LCs of both the burst transformed to the normal decay phase $\alpha_{o}$ = 0.75. As shown in panels 1 and 3 of Fig. \ref{fig:RS2}, the afterglow emission is well separated from the prompt BAT data. This implies that internal shock contamination is negligible in both X-ray and optical emission. Therefore, the early optical LCs of both GRBs are uncorrelated with the BAT and XRT LCs, and the decay index is consistent with the reverse shock component in a thin shell regime in an ISM-like medium. This indicates that the early optical decay has a reverse shock signature that does not affect the X-ray emission.\\

Further, the observed optical LCs of GRB 060729A and GRB 151027A consist of two bumps \citep[see also,][]{2009ApJ...702..489R, 2017A&A...598A..23N}, as shown in Fig. \ref{fig:RS_GRBs}. The first and second bumps for both GRBs are consistent with the prediction of the onset of forward shock and late energy injection from the central engine, as discussed in the sections \ref{sec:onset} and \ref{sec:LRB}. In addition to this, the optical LCs of both have early steep decay ($\alpha_{d}$ $\sim$ 2) preceding the early onset bump. The observed decay index is consistent with the reverse shock in the ISM-like medium, but it is hard to draw any conclusion with only 2-3 data points.\\

Similarly, the early optical emission from GRB 080714A, GRB 110205A \citep{2011ApJ...743..154C}, GRB 121011A \citep{2016RAA....16...12X}, and GRB 121024A \citep{2016A&A...589A..37V} decays steeply, consistent with the reverse shock in the ISM or wind-like medium. At the same time, the decay of X-ray LCs is normal throughout the afterglow emission phase. These suggest that, in these GRBs, optical emission has an early reverse shock signature, but the entire X-ray emission is from the forward shock.\\

Furthermore, in the case of GRB 160417A, an initial steep decay was observed in both X-ray and optical LCs and is well separated from the prompt emission phase. Decay indices of optical and X-ray LCs are similar to 3, consistent with the reverse shock emission in a Wind-like medium. Therefore, for GRB 160417A, both X-ray and optical LCs could have a reverse shock component in the early phase of afterglow emission.\\

Finally, the early optical LC of GRB 100316B has an initial steep decay before the plateau phase. The observed decay index $\alpha_{d} \sim 3$ of the initial steep decay phase is consistent with the reverse shock emission in the wind-like medium. The X-ray LC of the burst does not have sufficient observations to favour or discard the reverse or internal shock origin of the optical LC.

\subsection{Details of LCs with the signature of internal shock} \label{sec:IS_signature}
The decay of early optical LC of GRB 061121A is steep with $\alpha_{d}$ $\sim$ 5 and overlaps with flares observed in the X-ray and BAT LCs, as shown in Fig. \ref{fig:RS2}. Decay in the early optical LC is too steep from the prediction of external reverse shock, and simultaneous observations of flares by BAT and XRT indicate an internal origin. Following the early steep decay, the optical LC also consists of an onset bump. The observed X-ray LC is canonical, with the early steep decay followed by a plateau and then a normal decay phase. This implies that the early optical and X-ray emissions are dominated by the internal shock, and the late multi-wavelength LCs have the forward shock origin \citep{2005Natur.435..178V, 2006ApJ...638L..71B}.\\

In the case of GRB 080310A, the optical LC consists of a very early steep decay followed by two subsequent smooth bumps \citep{2012MNRAS.421.2692L}. The first bump in the optical LC overlaps with giant flares in the X-ray and $\gamma$-ray LCs, but the observed bump is smooth with parameters consistent with the onset of forward shock. The second bump observed in optical is again a smooth bump, possibly due to late re-brightening. The slope of the early decay ($\alpha \sim 4$) is too steep to be consistent with the external reverse. Further, the coexistence of flares in optical and $\gamma$-/X-ray indicates the internal origin of the flare.\\

GRB 110119A, GRB 120327A \citep{2017A&A...607A..29M}, and GRB 150910A \citep{2020ApJ...896....4X} have the first two data points decaying with slopes steeper than 7. Following the initial steep decay, the optical LCs are followed by either an onset bump or a normal decay phase. The observed X-ray LCs are canonical. The steep decay phase observed in optical LC overlaps with the flares observed in BAT and XRT, which is consistent with its internal origin. Similarly, the complete flares observed in the early optical LCs of GRB 121123A and GRB 161117A preceding the onset bump are found to overlap with the bright flares observed in BAT and XRT and have an internal origin. Further, the early optical LC of GRB 161214B is variable and follows the variability in X-ray LCs, which might have an internal origin. 

\subsection{Details of LCs with the signature of internal plateau} \label{sec:internal_plateau2}
GRB 060526A \citep{2009MNRAS.395..490O} have a plateau ($\alpha_{o}$ $\sim$ 0.4) in the optical LC followed by steep decay $\alpha_{o}$ $\sim$ 1.88, see panel 1 of Fig. \ref{fig:LRB_plots}. The X-ray LC of GRB 060526A does not have a plateau, but a giant flare is observed simultaneously in $\gamma$-ray, X-ray, and optical LCs. Therefore, the observed erratic flare is consistent with the internal shock emission. Ignoring the overlapping giant flares, the X-ray LC also follows a 1-break PL decay during the afterglow phase, which is consistent with the energy injection to the external forward shock model. Additionally, a late re-brightening is also observed in the optical LC starting around 10$^5$ s post-burst, but due to large error bars, the origin of late re-brightening is uncertain. Overall, the optical and X-ray emissions are achromatic. The observed optical emission could have an internal shock origin throughout the emission phase. Except for the observed giant flare, the entire X-ray emission has an external origin.\\

Similarly, for GRB 060614A \citep{2007ApJ...670..565L}, an early steep decay consistent with the tail of prompt emission is observed at all wavelengths from $\gamma$-X-rays to optical. Following the steep decay phase, both the X-ray and optical LC of GRB 060614A have a plateau $\alpha$ $\sim$ 0, see Fig. \ref{fig:plateau_GRBs1}. The plateau phase in the optical LC is further followed by a steep decay phase with decay indices $\alpha_{o}$ $\sim$ 1.91, which suggests the internal nature of the observed plateau \citep{2007ApJ...665..599T}. In X-ray LC, the observed plateau is followed by a normal decay phase with $\alpha_{x}$ $\sim$ 1.22. Therefore, the observed multi-wavelength LCs of GRB 060614A are achromatic and are consistent with the internal shock emission from a long-lasting central engine.\\

Optical LC in GRB 111209A \citep{2013ApJ...766...30G} has various flares overlapping the plateau and the subsequent normal decay phase. The decay slope following the plateau phase has a slightly steeper index $\alpha$ = 1.42 than the normal decay. Further, the late optical emission is deviating from the PL fit due to the observed late re-brightening. Being a ULGRB, its long active central engine causes various flares in the optical band throughout the emission phase. X-ray emission also shows an internal plateau followed by a steep decay. Therefore, in the case of GRB 111209A, an internal plateau is possibly observed in both optical and X-ray LCs. Further, both optical and X-ray overlapped by several flares. This indicates that for this burst, the central engine remained active for at least 10$^{5}$s. This confirms that a plateau followed by steep decay has its origin in internal shock.\\

The optical plateau observed in GRB 180618A \citep{2022MNRAS.516.1584S} is followed by a steep decay \(\alpha_{o} \sim 1.77\). Not only the optical LC but the entire X-ray LC is showing a steep decay phase with decay index $\alpha_{x} \sim 1.76$. Both the X-ray and optical LCs are inconsistent with the external origin.

\subsection{Details of LCs showing a plateau with achromatic breaks} \label{sec:external_plateau1}

The optical LC of GRB 100316B exhibits a plateau phase lasting until approximately 10\,ks. The available X-ray LC appears to show a plateau phase simultaneous with the observed plateau phase optical LC. However, due to the lack of sufficient UVOT and XRT observations, it is challenging to constrain a precise break time in the X-ray and optical LCs. Following the plateau phase, both the optical and X-ray LCs transit into a normal decay phase, which is consistent with the energy injection model \citep{2006ApJ...642..389N, 2006ApJ...642..354Z, 2018ApJ...869..155S}.
 
Similarly, as shown in Fig. \ref{fig:plateau_GRBs1}, GRB 110319A, GRB 111228A \citep{2016ApJ...817..152X}, GRB 141004A, GRB 160117B, 161219B \citep{2018ApJ...862...94L}, and 170705A have plateaus observed in both X-ray and optical LC with achromatic breaks, followed by a normal decay phase. Some initial flares and steep decay are observed in X-ray LC, but in most cases, no such features are present in optical LC. As we have discussed in the main text (see section \ref{sec:SPL}), the early flares and steep decay are the internal contamination present in the early X-ray LC. The achromatic breaks in the optical and X-ray LCs are consistent with the energy injection model. 

Further, in the case of GRB 130609B, both optical and X-ray LCs also have achromatic breaks following the plateau phase. The optical plateau is followed by a normal decay phase with decay index $\alpha_{o}$ $\sim$ 1, but the final decay in the X-ray LC is too steep with $\alpha_{x}$ $\sim$ 2.4. The late steep decay in the X-ray LC is either due to the late jet break or due to the internal origin of the plateau, as discussed in section \ref{sec:internal_plateau}.

\subsection{Details of LCs showing a plateau with chromatic breaks} \label{sec:external_plateau2}
\subsubsection{GRB with optical LCs shows only a shallow decay throughout the afterglow phase}
The optical LCs of GRB 060510A and GRB 151114A show a shallow decay phase ($\alpha_{x} \sim 0.44$) throughout the afterglow emission phase. However, the X-ray afterglow has a smooth bump fitted by a smoothly joined broken PL model (see Table \ref{tab:xpow1}). The parameters obtained from the fitting of XRT LC are consistent with the onset of afterglow, discussed in section \ref{fig:onset}. The onset bump is possible in these case because X-ray emission is well separated from internal shock emission, and the tail of the prompt emission is not affecting the X-ray LC of GRB 060510A and GRB 151114A. Following the bump, X-ray emission is followed by a normal decay phase with $\alpha_{x} \sim 1$. The observed chromatic behaviour in the afterglow LCs of GRB 060510A supports the presence of multiple jet components to account for the optical and X-ray decay \citep{2005NCimC..28..439P}.

For GRB 080605A ($\alpha_{o} \sim 0.45$) and GRB 090929B ($\alpha_{o} \sim 0.26$), the shallow decay phase is observed in the optical LC at the early time. The optical LC is nearly flat compared to the X-ray LC ($\alpha_{x} \sim 0.8 $), which shows a normal decay phase followed by a jet break. The energy injection from the central engine is not favoured by the observed chromaticity in X-ray and optical LCs. In addition to this, the X-ray afterglow LC exhibits a steep decay around 10\,ks and 24\,ks, respectively, for GRB 080605A and GRB 090929B, which might be due to the jet break, but the jet break model is not favoured by optical emission. This discrepancy also suggests that the optical emission may have a different origin compared to the X-ray emission. GRB 081126A has a plateau in both optical and X-ray LCs, but the plateau in X-ray LC lasts only for 2\,ks followed by a normal decay phase, whereas in optical LCs, the plateau phase remains throughout.

For GRB 151215A \citep{2022MNRAS.516.1584S}, the optical LC shows only a shallow decay phase ($\alpha_{o} \sim 0.5$) through the detection limit of \swift-UVOT. The X-ray LC, on the other hand, decays following a normal decay ($\alpha_o$ $\sim$ 0.92) consistent with the forward shock in the external medium. The chromatic behaviour observed in optical and X-ray LCs is against the energy injection scenario. Multiple jet components could be the possible origin of the shallow decay phase in optical LC.\\

Similarly, the long plateau observed in the optical LC of GRB 160425A does not seem to be related to early $\gamma$-ray and X-ray emissions. Both the BAT and XRT LCs of GRB 160425A feature a huge simultaneous flare due to the erratic behaviour of the central engine, which is consistent with an internal shock origin. After the initial flare, the \swift-XRT LC decays normally with the decay index $\alpha_x$ = -1.5. Similarly, the optical LC of GRB 111225A has a long shallow decay phase, but the X-ray LC at the same time displays a steep decay phase followed by a normal decay. The observed discrepancies between the decay indices of optical and X-ray LCs suggest their different origin.

\subsubsection{GRB with optical LCs exhibit at least a break before or after the shallow decay phase}
The optical LCs of GRB 061021A \citep{2009MNRAS.395..490O} and GRB 130725B consist of a normal decay phase followed by a short plateau phase. Following the plateau phase, both the LCs again transit to the normal or steep decay phase. Therefore, the observed optical LCs of GRB 061021A and GRB 130725B are canonical. In contrast to the optical, the X-ray LCs of both the GRBs show only a normal decay phase throughout the afterglow phase. The observed chromatic behaviour demands their different origin. Similarly, the optical LCs of GRB 130418A, GRB 140907A, GRB 161017A and GRB 170607A have a plateau followed by normal decay. However, as shown in Fig. \ref{fig:plateau_GRBs1}, X-ray LCs of all four GRBs have initial flares followed by normal decay.\\

The optical LC of GRB 060708A \citep{2009MNRAS.395..490O} has a plateau followed by a normal decay phase. However, the observed X-ray LCs show canonical behaviour where the initial steep decay is followed by a normal decay phase and a jet break. In these bursts, the observed break in optical and X-ray LCs are chromatic, suggesting their different origin.\\

Further, the optical LC of GRB 110420A displays multiple overlapping flares throughout the plateau phase that last up to 30 ks, which is then followed by a normal decay phase. The X-ray LC, in this case, exhibited an initial bump with several overlapping flares. We modelled the observed X-ray bump using a smoothly joined broken PL, and the resulting parameters are consistent with the onset of afterglow. After the plateau or bump, the X-ray and optical LCs transitioned into a normal decay phase. However, the break in the X-ray LC occurred significantly earlier than in the optical LC, indicating that the breaks are chromatic.

\subsection{Details of LCs with Chromatic Late-rebrightening} \label{sec:cromatic_lrb}

Initially, the observed optical LC of GRB 080413B decays normally with a decay slope $\sim 1$. Following the normal decay phase, a late bump was observed in the optical LC of GRB 080413B, peaking around 10$^4$ s. In contrast to the optical, the X-ray LC of the burst decays normally throughout the afterglow phase with decay index $\alpha_{x}$ = 0.95. Following the bump, the decay index of the optical LC is $\alpha_{o} \sim 2.73$. The late re-brightening in the multi-band optical/NIR LC of GRB 080413B was also observed by \cite{2011A&A...526A.113F}. The observed feature in the afterglow LCs of GRB 080413B favours the two jet components.\\

In the case of GRB 101023A, the optical LC has a late bump, peaking around 10$^4$ s. Moreover, the observed X-ray LC is canonical and consists of a steep decay phase followed by a normal and a late steep decay phase. No flare or plateau was observed in the X-ray LC of GRB 101023A. Overall, observed optical and X-ray LCs are chromatic.\\

Similarly, the optical LC of GRB 110213A \citep{2013ApJ...774...13L} has a double bump. The first bump is consistent with the onset of afterglow, and the second smooth bump requires a different origin. Additionally, the X-ray LC of GRB 110213A also has a bump. The peak time of the bump in the X-ray LC is not aligned with any of the bumps observed in the optical LC. Following the bump, the X-ray LC decays normally with the decay index $\alpha_{x} \sim 1$, but the decay in the optical LC $\alpha_{o} \sim 1.72$ is steeper. The observed breaks in the optical and X-ray LCs are chromatic, which discards the energy injection model and favours the two jet components.\\

The observed breaks in the optical and X-ray LCs of GRB 080413B, GRB 101023A, and GRB 110213A are chromatic, discarding the energy injection model. In addition to this, in the case of GRB 080413B and GRB 110213A, the decay index following the bump is steeper than 1.5, which favours the multiple jet model as the origin of the observed bump in these GRBs \citep{2011A&A...526A.113F}. For GRB 101023A, due to the larger error bars associated with the late optical LC (see 5 panel of Fig. \ref{fig:LRB_plots}), it is hard to draw any conclusion in favour or against the multiple jet model.

\label{lastpage}
\end{document}